\documentclass[%
 reprint,
balancelastpage,
 amsmath,amssymb,
 aps,
 pre,
 colorlinks
]{revtex4-2}

\usepackage{graphicx}
\usepackage{dcolumn}
\usepackage{bm}
\usepackage[usenames,dvipsnames]{color}
\usepackage{natbib}
\usepackage{amsmath}
\usepackage{amssymb}
\usepackage{ dsfont }
\usepackage[urlcolor=NavyBlue,citecolor=NavyBlue,linkcolor=NavyBlue]{hyperref}

\newcommand{\mean}[1]{\left < #1 \right >}

\renewcommand{\vec}[1]{\mathbf{ #1 }}

\makeatletter
\renewcommand{\thesection}{\Roman{section}}
\renewcommand{\thesubsection}{\Alph{subsection}}
\renewcommand{\p@subsection}{\thesection.}
\renewcommand{\p@subsubsection}{\thesection.\thesubsection.}
\makeatother

\graphicspath{{figs/}}

\begin{document}

\title{The random walk of intermittently self-propelled particles}

\author{Agniva Datta$^{\mbox{\href{https://orcid.org/0000-0002-9856-1652}{\includegraphics[scale=1.7]{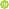}}}}$}
\author{Carsten Beta$^{\mbox{\href{https://orcid.org/0000-0002-0100-1043}{\includegraphics[scale=1.7]{orcid_id}}}}$}
\author{Robert Gro{\ss}mann$^{\mbox{\href{https://orcid.org/0000-0002-9817-2032}{\includegraphics[scale=1.7]{orcid_id}}}}$}
\email[Correspondence should be addressed to ]{rgrossmann@uni-potsdam.de, beta@uni-potsdam.de}

\affiliation{Institute of Physics and Astronomy, University of Potsdam, D-14476 Potsdam, Germany}

\begin{abstract}
Motivated by various recent experimental findings, we propose a dynamical model of intermittently self-propelled particles:~active particles that recurrently switch between two modes of motion, namely an active run-state and a turn state, in which self-propulsion is absent.
The durations of these motility modes are drawn from arbitrary waiting-time distributions.
We derive the expressions for exact forms of transport characteristics like mean-square displacements and diffusion coefficients to describe such processes.
Furthermore, the conditions for the emergence of sub- and superdiffusion in the long-time limit are presented. 
We give examples of some important processes that occur as limiting cases of our system, including run-and-tumble motion of bacteria, L\'{e}vy walks, hop-and-trap dynamics, intermittent diffusion and continuous time random walks. 
\end{abstract}

\date{\today}
\maketitle

\section{Introduction}

%
The motility and transport properties of active particles, that are characterized by their ability to self-propel their motion in a dissipative environment, has been subject to intense research over the last decade~\cite{romanczuk_active_2012,elgeti_physics_2015,bechinger2016active,shaebani_computational_2020}. 
Examples include diverse systems, such as bacteria~\cite{berg_chemotaxis_1972,berg_ecoli_2004}, algae~\cite{drescher_direct_2010,contino_microalgae_2015}, cells~\cite{schienbein1993langevin,mogilner_mathematics_2009,li_dicty_2011,aranson_physical_2016}, sperm~\cite{friedrich_chemotaxis_2007,jikeli_sperm_2015}, active colloids and Janus particles~\cite{paxton_catalytic_2004,howse_self_2007,buttinoni_active_2012,aranson_active_2013,muraveva2022interplay} as well as colloidal rollers driven by the Quincke effect~\cite{bricard2013emergence,pradillo2019quincke,kato2022active}. 
In many of these systems, the perpetual energy conversion into kinetic energy leads to persistent, uniform self-propelled motion in homogeneous environments.

%
Many active swimmers were reported to have multiple modes of motility:~the straight motion~(\textit{run motility}) of flagellated bacteria like \textit{Escherichia coli} is intermittently interrupted by sudden changes of their direction of motion~(\textit{tumbling})~\cite{berg_chemotaxis_1972,berg_ecoli_2004,santra_run_2020,kurzthaler_characterization_2024,lorentzgas}; other bacteria possess several motility modes depending on the flagellar arrangement with respect to the cell body~\cite{kuehn_bacteria_2017,hintsche2017polar,grognot_more_2021,thormann_wrapped_2022}.
Cells, that can recurrently repolarize thereby changing their direction of motion, were reported to spontaneously transition between different modes of motion, differing in their speed and persistence~\cite{moreno2020modeling,moldenhawer_spontaneous_2022,lepro_optimal_2022}. 
Recently, the chemokinetic response of mammalian sperm was found to switch between a chiral, active state and a hyperactive state characterized by random reorientations~\cite{zaferani2023biphasic}. 
%
%
At a macroscopic scale, intermittent active locomotion was suggested to emerge in flocks as a result of transient polarization~\cite{strefler2008swarming,grossmann2012active}---it may be observed in various species~\cite{kramer2015behavioral}, e.g.~grazing sheep~\cite{gomez2022intermittent} or schools of fish~\cite{tunstrom_collective_2013,lecheval_social_2018}.

%
Recently, there has been an increasing interest in active particle motility in disordered environments~\cite{bechinger2016active}. 
As a consequence of a (potentially randomly) structured medium, the motility pattern of self-propelled particles changes. 
Swimming bacteria, for example, were theoretically suggested~\cite{chepizhko2013diffusion} and experimentally observed to exhibit a hop-and-trap motility pattern in random media~\cite{bhattacharjee2019bacterial}:~while actively moving through porous spaces, they may get intermittently and transiently trapped. 
From an abstract point of view, this is an example of two-phase motion, in which periods of self-propulsion are interrupted by episodes without self-propulsion that lead to reorientations of the direction of motion. 
Similar motility patterns arise in different contexts, such as bacterial swimming close to surfaces~\cite{perez2019bacteria} and walls~\cite{figueroa20203d} as well as in microfluidic confinements~\cite{raza2022anomalous} or pillar arrays~\cite{raatz2015swimming,jakuszeit2024role}. 
%
Quincke rollers are also known to show this type of switching behaviour in presence of intermittent electric fields~\cite{karani2019tuning}.

%
From a theoretical point of view, the above-mentioned systems share a common feature:~several transport phases alternate---the motion in space is subordinated to a random switching process in time. 
An example of this compound random process is the so-called \textit{mobile-immobile model}~\cite{mora2018brownian,doerries2022rate}:~particles switching between~(non-thermal) diffusion and an immobile state. 
Its phenomenology is very rich:~it shows transient anomalous diffusion~\cite{doerries2022apparent} and Fickian yet non-Gaussian transport~\cite{mora2018brownian,doerries2022rate}, even in the case of exponentially distributed switching times~\cite{doerries2022apparent,doerries2023emergent}.
Due to its generality, it has many applications, ranging from solid-state physics~\cite{kurilovich2022nonmarkovian} to biophysics~\cite{igaev2014refined}.
The same process in the presence of drift shares structural similarities to active stop-and-go motion in one dimension~\cite{doerries2023emergent,peruani2023active}.

%
For the specific case of self-driven particles, descriptive models were developed to predict the large-scale transport of bacterial hop-and-trap dynamics in disordered environments based on experimental data~\cite{perez_impact_2021}. 
Experimental insights into the interaction of microswimmers with convex walls~\cite{raatz2015swimming,jakuszeit2024role} were moreover used in modeling to construct rate equations for the recurrent switching between free flights in cavities, sliding along walls and trapping~\cite{mattingly2023bacterial,saintillan2023dispersion}. 
More recently, active stop-and-go processes have been analyzed with regard to the question of optimal spatial exploration~\cite{peruani2023active}, similarly addressed in the context of driven polymers~\cite{kurzthaler2021geometric} as well as in simulations of bacterial locomotion~\cite{lohrmann_optimal_2023} in porous media. 
Several theoretical concepts, particularly applying renewal theory, were developed to understand the random transport of active particles with run-and-turn motility and its multi-state generalizations~\cite{portillo2011intermittent,thiel2012anomalous,zaburdaev_levy_2015,detcheverry2017generalized,kurzthaler2021geometric,salgado2022active,peruani2023active,jung2023hyperdiffusion,zhao_quantitative_2024,kurzthaler_characterization_2024,angelani2024anomalous}.

In this work, we propose a general model of self-propelled particles whose motility intermittently alters between two distinct modes of movement, namely an active, self-propelled state and a turn state. 
The two modes of motility may stand for different types of dynamics in actual systems, for example stop-and-go motility, run-and-tumble and hop-and-trap motility, or related forms of a general mobile-immobile dynamics as described above. 
Throughout the manuscript, we will oftentimes adopt the terminology of \textit{run} and \textit{turn} to be specific, keeping in mind, however, that the model applies to a more general type of intermittent, actively driven diffusion process.  
We describe the transport properties of intermittently self-propelled particles in terms of their velocity auto-correlation and, in particular, the mean-square displacement. 
The theoretical approach is routed in renewal theory---the times spent in the two motility modes are sampled from arbitrary waiting-time distributions---enabling us to derive these transport characteristics for arbitrary statistics of switching times between the active~(run) and turn state. 
This is relevant for several reasons.
We report how the statistics of lifetimes of these modes determines the type of long-time transport---in dependence on the tails of waiting-time distributions, either subdiffusion, superdiffusion, ballistic transport and also normal diffusion is observed in the long-time limit; for the case of normal diffusion, we specifically give an explicit formula for the effective long-time diffusion coefficient for arbitrary waiting-time distributions. 
Establishing the properties of intermittent self-propelled motion in homogeneous environments constitutes further the first step towards biased motion in external fields or chemical gradients. 
In the context of chemotaxis, it has already been shown for active particles with run-and-turn motility that not only the mean run-times but also the specific type of run-time distribution crucially determines the response to external chemical cues~\cite{celani_bacterial_2010,nava_markovian_2018,nava_novel_2020}.

%
This manuscript is structured as follows. 
The mathematical formulation of intermittently self-propelled particles
and its connection to related random transport processes is presented in Section~\ref{sec:model}.
Relevant quantities of interest for the characterization of the dynamics, such as the velocity auto-correlation function, the mean-square displacement and the diffusion coefficient are defined in Section~\ref{sec:def_msd} along with a general discussion on ergodicity breaking. 
In Section~\ref{sec:corr_func}, the central result, namely the correlation function is presented. 
The resulting mean-square displacement, both, under equilibrium and non-equilibrium conditions, a general expression for the effective long-time diffusion coefficient and the types of long-time dynamics, are discussed in Section~\ref{sec:analysis}.
Even though the main part of this manuscript considers intermittent self-propelled motion in two dimensions, the discussion is not limit to 2d---a generalization to arbitrary spatial dimensions is briefly shown in Section~\ref{sec:multidim}. 
Finally, we summarize the main conclusions of our work in Section~\ref{sec:summary}.

\section{Model}
\label{sec:model}

\paragraph*{Model dynamics}

The alternating temporal sequence of run and turn phases is described by a renewal process:~the run duration is drawn from a distribution denoted~$\psi_r(t)$; analogously, the duration of turn episodes is drawn from a waiting-time distribution~$\psi_t(t)$.

\begin{figure}
 \hspace*{-0.35cm}
 \includegraphics[width=0.974\columnwidth]{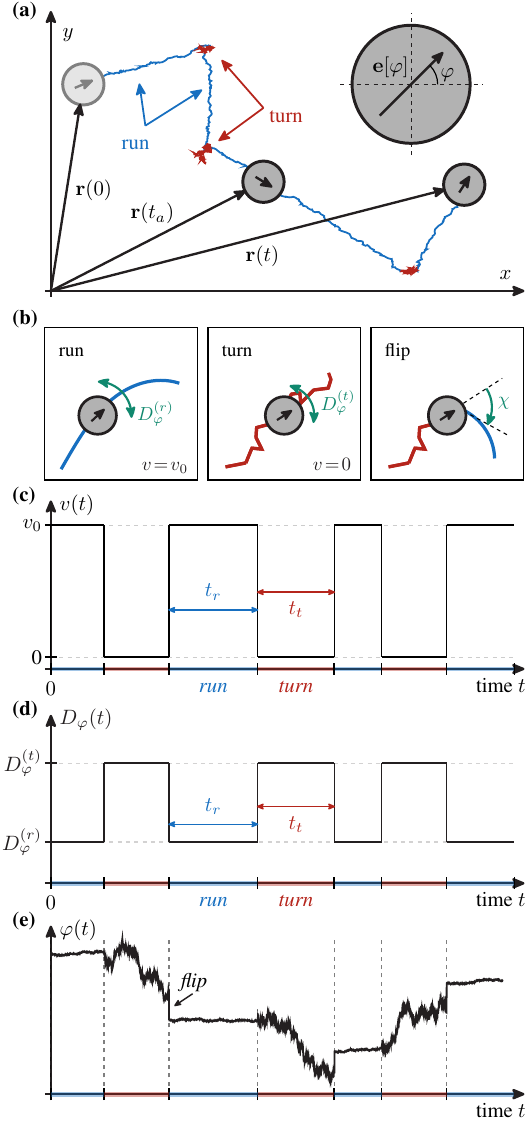}
 \vspace*{-0.3cm}
 \caption{Illustration of the model dynamics. In panel~(a), a schematic trajectory is drawn; run and turn episodes are highlighted in blue and red, respectively. The process is observed after the aging time~$t_a$. Panel~(b) shows different changes of the orientation vector~$\vec{e}$~[cf.~inset in panel~(a)]:~the rotational diffusion coefficient depends on the motility state~(first two panels); the last panel illustrates instantaneous reorientations by an angle~$\chi$~(flip). The random switching of speed and rotational diffusion is depicted in panels~(c) and~(d), respectively. The times~$t_r$ and~$t_t$ are drawn from the waiting-time distributions~$\psi_r(t)$ and~$\psi_t(t)$. Panel~(e) illustrates a time series of the direction of motion~$\varphi(t)$, where the effect of different rotational diffusion coefficients and sudden flips are visible by the variance of temporal fluctuations and instantaneous jumps. }
 \label{fig:cartoon}
\end{figure}

The spatial dynamics of particles is given by the following Langevin equations~(cf.~Fig.~\ref{fig:cartoon}): 
\begin{subequations}
\label{eq:runeq}
\begin{align}
\dot{\vec{r}}(t) & = v(t) \!  \begin{pmatrix} \cos \varphi(t) \\ \sin \varphi(t) \end{pmatrix} \!\!\: + \sqrt{2D} \, \boldsymbol{\xi}(t) , \label{eq:run_speed} \\  
\dot{\varphi}(t) &= \sqrt{2D_{\varphi}(t)} \, \eta (t) + \zeta_{\chi}(t). \label{eq:run_diffr}
\end{align} 
\end{subequations}
The position of a particle at time~$t$ is denoted by~$\vec{r}(t)$ in Eq.~\eqref{eq:run_speed}. 
The speed~$v(t)$ and rotational diffusion coefficient~$D_{\varphi}(t)$ generally depend on the mode of motility. 
Here, we focus on two dimensions---a generalization to arbitrary dimensions is straightforward as discussed Section~\ref{sec:multidim}.

In the active~(run) state, particles are assumed to be driven along a preferred body axis~$\vec{e} = (\cos \varphi, \sin \varphi)$ with a constant self-propulsion force, implying runs at a characteristic speed~$v(t) = v_0$, and the orientation of the body axis undergoes rotational diffusion~[Eq.~\eqref{eq:run_diffr}] with intensity~$D_{\varphi}(t) =D_{\varphi}^{(r)}$. 
We point out that rotational diffusion of self-propelled particles is not necessarily of thermal origin but depends on fluctuations of the driving force as well; the inverse rotational diffusion coefficient parametrizes the persistence time of runs~$\tau_r= 1/D_{\varphi}^{(r)}$.

In the turn state, self-propulsion is absent:~$v(t) = 0$.
Furthermore, we assume that the rotational diffusion coefficient in this state could differ from the run state: $D_{\varphi}(t) = D_{\varphi}^{(t)}$. 
Note that the spatial dynamics~$\dot{\vec{r}}(t)$ decouples from the stochastic rotations of the body axis~$\dot{\varphi}(t)$ in turn phases.

The dynamics of Eqs.~\eqref{eq:runeq} further includes isotropic Brownian motion with a diffusion coefficient~$D$. 
The independent random processes~$\boldsymbol{\xi}(t)$ and~$\eta(t)$ denote Gaussian white noise with zero mean and temporal $\delta$-correlations.

Additionally, abrupt \emph{reorientations~(flips)} are included in the angular dynamics by non-Poissonian shot-noise~$\zeta_{\chi}(t)$ in Eq.~\eqref{eq:run_diffr}~\cite{pohl2017inferring,seyrich2018statistical,kurzthaler_characterization_2024}:~whenever a new run phase begins, the particle is randomly reoriented with respect to the last orientation of the body axis,
\begin{align}
	\varphi \rightarrow \varphi + \chi, 
\end{align} 
where the angle~$\chi$ is drawn from the $2\pi$-periodic, symmetric distribution of angles~$p(\chi)$. 
In this way, the model includes two types of reorientation dynamics as limiting cases that were considered previously~\cite{schnitzer_theory_1993,saragosti_modeling_2012}, namely instantaneous flips without rotation~[$D_{\varphi}^{(t)}=0$ and arbitrary~$p(\chi)$], and reorientation by rotational diffusion without abrupt angular changes~[$p(\chi) = \delta(\chi)$ and $D_{\varphi}^{(t)}>0$]. 
In the latter case, there is a correlation of the reorientation angle and the duration of the turn phase. 
In any case, the modeling approach allows to include correlations of the direction of motion of two subsequent active episodes, e.g.~for a highly persistent dynamics if $p(\chi)$ is peaked around zero or anti-persistent motion~(velocity reversals~\cite{grossmann2016diffusion,fedotov2016single,giona2022extended}) when~$p(\chi)$ is peaked around~$\pm \pi$.
The correlation of run orientations is deleted by turns if the angle~$\chi$ is uniformly distributed or if~$D_{\varphi}^{(t)} \rightarrow \infty$.

As initial condition, we assume that a particle starts at~$t=0$ in the run state~(Fig.~\ref{fig:cartoon}). 
The observation of the process, however, begins at the so-called \textit{aging time}~$t_a$.
Note that the begin of an observation will generally neither coincide with the beginning of an run nor a turn episode, but the observation will rather start during one of these phases~(see Fig.~\ref{fig:cartoon} for an illustration and the Supplemental Material for technical implications~\cite{noteSI}). 
In this regard, we stress that the model dynamics is generally non-Markovian since the occurrence of transitions between the two motility modes is not a Poisson process~\cite{feller_an2_2008,klafter_first_2011}---the statistics for the forward waiting time, at which the next transition event will occur, depends generally also on the history of the process.

\begin{figure}[b]
\begin{center}
\includegraphics[width=\columnwidth]{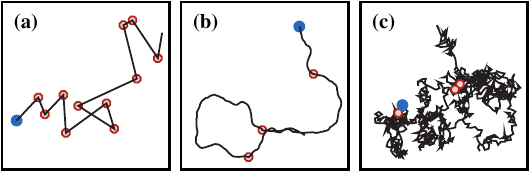}
\end{center}
\vspace*{-0.6cm}
\caption{The cartoons show three examples of different random walk patterns which are contained in the model~[Eq.~\eqref{eq:runeq}]. (a):~\textit{run-and-tumble} motion, characterized by highly persistent runs, interrupted by abrupt directional changes; (b):~\textit{self-propelled motion} at constant speed with a finite persistence time, interrupted by turns; (c):~\textit{mobile-immobile dynamics}~(see main text for a detailed description). In all cases, turn episodes are highlighted by red dots; the blue dot indicates the begin of a trajectory. } 
\label{fig:track_examples}
\end{figure}

\paragraph*{Relation to other random transport processes}

The model dynamics of intermittently self-propelled particles as introduced in the previous paragraph contains several well-known random transport processes as limiting cases, enabling a comparative study within one framework. 
The standard model of \textit{active Brownian particles} moving at a constant speed under the influence of rotational diffusion is recovered in the absence of the turn state, which is equivalent to infinitely short lifetime, $\psi_t(t) = \delta(t)$, and the absence of abrupt turns:~$p(\chi) = \delta (\chi)$~\cite{peruani2007self,romanczuk2011brownian,romanczuk_active_2012,sevilla_theory_2014}.

A second model class, contained in our framework, is \textit{active particles with run-and-turn} motility, commonly used to describe bacterial motility~\cite{berg_chemotaxis_1972,taylor_reversal_1974,johansen_variability_2002,berg_ecoli_2004,leonardy_reversing_2008,rashkov_model_2012,wu_periodic_2009,thutupalli_directional_2015,munoz2016myxobacteria,barbara_bacterial_2003,kuehn_bacteria_2017,beer_periodic_2013,duffy_turn_1997,davis_2d_2011,theves_bacterial_2013,hintsche2017polar,alirezaei_chemotaxis_2020,nava_novel_2020}. 
Typically, runs are considered to be highly persistent~(no rotational diffusion, $D_{\varphi}^{(r)} = 0$), and turn events are instantaneous:~$\psi_t(t) = \delta(t)$. 
If the run-time follows a power-law distribution with heavy tails, the resulting random process is a L\'{e}vy walk~\cite{zaburdaev_levy_2015}.

Our model is intimately related to models for intermittent active motion, such as bacteria in disordered environments showing \textit{hop-and-trap} behavior~\cite{bhattacharjee2019bacterial,bhattacharjee2021chemotactic,perez_impact_2021} or \textit{run-and-tumble}, where the small but finite tumble time is explicitly resolved~\cite{kurzthaler_characterization_2024,zhao_quantitative_2024}. 
Depending on the type of waiting time statistics and reorientation angles, the model dynamics reduces to L\'{e}vy walks, interrupted by stops~\cite{portillo2011intermittent} or isotropic Brownian diffusion~\cite{thiel2012anomalous}, as well as active stop-and-go motion~\cite{peruani2023active}. 
In the limit~$v_0 \rightarrow \infty$ and zero run-time~$t_r \rightarrow 0$, keeping, however, the stochastic run distance~$l = v_0 t_r$ finite, the particle dynamics becomes a \textit{continuous time random walk}~\cite{klafter_first_2011}.

Since the model~[Eq.~\eqref{eq:runeq}] includes both, isotropic diffusion of the center of mass as well as rotational diffusion, the modeling framework allows to study intermittent diffusion processes, i.e.~Brownian diffusion, during which the diffusion coefficient randomly changes between two different values~\cite{miyaguchi2016langevin,miyaguchi2019brownian}. 
This model is obtained for large rotational diffusion~$D_{\varphi}^{(r)}$ and high run speed, such that the run phase corresponds effectively to an active diffusion process with the diffusion coefficient~$D_{\text{eff}} = D + v_0^2/(2D_{\varphi})$. 
A particularly interesting dynamics emerges if the diffusivity~$D$ in the turn state vanishes, giving rise to isotropic diffusion with pauses, also referred to as \textit{mobile-immobile} dynamics~\cite{mora2018brownian,doerries2022rate,doerries2022apparent,doerries2023emergent}.

In short, the model introduced above provides a unifying framework that encloses various different types of random transport processes. 
As an example, three trajectories are compared in Fig.~\ref{fig:track_examples} for different settings.

\section{Definition of transport properties}
\label{sec:def_msd}

\paragraph*{Correlation function, mean-square displacement and long-time diffusion}

A central quantity to understand the characteristics of random transport is the ensemble-averaged mean-square displacement~(MSD), measuring the mean-squared distance which a particle traverses within a lag time~$\Delta$~\cite{metzler2014anomalous}:
\begin{align}
     \label{def:msd_time}
    m_{2}\! \left(t_a, \Delta\right) & = \mean{\left|\vec{r}\!\left(t_a+\Delta\right)-\vec{r}\! \left(t_a\right)\right|^{2}}. 
\end{align}
The brackets~$\langle \cdot \rangle$ denote ensemble averages over multiple realization of the stochastic process with identical initial conditions. 
Note that~$m_2$ will generally also be a function of the aging time~$t_a$. 
The MSD is directly related to the correlation function~$C \!\left(t_a,\Delta \right) = \mean{\dot{\vec{r}}(t_a) \cdot \dot{\vec{r}}(t_a + \Delta)}$ via
\begin{align}
    m_{2}\! \left(t_a, \Delta\right) 
    & = \int_{t_a}^{t_a + \Delta} \! d\xi \int_{t_a}^{t_a + \Delta}  \! d\xi' \mean{\dot{\vec{r}}(\xi) \cdot \dot{\vec{r}}(\xi')} \label{eqn:msd_time}\\
    & =  2 \!\int_{0}^{\Delta} d\xi \int_{0}^{\xi} d\xi' \, C \!\left(t_a+\xi', \xi-\xi'\right) \!. \nonumber 
\end{align}
In Laplace domain, the integrals are transformed into simpler algebraic expressions~\cite{doetsch_tabellen_1947,noteSI}
\begin{align}
    \widehat{\widehat{m}}_2 \left(s, u\right) = \frac{2}{u} \cdot \frac{\widehat{\!  \widehat{C}} (u, u ) - \widehat{\! \widehat{C}} (s, u )}{s-u} ,
    \label{eqn:msd}
\end{align}
in which~$s$ and~$u$ are the Laplace variables corresponding to~$t_{a}$ and~$\Delta$, respectively.

In the limit of long aging times~($t_a \rightarrow \infty$), the MSD is expected to become independent of the initial condition.
As discussed later in detail, this is, however, only the case if the waiting-time distributions for run and turn episodes, $\psi_r(t)$ and $\psi_t(t)$, possess a finite mean. 
Given this assumption, the process is said to equilibrate: 
\begin{align}
	m_2^{\text{(eq)}} (\Delta) = \lim_{t_{a} \rightarrow \infty} m_2 \! \left(t_{a}, \Delta \right) \! . \label{eqn:def_m2_eq}  
\end{align} 
The corresponding Laplace transform of the MSD under equilibrium conditions reads
\begin{align}
	\widehat{m}_2^{\text{(eq)}} (u) = \lim_{t_{a} \rightarrow \infty} \widehat{m}_2 \! \left(t_{a}, u\right) &= \frac{2}{u^{2}} \cdot \lim_{s \rightarrow 0} \left[s \cdot \,\widehat{\!\widehat{C}}\!\left(s, u\right)\right] \! .
     \label{eqn:msd_eq}
\end{align}
From the expression above, one can find the long-time diffusion coefficient~(in two dimensions) via
\begin{align}
	\mathcal{D} &= \frac{1}{4} \cdot \lim_{\Delta \rightarrow \infty} \frac{d m_2^{\text{(eq)}} \left(\Delta\right)}{d\Delta} = \frac{1}{4} \cdot \lim_{u \rightarrow 0} \left[u^{2} \cdot \widehat{m}_2^{\text{(eq)}}\! \left(u\right)\right] \!. 
    \label{eqn:diff_coeff}
\end{align} 
The discussion above underlines the relevance of the correlation function, from which the ensemble-averaged MSD and, in particular, the long-time diffusion coefficient, readily follow.

\paragraph*{Time-averaged MSD and ergodicity breaking}

Apart from the ensemble-averaged MSD, another important quantity that unravels transport properties is the time-averaged~MSD: 
\begin{align}
	\delta_T^2 \!\left(t_{a}, \Delta\right) = \frac{1}{T-\Delta} \int_{0}^{T-\Delta} \! d t \left | \vec{r}(t_a \! + t + \Delta) - \vec{r}(t_a \!+ t) \right |^2 \!. 
    \label{eqn:tamsd}
\end{align}
In the definition above, $T$ denotes the length of the observed trajectory. 
For each particle~(or realization), $\delta^2_T$ is a stochastic variable; the ensemble average of these reads
\begin{align}
    \mean{\delta^2_{T} \! \left(t_{a}, \Delta\right)} = \frac{1}{T-\Delta} \int_{0}^{T-\Delta} \! dt \, m_{2} \! \left(t_{a} +  t, \Delta\right) \!. 
    \label{eqn:ea_tamsd}
\end{align}
If the time-averaged MSD converges to the ensemble-averaged MSD in the limit~$\Delta/T \rightarrow 0$, the system is called ergodic~\cite{burov_aging_2010,barkai_2012_strange,metzler2014anomalous}. 
According to Eq.~\eqref{eqn:ea_tamsd}, time-averaged and ensemble-averaged MSD will generally differ---the dynamics of intermittently self-propelled particles is not ergodic. 
To gain physical intuition on the non-ergodicity, consider the situation that all particles are initially in the run state and~$t_a = 0$. The short-time ensemble-averaged MSD will scale as~$m_2 \simeq v_0^2 \Delta^2 + 4 D \Delta$~[cf.~Eq.~\eqref{eq:runeq}].  
In the course of time, each particle will also adopt the turn state; time-averages will also include these episodes, even for small lag times~$\Delta$, and consequently all time-averaged MSDs will be smaller than the ensemble-averaged MSD.

We conclude this section considering the specific case in which the equilibrium limit exists, implying that~$m_2(t_a,\Delta) \simeq m^{\text{(eq)}}_{2}(\Delta)$ for~$t_a \! \gtrsim t_{\text{eq}}$, where $t_{\text{eq}}$ denotes the equilibration time. 
The integration in Eq.~\eqref{eqn:ea_tamsd} can be split into two parts as follows, namely one over the interval~$t \in (0,t_{\text{eq}})$ and a second one for~$t \in (t_{\text{eq}},T - \Delta )$.  
The first integral is $T$-independent, whereas in the second integral, the integrand~$m_{2} \! \left(t_{a} +  t, \Delta\right)$ can be replaced by~$m^{\text{(eq)}}_{2}(\Delta)$ such that the integration becomes trivial: 
\begin{align}
\label{eqn:ergo}
    \mean{\delta^2_{T} \! \left(t_{a}, \Delta\right)} &\simeq \frac{1}{T-\Delta} \int_{0}^{t_{\text{eq}}} \! dt \, m_{2} \! \left(t_{a} +  t, \Delta\right) + m^{\text{(eq)}}_{2}(\Delta) \nonumber \\ 
    &=m^{\text{(eq)}}_{2}(\Delta) + \mathcal{O}(\Delta/T). 
\end{align}
The argument implies that the ensemble-average of the time-averaged MSDs~$\mean{\delta^2_{T} \! \left(t_{a}, \Delta\right)}$ equals the equilibrium MSD~$m^{\text{(eq)}}_{2}$ for sufficiently long trajectories~($T\rightarrow \infty$) given the equilibrium limit exists. 
Furthermore, this calculation reveals ergodicity in the equilibrium limit since the ensemble-averaged MSD~$m_2$ tends to the time-averaged MSD for~$t_a \rightarrow \infty$. 
In short, the dynamics of intermittently self-propelled particles is ergodic in equilibrium.

\section{The velocity auto-correlation function}
\label{sec:corr_func}

\begin{figure}[t]
\vspace*{0.6cm}
\begin{center}
 \includegraphics[width=\columnwidth]{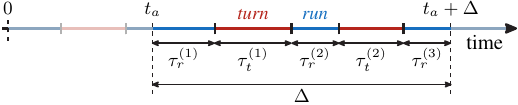}
\end{center}
\vspace*{-0.6cm}
\caption{The cartoon shows the temporal sequence of run and turn episodes for one particular realization of the diffusion process, as discussed in the context of the velocity auto-correlation function~[Eq.~\eqref{eqn:Cvv_time}]. The process starts at~$t=0$ and the observation begins at the aging time~$t_a$. Within the lag time~$\Delta$, there are~$N=2$ turn episodes. The total time spent in the two motility states are $\tau_r = \sum_{i=1}^{N+1} \tau_r^{(i)}$ and $\tau_t = \sum_{i=1}^{N} \tau_t^{(i)}$, respectively. The first and the last observed run is truncated~(censored). } 
\label{fig:cartoon_cvv}
\end{figure}

For the model dynamics described by Eq.~\eqref{eq:run_speed}, the correlation function~$C(t_a, \Delta) = \mean{\dot{\vec{r}}(t_a) \cdot \dot{\vec{r}}(t_a+\Delta)}$ is the sum of two contributions, 
\begin{align}
	C(t_a, \Delta) = C_{D}(t_a, \Delta) + C_{vv}(t_a, \Delta), 
\end{align}
one stemming from isotropic Brownian diffusion~$C_{D}(t_a,\Delta) = 4 D \delta \! \left( \Delta \right)$ and an active part from intermittent self-propulsion: 
\begin{align}
	\!\!C_{vv}(t_a, \Delta) \!=\! \mean{\!\!\: v(t_a) v(t_a \!+\! \Delta) \cos \! \Big [  \varphi (t_a \!+\! \Delta) \! - \! \varphi (t_a) \Big ] \! } \! . \!\! \!
\end{align}
In the tun state, the speed vanishes. 
Consequently, there is only a non-zero contribution to~$C_{vv}$ if a particle is in the active run state at both times,~$t_a$ and $t_a + \Delta$. 
Thus,~$C_{vv}(t_a,\Delta)$ is determined by the correlation of the orientation~$\varphi(t)$ and the times that a particle spends in the run and turn phase, respectively. 
Within a single, non-interrupted motility state, the orientational correlations decay exponentially~\cite{mikhailov_self_1997,peruani2007self,grossmann_anistropic_2015}: 
\begin{align}
	\mean{\cos \! \big [  \varphi (t + \tau) \! -  \varphi (t) \big ] } \! 
	&= e^{- D_{\varphi}^{(r,t)} \tau}. \label{eqn:decorrelation_single}
\end{align}
The full correlation function is obtained by averaging over all possible sequences of run and turn episodes: 
\begin{align}
    \label{eqn:Cvv_time}
	C_{vv}\! \left(t_a,\Delta\right) =\,\, & {v_0^2} \sum_{N=0}^\infty   
	\int_{0}^{\Delta} \! d\tau_{r} \! \int_{0}^{\Delta} \!d\tau_{t} \, \delta \! \left( \Delta - \left( \tau_r + \tau_t \right)  \right) \! \times \nonumber \\ 
	& \hspace*{-0.7cm} \times \! P_{N}\! \left(t_a, \tau_r, \tau_t\right)  e^{-D_{\varphi}^{(r)} \tau_r} e^{-D_{\varphi}^{(t)} \tau_t} \mean{\cos \chi}^{\!N} \!\! .
\end{align}
In the expression above, $P_{N}$ is the probability to observe exactly~$N$ turns in a time interval of length~$\Delta$ starting at time~$t_{a}$, during which the particle spends the time~$\tau_r$ running and the time~$\tau_t$ in the turn phase~(cf.~Fig.~\ref{fig:cartoon_cvv} for an illustration). 
This probability can be written in terms of the waiting time distributions in closed form in Laplace domain using renewal theory as shown in detail in the Supplemental Material~\cite{noteSI}, yielding an algebraic expression for the correlation function: 
\begin{widetext}
\begin{align}
    \label{eqn:Cvv_su}
    \widehat{\!\widehat{C}}_{vv}\! \left(s,u\right) \!\!\: = \!\!\: \frac{v_0^2}{\left(u+\!\!\:D_{\varphi}^{(r)}\!\!\:\right)} \!
\left [
	   \frac{1-\widehat{\psi}_{r} (s)}{s \! \left(1-\widehat{\psi}_{r}(s)\widehat{\psi}_{t} (s) \!\!\:\right )} 
	   \!-\! \frac{\widehat{\psi}_{r}\! \left(u+\!\!\:D_{\varphi}^{(r)}\!\!\:\right)-\widehat{\psi}_{r}(s)}{\left (s - u - \!\!\:D_{\varphi}^{(r)} \right) \! \left(1-\widehat{\psi}_{r}(s)\widehat{\psi}_{t}(s) \!\!\: \right)}
\! \cdot \! \frac{1- \mean{\cos \chi} \!\!\: \widehat{\psi}_{t} \! \left(u+\!\!\:D_{\varphi}^{(t)}\!\!\:\right)}{1-\mean{\cos \chi} \!\!\: \widehat{\psi}_{r} \!\left(u+\!\!\:D_{\varphi}^{(r)}\!\!\:\right) \! \widehat{\psi}_{t}\!\left(u+\!\!\:D_{\varphi}^{(t)}\!\!\:\right)}
\right] \! .
\end{align}
\end{widetext}
The derivation assumes that the diffusion process started with a run at~$t=0$~(Fig.~\ref{fig:cartoon_cvv}). 
If the particle was initially in the turn state, Eq.~\eqref{eqn:Cvv_su} is only modified by a multiplicative factor of~$\widehat{\psi}_t \!\left(s\right)$~\cite{noteSI}.

Note that not all parameters are equally relevant. The speed~$v_0$, for example, solely yields a multiplicative prefactor to the velocity auto-correlation function~$C_{vv}$.
Morever, Eq.~\eqref{eqn:Cvv_su} further reveals that the explicit form of the distribution of angles~$p(\chi)$ is of minor relevance---only the mean cosine of the angles~$\langle \cos \chi \rangle \in [-1,1]$~(first Fourier mode) enters the correlation function, exclusively via the last term. 
Physically speaking, it only matters whether particles have a tendency to reverse their direction of motion~($ \langle \cos \chi \rangle < 0$), keep on moving in the same direction as before~($\langle \cos \chi \rangle > 0$) or ``forget'' their previous direction of motion~($\langle \cos \chi \rangle = 0$); whether the underlying distribution~$p(\chi)$ is uni- or multimodal is not significant for the diffusion properties~\footnote{Higher order moments, like the kurtosis of the displacement distribution, will, in contrast, depend on details of the distribution of reorientation angles; similar arguments apply to the structure of trajectories. }.

At a technical level, Eq.~\eqref{eqn:Cvv_su} constitutes the main result of the paper. 
From the velocity auto-correlation function in Laplace domain, one can immediately derive the MSD, discuss the existence of the equilibrium limit and find the long-time diffusion coefficient by application of Eqs.~(\ref{eqn:msd}-\ref{eqn:ea_tamsd}) for arbitrary run- and turn-time distributions as discussed in Section~\ref{sec:analysis}.

In order to make practical use of Eq.~\eqref{eqn:Cvv_su} in terms of predictions for real-world systems, the four model parameters~(speed~$v_0$, mean angle~$\langle \cos \chi \rangle$ and two rotational diffusion coefficients~$D_{\varphi}^{(r,t)}$) as well as the waiting-time distributions~$\psi_{r,t}(t)$, entering via their Laplace transforms~$\widehat{\psi}_{r,t}$ evaluated at~$D_{\varphi}^{(r,t)}$, need to be determined from experimental data. 
Based on single-cell tracking and a subsequent classification of tracked positions into run and turn episodes, all these parameters can indeed be measured.

\section{Predictions} 
\label{sec:analysis}

In the preceding Sections, we established the dependence of the MSD on the correlation function~(Section~\ref{sec:def_msd}) and calculated the velocity auto-correlation function for intermittently self-propelled particles~(Section~\ref{sec:corr_func}). 
In this section, we use these results to characterize and predict the stochastic dynamics in terms of the MSD. 
We first rederive known results for active Brownian motion as well as classical models for run-and-tumble motility from the general expression of the correlation function in Section~\ref{subsec:resA}. 
Afterwards, general transport properties of intermittently self-propelled particles are discussed in Sections~\ref{sec:msd_eq}-\ref{sec:msd_short}. 
Two fundamentally different cases are distinguished. 
If both probability distributions~$\psi_{r,t}(t)$ possess a finite mean, the process equilibrates, i.e.~long-time properties are independent of the initial condition and the limit of long aging times~$t_a \rightarrow \infty$ exists. 
This case is discussed in Section~\ref{sec:msd_eq}. 
In contrast, this limit does not exist if one of the waiting-time distributions possesses power-law tails such that the mean value diverges---the consequence is anomalous diffusion in the long-time limit~(see Section~\ref{sec:msd_noneq}). 
If power-laws are tempered, anomalous diffusion is not expected in the long-time limit but may be observed at intermediate timescales as discussed in Section~\ref{sec:msd_noneq_int}. 
Eventually, we briefly comment on the short-time properties of the MSD in Section~\ref{sec:msd_short}.

\subsection{Active Brownian motion and run-and-tumble motility}
\label{subsec:resA}

The model~[Eq.~\eqref{eq:runeq}] contains, as discussed in the Section~\ref{sec:model}, many well-known types of~(active) diffusion models~(see also Fig.~\ref{fig:track_examples}). 
In the following, we particularly link Eq.~\eqref{eqn:Cvv_su} to standard models for self-propelled motion and run-and-tumble particles. 
For that purpose, we choose the run-times to be distributed exponentially, 
\begin{align}
	\label{eqn:exponential_wtpdf}
	\psi_{r} (t) = \lambda e^{-\lambda t}, 
\end{align}
and the time intervals of turn episodes vanish:~$\psi_t (t) = \delta \! \left(t\right)$. 
Using the Laplace transform of these waiting time distributions, 
\begin{align}
	\widehat{\psi}_{r} (s) = \frac{\lambda}{\lambda + s}, \quad  \widehat{\psi}_{t} (s) = 1,
\end{align}
the correlation function~[Eq.~\eqref{eqn:Cvv_su}] takes the simple form 
\begin{align}
	\widehat{\! \widehat{C}} \left(s, u\right)
	&= \frac{1}{s} \! \cdot \! \left [ 2 D +  \frac{v_0^2}{u + \lambda \! \left(1 - \langle \cos \chi \rangle \right) + D_{\varphi}^{(r)}} \right ] \! .
    \label{Eqn:Cvv_su_rtp}
\end{align}
In time domain, correlations are thus given by the sum of a~$\delta$-function~(Brownian diffusion) and an exponential decay~(self-propelled motion), 
\begin{align}
	C \! \left(t_a, \Delta \right) = 4 D \delta \! \left(  \Delta \right) +  v_0^2 e^{- \kappa \Delta} ,
\end{align}
with the decay rate~$\kappa = \lambda \! \left(1 - \langle \cos \chi \rangle \right) + D_{\varphi}^{(r)}$. 
It is interesting to see that the correlation function does not depend on the aging time~$t_{a}$.
This is a consequence of the Markovian nature of the dynamics in this limit~\cite{gardiner_stochastic_2009,klafter_first_2011}. 
The occurrence of tumbles in time is a Poisson process~\cite{lovely1975statistical,schnitzer_theory_1993}.  
Likewise, the mean-square displacement is independent of the aging time and the dynamics is strictly ergodic~[cf.~Eq.~\eqref{eqn:ea_tamsd}].

Following Eq.~\eqref{eqn:msd_time}, the MSD is given by F\"urths formula~\cite{fuerth_brownsche_1920}
\begin{align}
	\! m_2 \!=\! m_2^{\text{(eq)}} \! \left( \Delta \right) \! = 4 D \Delta +  \frac{2v_0^2}{\kappa^2} \! \cdot \! \Big [ \!\left ( \kappa \Delta - 1 \right ) + e^{- \kappa \Delta} \Big ] \!\!
\end{align}
with the long-time diffusion coefficient
\begin{align}
	\mathcal{D} = D + \frac{v_0^2}{ 2 \kappa} = D + \frac{v_0^2}{2 \Big[ \lambda \big (1 - \langle \cos \chi \rangle \big) + D_{\varphi}^{(r)} \Big]} .
\end{align}
This finding is consistent with existing results for standard models:~catalytic microswimmers~\cite{bechinger2016active}, for example, are usually described as active Brownian or self-propelled particles, i.e.~non-tumbling particles with a constant speed undergoing rotational diffusion, obtained for~$\langle \cos \chi \rangle = 1$.
Moreover, a well-known dynamical model for the motility of flagellated bacteria like \textit{E.~coli}~\cite{berg_ecoli_2004} is a \textit{run-and-tumble particle} with straight runs at a constant speed without rotational diffusion~($D_{\varphi}^{(r)} = 0$), that change their direction of motion abruptly~(tumbles), where, in general, $\langle \cos \chi \rangle < 1$. 
Other types of bacteria, like \textit{Myxococcus xanthus} for example, are known to instantaneously reverse their direction of motion repeatedly, in between their runs at constant speed~\cite{munoz2016myxobacteria}.
This type of motility pattern, as discussed in detail in Ref.~\cite{grossmann2016diffusion}, is obtained for~$\langle \cos \chi \rangle = -1$.

\subsection{Equilibrium limit}
\label{sec:msd_eq}

We now discuss the MSD for arbitrary distributions~$\psi_{r,t}$ in the limiting case of large aging times~$t_a \rightarrow \infty$, \textit{i.e.}, the observation of the process begins a long time after the process started~(equilibrium limit). 
The limit of large aging times corresponds to the limit~$s \rightarrow 0$, cf. Eqs.~(\ref{eqn:def_m2_eq}-\ref{eqn:diff_coeff}). 
In this regime, all results will be independent of the initial condition.

The equilibrium limit only exists if both waiting-time distributions decay sufficiently fast, such that the first moment~(mean) is finite. 
Given this assumption, one can express the distributions~$\widehat{\psi}_{r,t}(s)$ for small arguments of the Laplace variable~$s$ by a series, independent of the specific functional form:
\begin{align}
	\label{eqn:laplace_scaling_normal_psi}
	\widehat{\psi}_{r,t}(s) = 1 - \mean{t_{r,t}} \!s + \text{h.o.t.}
\end{align}
Using the definition of the MSD in equilibrium~[Eq.~\eqref{eqn:msd_eq}] and taking the corresponding limit of the correlation function~[Eq.~\eqref{eqn:Cvv_su}], we obtain the expression 
\begin{widetext}
\begin{align}
	\label{eqn:m2_eq}
	\widehat{m}_{2}^{(\text{eq})} (u) = \frac{4D}{u^2} + \frac{2v_0^2}{u^2} \! \cdot \! \frac{1}{u + D_{\varphi}^{(r)}} \! \cdot \! \frac{\mean{t_r}}{\mean{t_r} + \mean{t_t}} \! \cdot \! \left[ 1 - \frac{1 - \widehat{\psi}_r \! \left ( u + D_{\varphi}^{(r)} \!\!\: \right )}{\mean{t_r} \! \left( u + D_{\varphi}^{(r)} \!\!\: \right)} \! \cdot \! \frac{1- \mean{\cos \chi} \!\!\: \widehat{\psi}_{t} \! \left(u+\!\!\:D_{\varphi}^{(t)}\!\!\:\right)}{1-\mean{\cos \chi} \!\!\: \widehat{\psi}_{r} \!\left(u+\!\!\:D_{\varphi}^{(r)}\!\!\:\right) \! \widehat{\psi}_{t}\!\left(u+\!\!\:D_{\varphi}^{(t)}\!\!\:\right)} \right] \! .
\end{align}
\end{widetext}
In the short time limit, only leading orders in large~$u$ matter. 
In particular, to leading order, the term in brackets reduces to~$1$. 
Therefore, one obtains
\begin{align}
	\widehat{m}_{2}^{(\text{eq})} (u) \simeq \frac{4D}{u^2} + \frac{2v_0^2}{u^3} \! \cdot \! \frac{\mean{t_r}}{\mean{t_r} + \mean{t_t}} \! ,
\end{align}
corresponding to the following MSD in time domain: 
\begin{align}
	\label{eqn:msd:time:shorttime:eq}
	m_{2}^{(\text{eq})} \! \left( \Delta \right) \simeq 4 D \Delta +  \frac{\mean{t_r}}{\mean{t_r} + \mean{t_t}} \! \cdot \! v_0^2 \Delta^{\!2}. 
\end{align}
The first term reflects isotropic Brownian diffusion; the second term indicates ballistic scaling due to intermittent self-propulsion. 
The ratio of particles that are in the active run mode at a random time in equilibrium is determined by~$\langle t_r \rangle / ( \langle t_r \rangle + \langle t_t \rangle )$ and~$v_0^2 \Delta^2$ is their ballistic displacement.

In the long time limit, the MSD is found to scale linearly in time~(normal diffusion),
\begin{align}
	m_{2}^{(\text{eq})} \! \left( \Delta \right) \simeq 4 \mathcal{D} \Delta,
\end{align}
with the diffusion coefficient derived via Eq.~\eqref{eqn:diff_coeff}: 
\begin{align}
	\label{eqn:diff_coeff_general}
	\mathcal{D} & = D + \frac{v_0^2}{2 D_{\varphi}^{(r)}} \! \cdot \! \frac{\mean{t_r}}{\mean{t_r} + \mean{t_t}} \times \\
	& \:\; \times \! \left[ 1 - \frac{1 - \widehat{\psi}_r \! \left ( D_{\varphi}^{(r)} \!\!\: \right )}{\mean{t_r} \! D_{\varphi}^{(r)} } \! \cdot \!  \frac{1- \mean{\cos \chi} \!\!\: \widehat{\psi}_{t} \! \left(D_{\varphi}^{(t)}\!\!\:\right)}{1-\mean{\cos \chi} \!\!\: \widehat{\psi}_{r} \!\left(D_{\varphi}^{(r)}\!\!\:\right) \! \widehat{\psi}_{t}\!\left(D_{\varphi}^{(t)}\!\!\:\right)} \right] \! . \nonumber 
\end{align}
We thereby provide a closed formula for the diffusion coefficient of intermittently self-propelled particles for arbitrary waiting time distributions. 
The diffusivity depends on the speed via a prefactor~$v_0^2$, the persistence length of trajectories in the run state determined by~$v_0 / D_{\varphi}^{(r)}$, the mean angular reorientation~$\langle \cos \chi \rangle$, as well as the waiting-distributions~$\psi_{r,t}$ via their Laplace transforms~$\widehat{\psi}_{r,t}$ evaluated at~$D_{\varphi}^{(r,t)}$, respectively.

The distribution of turn-times enters in two ways into the expression of the diffusion coefficient~$\mathcal{D}$:~first via its mean value~$\langle t_t \rangle$ and secondly via the Laplace transform~$\widehat{\psi}_{t}$. 
The latter contribution, which always appears along with a factor~$\langle \cos \chi \rangle$, is due to reorientations.
The factor~$\widehat{\psi} \! \left ( D_{\varphi}^{(t)} \right )$ may be understood as follows. 
The mean cosine of the angle change during a turn event of duration~$\tau_t$ reads~$\langle \cos \Delta \varphi \rangle = e^{-D_{\varphi}^{(t)} \tau_t}$, cf.~Eq.~\eqref{eqn:decorrelation_single}, and averaging over~$\tau_t$ yields
\begin{align}
	\! \int_0^{\infty} \!\! d\tau_t \, \langle \cos \Delta \varphi \rangle \psi_t(\tau_t) = \!\!\int_0^{\infty} \!\! d\tau_t \, e^{-D_{\varphi}^{(t)} \tau_t} \psi_t(\tau_t) = \widehat{\psi} \! \left ( D_{\varphi}^{(t)} \right ) \!. 
\end{align}
The average is formally identical to the Laplace transform of~$\psi_t$, evaluated at~$D_{\varphi}^{(t)}$.

For the effective diffusivity~$\mathcal{D}$, the exact shape of the distribution of turn-times is only relevant if there is rotational diffusion, $D_{\varphi}^{(t)} > 0$, since there will be a correlation of the duration and the change of the direction of motion. 
If, in contrast, rotational diffusion during turns is absent~($D_{\varphi}^{(t)} = 0$), the last factor in brackets of Eq.~\eqref{eqn:diff_coeff_general} equals one.
In this case, only the mean turn-time matters for the long-time diffusion coefficient. 
Likewise, the shape of the waiting-time distribution~$\psi_t(t)$ is irrelevant if the reorientation dynamics is such that the direction of motion decorrelates within a single turn event, e.g.~for $\langle \cos \chi \rangle = 0$ or $D_{\varphi}^{(t)} \rightarrow \infty$.

As stated earlier in the context of Eqs.~(\ref{eqn:tamsd}-\ref{eqn:ergo}), the dynamics is ergodic in the equilibrium regime; the time-averaged MSD calculated from a single trajectory will eventually converge to the expressions discussed above for sufficiently long trajectories~$\Delta/T \rightarrow 0$.

In summary, the equilibrium MSD of an intermittently self-propelled particle scales according to Eq.~\eqref{eqn:msd:time:shorttime:eq} for short lag times~$\Delta$, and diffusively in the long-time limit with the effective diffusion coefficient~$\mathcal{D}$ given by Eq.~\eqref{eqn:diff_coeff_general}.

\paragraph*{The limit of straight runs}

There is one special case that needs particular attention, namely if runs are perfectly straight~($D_{\varphi}^{(r)} = 0$). 
The underlying random transport model is an intermittent L\'{e}vy walk with pauses during which a particle does not self-propelled but undergoes rotational diffusion~\cite{zaburdaev_levy_2015}. 
Given the run-time distributions possess a finite variance, the expression for the diffusion coefficient~[Eq.~\eqref{eqn:diff_coeff_general}] can be simplified to
\begin{align}
	\label{eqn:msd_eq_simpl:dr0}
	\mathcal{D} \xrightarrow{ \; D_{\varphi}^{(r)} \rightarrow 0 \;\; } D &+ \frac{v_0^2 \langle t_r \rangle^2 }{2} \! \cdot \! \frac{1}{\langle t_r \rangle + \langle t_t \rangle} \times \\ 
	& \times \left[ \frac{1}{2} \frac{\langle t_r^2\rangle}{\langle t_r \rangle^2} + \frac{\mean{\cos \chi} \!\!\: \widehat{\psi}_{t} \! \left(D_{\varphi}^{(t)}\!\!\:\right)}{1- \mean{\cos \chi} \!\!\: \widehat{\psi}_{t} \! \left(D_{\varphi}^{(t)}\!\!\:\right)} \right] \nonumber 
\end{align}
in this limit. 
If, on the other hand, the run-times follow a power-law distribution $\psi_r(t) \sim t^{-1-\beta}$ with~$\beta \in (1,2)$, such that its mean is finite but the variance of run-times diverges, Eq.~\eqref{eqn:diff_coeff_general} is not applicable.
The reason is that particles move superdiffusively in this case as shown below. 
For this type of distribution, the Laplace transform for small arguments~$s$ scales according to
\begin{align}
	\label{eqn:scaling_psi_laplace}
	\widehat{\psi}_r(s) \simeq 1 - \mean{t_r}\! s + A s^{\beta} + \text{h.o.t.}
\end{align}
with~$A=\text{const.}$
Reconsidering the MSD~[Eq.~\eqref{eqn:m2_eq}] for this type of distribution in the time limit of large lag times~(small~$u$), yields 
\begin{align}
	\label{eqn:msd:eq:super:b12}
	\widehat{m}_{2}^{(\text{eq})} (u) \simeq \frac{4D}{u^2} + \frac{2v_0^2}{u^3} \! \cdot \! \frac{\mean{t_r}}{\mean{t_r} + \mean{t_t}} \! \cdot \! \left[ \frac{A}{\mean{t_r}}u^{\beta - 1} \right] \! ,
\end{align}
implying superdiffusion~$m_{2}^{(\text{eq})} \simeq \Delta^{\!\!\: 3 - \beta}$ for~$\beta \in (1,2)$ to leading order. 
That is why the diffusion coefficient~$\mathcal{D}$ is not defined regime and Eq.~\eqref{eqn:diff_coeff_general} looses its applicability in this parameter regime. 
Surprisingly, Eq.~\eqref{eqn:msd:eq:super:b12} is independent of the reorientation dynamics---the scaling of the MSD does neither dependent on the mean flip angle~$\langle \cos \chi \rangle$ nor on the explicit type of turn-time distribution.

\subsection{Anomalous active diffusion}
\label{sec:msd_noneq}

In the previous paragraph, we considered run- and turn-time distributions which have a finite mean such that the diffusion process equilibrates. 
To complete the analysis, we now describe the case in which at least one of the distributions does not have a finite mean. 
As a specific example, one could think of the following distributions: 
\begin{align}
	\label{eqn:psi_power_laws}
	\!\psi_t(t) \!=\!  \frac{\alpha}{\tau_{\alpha}\! \left( 1 + t/\tau_{\alpha} \right)^{1 + \alpha}}, \;
	\psi_r(t) \!=\! \frac{\beta}{\tau_{\beta}\! \left( 1 + t/\tau_{\beta} \right)^{1 + \beta}}. \!\!
\end{align}
The specific functional dependence is, however, irrelevant for the following discussion---the power-law decay~$\psi_t(t) \sim t^{1+\alpha}$ and~$\psi_r(t) \sim t^{1+\beta}$ will determine the long-time behavior of the MSD. 
The respective means~$\langle t_t \rangle $ and~$\langle t_r \rangle$ of these distributions do not exist for~$\alpha \in (0,1)$ and~$\beta \in (0,1)$.

\begin{figure}[t]
\begin{center}
	\includegraphics[width=\columnwidth]{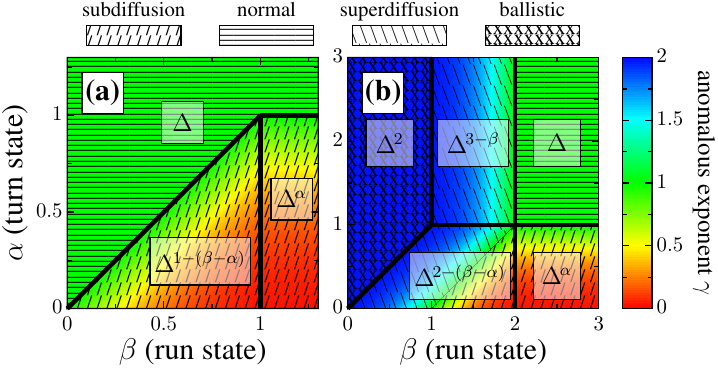}
	\vspace*{-0.65cm}
	\caption{Anomalous diffusion exponent~$\gamma$ of the ensemble-averaged MSD, $m_2(t_a=0,\Delta) \sim \Delta^{\!\!\:\gamma}$, shown in color as predicted by Eq.~\eqref{eqn:m2_nonequilibrium}, for run- and turn-time distributions with power-law decay~[cf.~Eq.~\eqref{eqn:psi_power_laws}]. Panel~(a) shows the scaling for runs with a finite persistence time~($D_{\varphi}^{(r)} > 0$); if runs are perfectly straight~($D_{\varphi}^{(r)} = 0$), the phase behavior is richer as shown in~(b). The background pattern indicates the type of diffusion process:~subdiffusion~($\gamma < 1$), normal diffusion~($\gamma = 1$), superdiffusion~($1 <\gamma < 2 $) and ballistic motion~($\gamma = 2$). The diffusion process equilibrates, as discussed in the context of Eq.~\eqref{eqn:m2_eq}, if~$\alpha > 1$ and~$\beta > 1$---the waiting time distributions possess a finite mean in this case. Note that Brownian diffusion was neglected~($D=0$) to produce the plots above; if isotropic Brownian motion is present during the turn state, subdiffusion is not observed but the corresponding regions are replaced by normal diffusion~($\gamma = 1$). }
	\label{fig:long_time_exponents}
\end{center}
\end{figure}

If one of the waiting-time distributions does not possess a finite mean, the equilibrium limit $t_a \rightarrow \infty$ does not exist. 
We will focus on the special case~$t_a=0$ below. 
From the general relation between MSD~[cf.~Eqs.~(\ref{def:msd_time}-\ref{eqn:msd})] and correlation function~[Eq.~\eqref{eqn:Cvv_su}], we derive the following MSD in Laplace domain: 
\begin{widetext}
\begin{align}
	\label{eqn:m2_nonequilibrium}
	\widehat{m}_{2} (t_a=0,u) = & \frac{4D}{u^2} + \frac{2v_0^2}{u^2} \! \cdot \! \frac{1}{u + D_{\varphi}^{(r)}} \! \cdot \! \frac{1 - \widehat{\psi}_r(u)}{1 - \widehat{\psi}_r(u) \widehat{\psi}_t(u)} \times  \\ 
	& \times \! \left[ 1 - \frac{u}{1 - \widehat{\psi}_r(u)} \cdot \frac{\widehat{\psi}_r(u) - \widehat{\psi}_r \! \left ( u + D_{\varphi}^{(r)} \!\!\: \right )}{ D_{\varphi}^{(r)} } \! \cdot \! \frac{1- \mean{\cos \chi} \!\!\: \widehat{\psi}_{t} \! \left(u+\!\!\:D_{\varphi}^{(t)}\!\!\:\right)}{1-\mean{\cos \chi} \!\!\: \widehat{\psi}_{r} \!\left(u+\!\!\:D_{\varphi}^{(r)}\!\!\:\right) \! \widehat{\psi}_{t}\!\left(u+\!\!\:D_{\varphi}^{(t)}\!\!\:\right)} \right] \! . \nonumber 
\end{align}
\end{widetext}
The long-time limit of the MSD is derived from the expression above by considering the scaling for small~$u$~\cite{doetsch_tabellen_1947}. 
If the waiting-time distributions do not possess a finite mean, their Laplace transforms scale according to 
\begin{align}
	\!\!\! 
	\widehat{\psi}_t(u) \simeq 1 - A u^{\alpha} \! + \mathcal{O}(u), \; 
	\widehat{\psi}_r(u) \simeq 1 - B u^{\beta} \! + \mathcal{O}(u),  \!
\end{align}
in Laplace domain, which is notably different from Eqs.~(\ref{eqn:laplace_scaling_normal_psi},~\ref{eqn:scaling_psi_laplace}). 
Analogously to the discussion in the previous Section, two regimes have to be distinguished, namely~$D_{\varphi}^{(r)} > 0$ and~$D_{\varphi}^{(r)} = 0$, highlighting the importance of rotational fluctuations. 
We summarize the possible long-time behaviors~$m_2(t_a=0,\Delta) \sim \Delta^{\!\!\: \gamma}$ of the MSD in Fig.~\ref{fig:long_time_exponents}. 
Depending on the exponents~$\alpha$ and~$\beta$, subdiffusion~($\gamma < 1$), normal diffusion~($\gamma = 1$), superdiffusion~($1 <\gamma < 2 $) and even ballistic spreading~($\gamma = 2$) is predicted:~the tendency towards subdiffusion is due to turns during which self-propulsion is absent, whereas superdiffusion is due to power-laws in the run-time distribution. 
Note that only those terms of the MSD were taken into account which are due to active motion---if Brownian motion is relevant, subdiffusion is not the leading order behavior but the MSD will increase linearly~(normal) for long times~\cite{thiel2012anomalous}.

We underline that the superdiffusive scaling of the MSD in the long-time limit is not robust with respect to rotational noise, in the sense that the existence of superdiffusion requires perfectly straight runs~($D_{\varphi}^{(r)} = 0$):~if runs are straight, like it is assumed for L\'{e}vy walks~\cite{zaburdaev_levy_2015}, and the run-time distributions decays sufficiently slow, a particle will tend to stay in the run state for longer times the longer the process evolves, leading to superdiffusion or even ballistic scaling. 
If, however, rotational noise during runs is present, runs will have a finite persistence time, thereby again inducing normal diffusion in the limit~$\Delta \rightarrow \infty$.

Note that the special case without noise, $D_{\varphi}^{(r,t)} = 0$, and~$\langle \cos \chi \rangle = 0$ has already been discussed in Ref.~\cite{portillo2011intermittent}, however, their prediction regarding the long-time behavior differs from ours.

\subsection{Intermediate anomalous diffusion---tempered power-laws}
\label{sec:msd_noneq_int}

\begin{figure}[b]
\includegraphics[width=\columnwidth]{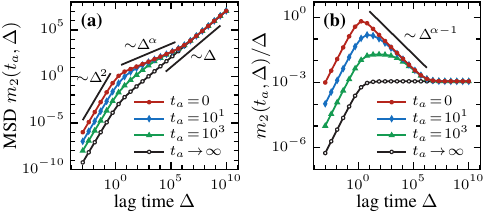}
\vspace{-0.55cm}
\caption{MSD of intermittently self-propelled particles with exponentially distributed run-times~[Eq.~\eqref{eqn:exponential_wtpdf}] and turn-times that follow a tempered power-law distribution~[Eq.~\eqref{eqn:temp_power_law}], obtained by numerical inverse Laplace transform~\cite{noteSI} of Eqs.~(\ref{eqn:msd},~\ref{eqn:Cvv_su}). Panel~(a): ensemble-averaged MSD~$m_2(t_a,\Delta)$ as a function of the lag time~$\Delta$ for various aging times~$t_a$. For small aging times, the MSD shows three regimes:~ballistic motion---the processes starts with a run---followed by subdiffusion as predicted by~Fig.~\ref{fig:long_time_exponents}; in the long-time limit, there is a crossover to normal diffusion due to the exponential cutoff of waiting times in the distribution~$\psi_t(t)$. Panel~(b): For the same data, the MSD divided by the lag time~$\Delta$ is shown, underlining the sub-linear scaling at intermediate timescales and linear scaling of the MSD for large lag times. The MSD~$m_2$ for large aging time~($t_a \rightarrow \infty$) are identical to the ensemble-average of the time-averaged MSD~$\langle \delta^2_T(t_a,\Delta) \rangle$, cf.~Eq.~\eqref{eqn:ergo}. Parameters: mean run-time~$1/\lambda = 1$, timescale~$\tau_{\alpha} = 1$, cutoff timescale~$\tau_c = 10^6$, exponent~$\alpha = 1/2$, rotational diffusion coefficients~$D_{\varphi}^{(r)} = 10^{-5}$, $D_{\varphi}^{(t)} = 0$, speed~$v_0 = 1$, $\langle \cos \chi \rangle = 0$, isotropic diffusion coefficient~$D = 0$. }
\label{fig:non-poisson-traps}
\end{figure}

In Section~\ref{sec:msd_noneq}, we considered the long-time behavior of the MSD~($\Delta \rightarrow \infty$) and found that anomalous diffusion may be observed if the underlying waiting-time distributions for run or turn episodes~$\psi_{r,t}$ possess heavy tails~(cf.~Fig.~\ref{fig:long_time_exponents}). 
In experimental situations, however, these heavy tails may be tempered, i.e.~power-laws are cut off beyond some characteristic timescale~$\tau_c$. 
In this case, the MSD will scale linearly with the lag time~$\Delta$ for large times, and anomalous scaling can only be observed at intermediate timescales. 
We illustrate this using the example of self-propelled particles switching between run- and turn phases, where the run-time distribution is exponential~[Eq.~\eqref{eqn:exponential_wtpdf}] and turn-times follow a tempered power-law distribution 
\begin{align}
	\psi_t(t) = - \frac{d}{dt} \! \left[ \frac{e^{- t/\tau_c }}{\left(1+ t/\tau_{\alpha} \right)^{\alpha}} \right] 
	\label{eqn:temp_power_law}
\end{align}
with an exponent~$\alpha \in (0,1)$; the cutoff timescale~$\tau_c$ shall satisfy~$\tau_c \gg \tau_{\alpha}$. 
The distribution is written in such a way that the term in brackets corresponds to the cumulative distribution~$\Psi_t(t) \!=\! \int_{t}^{\infty} dt' \, \psi(t')$. 
In the limit of vanishing cutoff~$\tau_c \rightarrow \infty$, the distribution is identical to Eq.~\eqref{eqn:psi_power_laws} and, hence, anomalous diffusion is expected~(Fig.~\ref{fig:long_time_exponents}). 
The MSD for the dynamics described above, obtained by numerically inverting the Laplace transform~(Eq.~\eqref{eqn:m2_nonequilibrium}, see also SM for technical details~\cite{noteSI}), is presented in Fig.~\ref{fig:non-poisson-traps}. 
The ensemble-averaged MSD~$m_2(t_a,\Delta)$ shows an intermediate subdiffusive scaling. 
However, subdiffusive motion crosses over to normal diffusion in the limit~$\Delta \gg \tau_c$ due to tempering.

Since both, the mean run- and turn-times exist as a consequence of tempering, the process will equilibrate~(cf.~the general discussion on the equilibrium limit in Section~\ref{sec:msd_eq}). 
For large aging times~$t_a$, the ensemble-averaged MSD~$m_2$ hence equals the ensemble-average of time-averaged MSDs~$\langle \delta^2_T \rangle$ as the system attains equilibrium~(black lines in Fig.~\ref{fig:non-poisson-traps}).  
Since the processes starts with a run (non-equilibrated initial condition), time-averages differ from ensemble averages~(cf.~Fig.~\ref{fig:non-poisson-traps}).  
Note in particular that the ensemble-average of time-averaged MSDs~$\langle \delta_T^2 \rangle$ does not show an intermediate subdiffusive regime, which can only be observed at the level of the ensemble-averaged MSD~$m_2$.

\subsection{Short-time scaling of the MSD}
\label{sec:msd_short}

We conclude the analysis of the transport properties of intermittently self-propelled particles by briefly commenting on the MSD in the short-time limit, i.e.~for small lag times~$\Delta$.
In general, the short-time behavior of the MSD~$m_2$ depends on the initial condition~(process starting with a run or turn), as well as the aging time~$t_a$; in this sense, it is not universal but depends details of the model and the process.
For simplicity, we limit the following discussion to the case~$t_a=0$. 
The short-time behavior of~$m_2(t_a=0,\Delta)$ can be readily determined from its Laplace transform~$\widehat{m}_2(t_a=0,u)$ by considering leading order terms in~$1/u$ for~$u \rightarrow \infty$.

The MSD, determined by Eq.~\eqref{eqn:m2_nonequilibrium}, was derived under the assumption that the process started with a run, which we reconsider now. 
For short lag times, the particle is therefore certainly in the run state and, consequently, the MSD is expected to contain a ballistic term due to active motion. 
This is directly confirmed by taking the leading order terms in inverse~$u$ of Eq.~\eqref{eqn:m2_nonequilibrium}: $\widehat{m}_2(t_a\!=\!0,u) \simeq 4 D /u^2 + 2 v_0^2/u^3$. 
The corresponding expression in time domain reads 
\begin{align}
	\label{eqn:short_time_msd:active}
	m_2(t_a=0,\Delta) \simeq 4 D \Delta \!+\! v_0^2\Delta^2 = 
	\begin{cases}
		4 D \Delta, \!\!\! & \!\! \Delta \! \ll \! D/v_0^2, \\ v_0^2 \Delta^2, \!\!\! & \!\! \Delta \! \gg \! D/v_0^2. 
	\end{cases}
\end{align}
As indicated in the equation above, isotropic Brownian diffusion plays a nontrivial role:~the ballistic scaling due to active motion is only visible on timescales larger than~$D/v_0^2$, whereas Brownian diffusion dominates at the shortest timescales~\footnote{This argument presumes small rotational diffusion~$D_{\varphi}^{(r)}$. }.

If the process started with a turn phase, the MSD differs from Eq.~\eqref{eqn:m2_nonequilibrium} solely by a multiplicative factor~$\widehat{\psi}_t(u)$~\cite{noteSI}. 
Accordingly, the MSD in Laplace domain in the limit~$1/u\rightarrow 0$ reads
\begin{align}
	\label{eqn:short_time_msd:passive:lapl}	
	\widehat{m}_{2} (t_a=0,u) \simeq \frac{4D}{u^2} + \frac{2v_0^2}{u^3} \cdot \widehat{\psi}_t(u). 
\end{align}
Hence, the MSD depends on the explicit scaling of the waiting time distribution~$\psi_t(t)$. 
As an example, we consider durations of turn events that follow a Gamma-distribution, 
\begin{align}
	\psi_t(t) = \frac{\nu^{\kappa}}{\Gamma(\kappa)} t^{\kappa-1} e^{-\nu t},
	\label{eqn:gamma_pdf}
\end{align}
with shape parameter~$\kappa$ and rate~$\nu$, implying the mean value~$\kappa / \nu $.  
Its Laplace transform can be obtained in closed form:~$\widehat{\psi}_t(u) = \big[ \nu / (\nu + u)  \big]^{\kappa}$.
Combining Eqs.~(\ref{eqn:short_time_msd:passive:lapl},~\ref{eqn:gamma_pdf}), the MSD is expected to scale according to
\begin{align}
	m_2(t_a=0,\Delta) \simeq 4 D \Delta + c_{\kappa} v_0^2\Delta^2 \! \left( \frac{\Delta}{\langle t_t \rangle} \right)^{\!\!\kappa} 
	\label{eqn:short_time_msd:passive}
\end{align}
with a constant~$c_{\kappa}$ that solely depends on~$\kappa$. 
Accordingly, the MSD increases faster than ballistic for all~$\kappa > 0$. 
Similar arguments were put forward by Jung in Ref.~\cite{jung2023hyperdiffusion}. 
A related scaling behaviour has been reported recently by Doerries et.~al.~\cite{doerries2023emergent} in a mobile-immobile model of diffusing particles with advection, where exponential waiting times were considered, implying~$\kappa = 1$ and $m_2 \simeq \Delta^{\!\!\: 3}$.

Intuitively, the short-time scaling of the MSD may be understood as follows. 
Active motion leads to ballistic motion~(factor~$v_0^2 \Delta^{\!\!\:2}$) if particles are in the run state. 
The following short-time MSD is therefore expected 
\begin{align}
	m_2(t_a=0,\Delta) \simeq 4 D \Delta + P_a(\Delta) v_0^2 \Delta^2, 
\end{align}
where the first term accounts for Brownian diffusion and~$P_a(\Delta)$ is the probability of a particle to be in the run state at lag time~$\Delta$. 
If the process started with a run,~$P_a(\Delta) \simeq 1$ to leading order and, hence, Eq.~\eqref{eqn:short_time_msd:active} is obtained.  
If, on the other hand, the process started in the turn phase, the probability~$P_a(\Delta)$ is a non-trivial function of the lag time~$\Delta$ to leading order, since the number of particles that run increases over time. 
For the example discussed above~[Eq.~\eqref{eqn:gamma_pdf}], we derived~$P_a(\Delta) \sim \Delta^{\kappa}$, eventually rationalizing Eq.~\eqref{eqn:short_time_msd:passive}. 
We underline that the scaling discussed above crucially depends on the initial condition and the aging time~$t_a = 0$. 
Hyper-ballistic scaling~$m_2 \sim \Delta^{\!\!\:\gamma}$ with exponents~$\gamma > 2$ is absent in the equilibrium limit~$t_a \rightarrow \infty$---given this limit exists---as the initial condition is forgotten for long lag times~[cf.~Eq.~\eqref{eqn:m2_eq}]. 
If the equilibrium exists, the probability to be in the run state is constant:~$P_a(\Delta) = \langle  t_r \rangle / \left( \langle t_r \rangle + \langle t_t \rangle \right)$, cf.~Eq.~\eqref{eqn:msd:time:shorttime:eq}.

\section{Generalization to arbitrary spatial dimensions}
\label{sec:multidim}

The results discussed so far can directly by extended to arbitrary spatial dimensions~$d$. 
Generally, the effective, long-time diffusion coefficient~$\mathcal{D}$ is defined via~$m_2^{\text{(eq)}} \! \left( \Delta \right) \sim 2 d \mathcal{D} \Delta$, such that first the prefactor in Eq.~\eqref{eqn:diff_coeff} needs to be adapted: 
\begin{align}
	\mathcal{D} &= \frac{1}{2d} \cdot \lim_{\Delta \rightarrow \infty} \frac{d m_2^{\text{(eq)}} \left(\Delta\right)}{d\Delta} . 
\end{align} 
Predictions on the one-dimensional dynamics~\cite{portillo2011intermittent} directly follow from the two-dimensional case~[cf.~Eq.~\eqref{eqn:Cvv_su}] by setting the rotational diffusion coefficients to zero, $D_{\varphi}^{(r,t)} = 0$, such that the dynamics takes place on a line. 
Moreover, the distribution of reorientation angles is the sum of two~$\delta$-peaks in one dimension, $p(\chi) = q \delta(\chi) + (1-q) \delta (\chi - \pi)$, where~$q$ denotes the probability to continue a run after a turn in the same direction as before. 
This implies~$\langle \cos \chi \rangle = 2q-1 $, including the two relevant special cases of reversal~\cite{grossmann2016diffusion,fedotov2016single,giona2022extended}, $q=0$ implying~$\langle \cos \chi \rangle = -1$, and randomization of the direction of motion, $q=0.5$ implying~$\langle \cos \chi \rangle = 0$~\cite{peruani2023active}.

The generalization to three two-dimensions is slightly more involved, as it requires to replace the orientational dynamics of~Eq.~\eqref{eq:runeq} as follows~\cite{noteStrato,grossmann_anistropic_2015}: 
\begin{subequations}
\begin{align}
	\dot{\vec{r}}(t) &= v(t) \vec{e}(t) + \!\!\: \sqrt{2D} \, \boldsymbol{\xi}(t) \\ 
	\dot{\vec{e}}(t) &= \sqrt{2 D_{\varphi}(t)} \Big[ \mathds{1} - \vec{e}(t) \!\!\: \otimes \vec{e}(t) \Big ] \! \cdot \!  \boldsymbol{\eta}(t) + \boldsymbol{\zeta}_{\chi}(t) .  \label{eqn:e_nd}
\end{align}
\end{subequations}
The unit vector~$\vec{e}$ denotes the direction of motion in three dimensions. 
Sudden flips, symbolized by the non-Poissonian shot noise~$\boldsymbol{\zeta}_{\chi}(t)$, are implemented via stochastic rotations~$\vec{e} \rightarrow \vec{e}' = \mathcal{R} \cdot \vec{e}$ with a rotation matrix~$\mathcal{R}$, such that~$\vec{e}'$ lies on a cone centered around~$\vec{e}$ with~$\vec{e} \cdot \vec{e}' = \cos \chi$, where the angle~$\chi$ is drawn from~$p(\chi)$ as before. 
The orientational dynamics of Eq.~\eqref{eqn:e_nd} implies the following correlation in a single motility mode: 
\begin{align}
	\mean{ \vec{e}(t + \tau) \cdot \vec{e}(t) } \!  &= e^{- (d-1) D_{\varphi}^{(r,t)} \tau}. 
\end{align}
This is formally identical to Eq.~\eqref{eqn:decorrelation_single}---orientational correlations decay exponentially---when the noise amplitudes are formally replaced by 
\begin{align}
	\label{eqn:diffrot_general}
	D_{\varphi}^{(r,t)} \rightarrow (d-1) D_{\varphi}^{(r,t)} .
\end{align}
Hence, all results presented above are also applicable to three~(and higher) dimensions if the substitution of rotational diffusion coefficients~[Eq.~\eqref{eqn:diffrot_general}] is made in the correlation function.

\section{Summary \& Discussion}
\label{sec:summary}

In this work, we proposed and discussed a general dynamical model of intermittently self-propelled particles. 
The model takes two distinct states of motility into account, namely an active run-phase and a turn state in which self-propulsion is absent. 
The switching between those two states is described by a renewal process. 
The model does not assume a specific switching statistics, but the stochastic life times of the two modes of motility are drawn from arbitrary waiting-time distributions.

For suitable choices of the waiting-time statistics and parameters, our model reduces to well-studied stochastic transport models such as run-and-tumble, active Brownian motion, L\'{e}vy walks, and continuous time random walks or to the mobile-immobile model, thereby bridging different model classes and enabling us to study short-time and long-time transport properties analytically within one framework. 
We focused the analysis of the model on the properties of the mean-square displacement~(MSD).

%
At short timescales, the MSD is generally found to scale diffusively due to isotropic Brownian diffusion, followed by a crossover to ballistic scaling as a consequence of active motion. 
Hyper-ballistic scaling~$m_2 \sim \Delta^{\!\!\: \gamma}$ with scaling exponents~$\gamma$ larger than two may be observed for certain initial conditions, specifically if an observation starts in the turn state:~mobile particles will contribute a factor~$\Delta^2$ to the MSD, but the number of particles in the run state will also increase with the lag time~$\Delta$, inducing hyper-ballistic scaling~\cite{doerries2023emergent,jung2023hyperdiffusion}.

%
Typically, the MSD of intermittently self-propelled particles scales linearly with the lag time~(normal diffusion) in the long-time limit. 
More precisely, this is the case if the mean life times of the motility modes are finite, and runs have a finite persistence. 
A closed formula for the diffusion coefficient of intermittently self-propelled particles was derived for arbitrary waiting-time distributions~[Eq.~\eqref{eqn:diff_coeff_general}]; we furthermore discussed how to extend this formula to any spatial dimension. 
However, if runs are infinitely persistent and the variance of run-times diverges, superdiffusion is expected for long times. 
Moreover, the MSD may show subdiffusion, superdiffusion and also ballistic motion in dependence on the switching statistics between the run and turn state, if their mean lifetimes diverge, which is the case for waiting-time distributions with heavy tails. 
A summary of scaling exponents~$\gamma$ of the MSD~$m_2 \sim \Delta^{\!\!\: \gamma}$ in the limit of large lag times~$\Delta$ is given in Fig.~\ref{fig:long_time_exponents}. 
We also addressed the robustness of theoretical predictions with respect to noise, in particular the observability of anomalous scaling in the presence of isotropic diffusion, rotational diffusion~(limiting the persistence length of runs) and tempered power-laws. 
In the latter case, i.e.~if power-law waiting-time distributions are cut off, we show that anomalous transport is solely expected at intermediate timescales.

The general structure of our model makes it convenient for application to various experimental settings that involve switching behaviour of self-propelled particles, e.g.~bacteria in heterogeneous media~\cite{bhattacharjee2019bacterial,beier2024mot}.
We expect single-particle tracking to enable measurements of both, the waiting-time distributions, the speed of active motion as well as the structural properties of trajectories like their persistence and turn angle distributions. 
The presented model reveals that the statistics of lifetimes of motility modes, described by the waiting-time distribution~$\psi_{r,t}$, crucially determine the long-time transport of intermittently self-propelled particles and should be therefore of particular interest experimentally.  
We applied this model successfully to describe the motility of the soil bacterium \textit{Pseudomonas putida} in agar gel, in which it shows a switching behaviour of self-propulsion and trapping with power-law distributed trapping times and, consequently, anomalous scaling of the MSD~\cite{beier2024mot}.

We assumed the speed of self-propulsion to be constant throughout this manuscript; this may not be the case in an experimental situation. 
If the speed varies between active episodes or temporal fluctuations are present~\cite{peruani2007self,romanczuk2011brownian,romanczuk_active_2012} but mild, the predictions of the here-presented theory may still yield reasonable results; the factors of~$v_0^2$ would have to be replaced by an average value.

The presented modeling approach to intermittent self-propelled motion can be extended in several directions in the future. 
Heterogeneous particle properties may change the long-time scaling of the MSD---a problem that is specifically relevant for the analysis of biological tracking data~\cite{zhao_quantitative_2024}. 
Moreover, it is interesting to look at biased motion, e.g.~active particles in external fields or chemical gradients~\cite{alirezaei_chemotaxis_2020,bhattacharjee2021chemotactic}.  
At a theoretical level, the presented framework should be extended to study the displacement statistics beyond the MSD, with a particular focus on non-Gaussian effects~\cite{metzler2014anomalous,he_dynamic_2016,lamp_cytoplasmic_2017,grossmann2024nongaussian}. 
Already in the case of Poissonian switching statistics between mobile and immobile states, the displacement distribution was shown to undergo non-trivial transitions and pronounced non-Gaussianity~\cite{mora2018brownian,doerries2023emergent}. 
Apart from that, it will be interesting to study self-propelled particles with more than two modes of motility and extend the theoretical framework accordingly.

\section*{Data availability}

The data and code that support the findings of this study are available from the corresponding author upon reasonable request. 

\begin{acknowledgments}

This research has been partially funded by Deutsche Forschungsgemeinschaft~(DFG): Project-ID 318763901~--~SFB1294. 

\end{acknowledgments}


\begin{thebibliography}{122}%
\makeatletter
\providecommand \@ifxundefined [1]{%
 \@ifx{#1\undefined}
}%
\providecommand \@ifnum [1]{%
 \ifnum #1\expandafter \@firstoftwo
 \else \expandafter \@secondoftwo
 \fi
}%
\providecommand \@ifx [1]{%
 \ifx #1\expandafter \@firstoftwo
 \else \expandafter \@secondoftwo
 \fi
}%
\providecommand \natexlab [1]{#1}%
\providecommand \enquote  [1]{``#1''}%
\providecommand \bibnamefont  [1]{#1}%
\providecommand \bibfnamefont [1]{#1}%
\providecommand \citenamefont [1]{#1}%
\providecommand \href@noop [0]{\@secondoftwo}%
\providecommand \href [0]{\begingroup \@sanitize@url \@href}%
\providecommand \@href[1]{\@@startlink{#1}\@@href}%
\providecommand \@@href[1]{\endgroup#1\@@endlink}%
\providecommand \@sanitize@url [0]{\catcode `\\12\catcode `\$12\catcode
  `\&12\catcode `\#12\catcode `\^12\catcode `\_12\catcode `\%12\relax}%
\providecommand \@@startlink[1]{}%
\providecommand \@@endlink[0]{}%
\providecommand \url  [0]{\begingroup\@sanitize@url \@url }%
\providecommand \@url [1]{\endgroup\@href {#1}{\urlprefix }}%
\providecommand \urlprefix  [0]{URL }%
\providecommand \Eprint [0]{\href }%
\providecommand \doibase [0]{https://doi.org/}%
\providecommand \selectlanguage [0]{\@gobble}%
\providecommand \bibinfo  [0]{\@secondoftwo}%
\providecommand \bibfield  [0]{\@secondoftwo}%
\providecommand \translation [1]{[#1]}%
\providecommand \BibitemOpen [0]{}%
\providecommand \bibitemStop [0]{}%
\providecommand \bibitemNoStop [0]{.\EOS\space}%
\providecommand \EOS [0]{\spacefactor3000\relax}%
\providecommand \BibitemShut  [1]{\csname bibitem#1\endcsname}%
\let\auto@bib@innerbib\@empty
\bibitem [{\citenamefont {Romanczuk}\ \emph {et~al.}(2012)\citenamefont
  {Romanczuk}, \citenamefont {B{\"a}r}, \citenamefont {Ebeling}, \citenamefont
  {Lindner},\ and\ \citenamefont {Schimansky-Geier}}]{romanczuk_active_2012}%
  \BibitemOpen
  \bibfield  {author} {\bibinfo {author} {\bibfnamefont {P.}~\bibnamefont
  {Romanczuk}}, \bibinfo {author} {\bibfnamefont {M.}~\bibnamefont {B{\"a}r}},
  \bibinfo {author} {\bibfnamefont {W.}~\bibnamefont {Ebeling}}, \bibinfo
  {author} {\bibfnamefont {B.}~\bibnamefont {Lindner}},\ and\ \bibinfo {author}
  {\bibfnamefont {L.}~\bibnamefont {Schimansky-Geier}},\ }\bibfield  {title}
  {\bibinfo {title} {Active {B}rownian particles},\ }\href
  {https://doi.org/10.1140/epjst/e2012-01529-y} {\bibfield  {journal} {\bibinfo
   {journal} {Eur. Phys. J.: Spec. Top.}\ }\textbf {\bibinfo {volume} {202}},\
  \bibinfo {pages} {1} (\bibinfo {year} {2012})}\BibitemShut {NoStop}%
\bibitem [{\citenamefont {Elgeti}\ \emph {et~al.}(2015)\citenamefont {Elgeti},
  \citenamefont {Winkler},\ and\ \citenamefont
  {Gompper}}]{elgeti_physics_2015}%
  \BibitemOpen
  \bibfield  {author} {\bibinfo {author} {\bibfnamefont {J.}~\bibnamefont
  {Elgeti}}, \bibinfo {author} {\bibfnamefont {R.~G.}\ \bibnamefont
  {Winkler}},\ and\ \bibinfo {author} {\bibfnamefont {G.}~\bibnamefont
  {Gompper}},\ }\bibfield  {title} {\bibinfo {title} {Physics of
  microswimmers—single particle motion and collective behavior: {A} review},\
  }\href {https://doi.org/10.1088/0034-4885/78/5/056601} {\bibfield  {journal}
  {\bibinfo  {journal} {Rep. Prog. Phys.}\ }\textbf {\bibinfo {volume} {78}},\
  \bibinfo {pages} {056601} (\bibinfo {year} {2015})}\BibitemShut {NoStop}%
\bibitem [{\citenamefont {Bechinger}\ \emph {et~al.}(2016)\citenamefont
  {Bechinger}, \citenamefont {Di~Leonardo}, \citenamefont {L\"owen},
  \citenamefont {Reichhardt}, \citenamefont {Volpe},\ and\ \citenamefont
  {Volpe}}]{bechinger2016active}%
  \BibitemOpen
  \bibfield  {author} {\bibinfo {author} {\bibfnamefont {C.}~\bibnamefont
  {Bechinger}}, \bibinfo {author} {\bibfnamefont {R.}~\bibnamefont
  {Di~Leonardo}}, \bibinfo {author} {\bibfnamefont {H.}~\bibnamefont
  {L\"owen}}, \bibinfo {author} {\bibfnamefont {C.}~\bibnamefont {Reichhardt}},
  \bibinfo {author} {\bibfnamefont {G.}~\bibnamefont {Volpe}},\ and\ \bibinfo
  {author} {\bibfnamefont {G.}~\bibnamefont {Volpe}},\ }\bibfield  {title}
  {\bibinfo {title} {Active particles in complex and crowded environments},\
  }\href {https://doi.org/10.1103/RevModPhys.88.045006} {\bibfield  {journal}
  {\bibinfo  {journal} {Rev. Mod. Phys.}\ }\textbf {\bibinfo {volume} {88}},\
  \bibinfo {pages} {045006} (\bibinfo {year} {2016})}\BibitemShut {NoStop}%
\bibitem [{\citenamefont {Shaebani}\ \emph {et~al.}(2020)\citenamefont
  {Shaebani}, \citenamefont {Wysocki}, \citenamefont {Winkler}, \citenamefont
  {Gompper},\ and\ \citenamefont {Rieger}}]{shaebani_computational_2020}%
  \BibitemOpen
  \bibfield  {author} {\bibinfo {author} {\bibfnamefont {M.~R.}\ \bibnamefont
  {Shaebani}}, \bibinfo {author} {\bibfnamefont {A.}~\bibnamefont {Wysocki}},
  \bibinfo {author} {\bibfnamefont {R.~G.}\ \bibnamefont {Winkler}}, \bibinfo
  {author} {\bibfnamefont {G.}~\bibnamefont {Gompper}},\ and\ \bibinfo {author}
  {\bibfnamefont {H.}~\bibnamefont {Rieger}},\ }\bibfield  {title} {\bibinfo
  {title} {Computational models for active matter},\ }\href
  {https://doi.org/10.1038/s42254-020-0152-1} {\bibfield  {journal} {\bibinfo
  {journal} {Nat. Rev. Phys.}\ }\textbf {\bibinfo {volume} {2}},\ \bibinfo
  {pages} {181} (\bibinfo {year} {2020})}\BibitemShut {NoStop}%
\bibitem [{\citenamefont {Berg}\ and\ \citenamefont
  {Brown}(1972)}]{berg_chemotaxis_1972}%
  \BibitemOpen
  \bibfield  {author} {\bibinfo {author} {\bibfnamefont {H.~C.}\ \bibnamefont
  {Berg}}\ and\ \bibinfo {author} {\bibfnamefont {D.~A.}\ \bibnamefont
  {Brown}},\ }\bibfield  {title} {\bibinfo {title} {Chemotaxis in
  \textit{{E}scherichia coli} analysed by three-dimensional tracking},\ }\href
  {https://doi.org/10.1038/239500a0} {\bibfield  {journal} {\bibinfo  {journal}
  {Nature}\ }\textbf {\bibinfo {volume} {239}},\ \bibinfo {pages} {500}
  (\bibinfo {year} {1972})}\BibitemShut {NoStop}%
\bibitem [{\citenamefont {Berg}(2004)}]{berg_ecoli_2004}%
  \BibitemOpen
  \bibfield  {author} {\bibinfo {author} {\bibfnamefont {H.~C.}\ \bibnamefont
  {Berg}},\ }\href {https://doi.org/10.1007/b97370} {\emph {\bibinfo {title}
  {E. coli in Motion}}}\ (\bibinfo  {publisher} {Springer},\ \bibinfo {year}
  {2004})\BibitemShut {NoStop}%
\bibitem [{\citenamefont {Drescher}\ \emph {et~al.}(2010)\citenamefont
  {Drescher}, \citenamefont {Goldstein}, \citenamefont {Michel}, \citenamefont
  {Polin},\ and\ \citenamefont {Tuval}}]{drescher_direct_2010}%
  \BibitemOpen
  \bibfield  {author} {\bibinfo {author} {\bibfnamefont {K.}~\bibnamefont
  {Drescher}}, \bibinfo {author} {\bibfnamefont {R.~E.}\ \bibnamefont
  {Goldstein}}, \bibinfo {author} {\bibfnamefont {N.}~\bibnamefont {Michel}},
  \bibinfo {author} {\bibfnamefont {M.}~\bibnamefont {Polin}},\ and\ \bibinfo
  {author} {\bibfnamefont {I.}~\bibnamefont {Tuval}},\ }\bibfield  {title}
  {\bibinfo {title} {Direct measurement of the flow field around swimming
  microorganisms},\ }\href {https://doi.org/10.1103/PhysRevLett.105.168101}
  {\bibfield  {journal} {\bibinfo  {journal} {Phys. Rev. Lett.}\ }\textbf
  {\bibinfo {volume} {105}},\ \bibinfo {pages} {168101} (\bibinfo {year}
  {2010})}\BibitemShut {NoStop}%
\bibitem [{\citenamefont {Contino}\ \emph {et~al.}(2015)\citenamefont
  {Contino}, \citenamefont {Lushi}, \citenamefont {Tuval}, \citenamefont
  {Kantsler},\ and\ \citenamefont {Polin}}]{contino_microalgae_2015}%
  \BibitemOpen
  \bibfield  {author} {\bibinfo {author} {\bibfnamefont {M.}~\bibnamefont
  {Contino}}, \bibinfo {author} {\bibfnamefont {E.}~\bibnamefont {Lushi}},
  \bibinfo {author} {\bibfnamefont {I.}~\bibnamefont {Tuval}}, \bibinfo
  {author} {\bibfnamefont {V.}~\bibnamefont {Kantsler}},\ and\ \bibinfo
  {author} {\bibfnamefont {M.}~\bibnamefont {Polin}},\ }\bibfield  {title}
  {\bibinfo {title} {Microalgae scatter off solid surfaces by hydrodynamic and
  contact forces},\ }\href {https://doi.org/10.1103/PhysRevLett.115.258102}
  {\bibfield  {journal} {\bibinfo  {journal} {Phys. Rev. Lett.}\ }\textbf
  {\bibinfo {volume} {115}},\ \bibinfo {pages} {258102} (\bibinfo {year}
  {2015})}\BibitemShut {NoStop}%
\bibitem [{\citenamefont {Schienbein}\ and\ \citenamefont
  {Gruler}(1993)}]{schienbein1993langevin}%
  \BibitemOpen
  \bibfield  {author} {\bibinfo {author} {\bibfnamefont {M.}~\bibnamefont
  {Schienbein}}\ and\ \bibinfo {author} {\bibfnamefont {H.}~\bibnamefont
  {Gruler}},\ }\bibfield  {title} {\bibinfo {title} {Langevin equation,
  {F}okker-{P}lanck equation and cell migration},\ }\href
  {https://doi.org/10.1007/BF02460652} {\bibfield  {journal} {\bibinfo
  {journal} {Bull. Math. Biol.}\ }\textbf {\bibinfo {volume} {55}},\ \bibinfo
  {pages} {585} (\bibinfo {year} {1993})}\BibitemShut {NoStop}%
\bibitem [{\citenamefont {Mogilner}(2009)}]{mogilner_mathematics_2009}%
  \BibitemOpen
  \bibfield  {author} {\bibinfo {author} {\bibfnamefont {A.}~\bibnamefont
  {Mogilner}},\ }\bibfield  {title} {\bibinfo {title} {Mathematics of cell
  motility: {H}ave we got its number?},\ }\href
  {https://doi.org/10.1007/s00285-008-0182-2} {\bibfield  {journal} {\bibinfo
  {journal} {J. Math. Biol.}\ }\textbf {\bibinfo {volume} {58}},\ \bibinfo
  {pages} {105} (\bibinfo {year} {2009})}\BibitemShut {NoStop}%
\bibitem [{\citenamefont {Li}\ \emph {et~al.}(2011)\citenamefont {Li},
  \citenamefont {Cox},\ and\ \citenamefont {Flyvbjerg}}]{li_dicty_2011}%
  \BibitemOpen
  \bibfield  {author} {\bibinfo {author} {\bibfnamefont {L.}~\bibnamefont
  {Li}}, \bibinfo {author} {\bibfnamefont {E.~C.}\ \bibnamefont {Cox}},\ and\
  \bibinfo {author} {\bibfnamefont {H.}~\bibnamefont {Flyvbjerg}},\ }\bibfield
  {title} {\bibinfo {title} {‘dicty dynamics’: \textit{Dictyostelium}
  motility as persistent random motion},\ }\href
  {https://doi.org/10.1088/1478-3975/8/4/046006} {\bibfield  {journal}
  {\bibinfo  {journal} {Phys. Biol.}\ }\textbf {\bibinfo {volume} {8}},\
  \bibinfo {pages} {046006} (\bibinfo {year} {2011})}\BibitemShut {NoStop}%
\bibitem [{\citenamefont {Aranson}(2016)}]{aranson_physical_2016}%
  \BibitemOpen
  \bibinfo {editor} {\bibfnamefont {I.~S.}\ \bibnamefont {Aranson}},\ ed.,\
  \href {https://doi.org/10.1007/978-3-319-24448-8} {\emph {\bibinfo {title}
  {Physical models of cell motility}}}\ (\bibinfo  {publisher} {Springer},\
  \bibinfo {year} {2016})\BibitemShut {NoStop}%
\bibitem [{\citenamefont {Friedrich}\ and\ \citenamefont
  {J\"ulicher}(2007)}]{friedrich_chemotaxis_2007}%
  \BibitemOpen
  \bibfield  {author} {\bibinfo {author} {\bibfnamefont {B.~M.}\ \bibnamefont
  {Friedrich}}\ and\ \bibinfo {author} {\bibfnamefont {F.}~\bibnamefont
  {J\"ulicher}},\ }\bibfield  {title} {\bibinfo {title} {Chemotaxis of sperm
  cells},\ }\href {https://doi.org/10.1073/pnas.0703530104} {\bibfield
  {journal} {\bibinfo  {journal} {Proc. Natl. Acad. Sci. USA}\ }\textbf
  {\bibinfo {volume} {104}},\ \bibinfo {pages} {13256} (\bibinfo {year}
  {2007})}\BibitemShut {NoStop}%
\bibitem [{\citenamefont {Jikeli}\ \emph {et~al.}(2015)\citenamefont {Jikeli},
  \citenamefont {Alvarez}, \citenamefont {Friedrich}, \citenamefont {Wilson},
  \citenamefont {Pascal}, \citenamefont {Colin}, \citenamefont {Pichlo},
  \citenamefont {Rennhack}, \citenamefont {Brenker},\ and\ \citenamefont
  {Kaupp}}]{jikeli_sperm_2015}%
  \BibitemOpen
  \bibfield  {author} {\bibinfo {author} {\bibfnamefont {J.~F.}\ \bibnamefont
  {Jikeli}}, \bibinfo {author} {\bibfnamefont {L.}~\bibnamefont {Alvarez}},
  \bibinfo {author} {\bibfnamefont {B.~M.}\ \bibnamefont {Friedrich}}, \bibinfo
  {author} {\bibfnamefont {L.~G.}\ \bibnamefont {Wilson}}, \bibinfo {author}
  {\bibfnamefont {R.}~\bibnamefont {Pascal}}, \bibinfo {author} {\bibfnamefont
  {R.}~\bibnamefont {Colin}}, \bibinfo {author} {\bibfnamefont
  {M.}~\bibnamefont {Pichlo}}, \bibinfo {author} {\bibfnamefont
  {A.}~\bibnamefont {Rennhack}}, \bibinfo {author} {\bibfnamefont
  {C.}~\bibnamefont {Brenker}},\ and\ \bibinfo {author} {\bibfnamefont {U.~B.}\
  \bibnamefont {Kaupp}},\ }\bibfield  {title} {\bibinfo {title} {Sperm
  navigation along helical paths in 3d chemoattractant landscapes},\ }\href
  {https://doi.org/10.1038/ncomms8985} {\bibfield  {journal} {\bibinfo
  {journal} {Nat. Commun.}\ }\textbf {\bibinfo {volume} {6}},\ \bibinfo {pages}
  {7985} (\bibinfo {year} {2015})}\BibitemShut {NoStop}%
\bibitem [{\citenamefont {Paxton}\ \emph {et~al.}(2004)\citenamefont {Paxton},
  \citenamefont {Kistler}, \citenamefont {Olmeda}, \citenamefont {Sen},
  \citenamefont {St.~Angelo}, \citenamefont {Cao}, \citenamefont {Mallouk},
  \citenamefont {Lammert},\ and\ \citenamefont
  {Crespi}}]{paxton_catalytic_2004}%
  \BibitemOpen
  \bibfield  {author} {\bibinfo {author} {\bibfnamefont {W.~F.}\ \bibnamefont
  {Paxton}}, \bibinfo {author} {\bibfnamefont {K.~C.}\ \bibnamefont {Kistler}},
  \bibinfo {author} {\bibfnamefont {C.~C.}\ \bibnamefont {Olmeda}}, \bibinfo
  {author} {\bibfnamefont {A.}~\bibnamefont {Sen}}, \bibinfo {author}
  {\bibfnamefont {S.~K.}\ \bibnamefont {St.~Angelo}}, \bibinfo {author}
  {\bibfnamefont {Y.}~\bibnamefont {Cao}}, \bibinfo {author} {\bibfnamefont
  {T.~E.}\ \bibnamefont {Mallouk}}, \bibinfo {author} {\bibfnamefont {P.~E.}\
  \bibnamefont {Lammert}},\ and\ \bibinfo {author} {\bibfnamefont {V.~H.}\
  \bibnamefont {Crespi}},\ }\bibfield  {title} {\bibinfo {title} {Catalytic
  nanomotors: {A}utonomous movement of striped nanorods},\ }\href
  {https://doi.org/10.1021/ja047697z} {\bibfield  {journal} {\bibinfo
  {journal} {J. Am. Chem. Soc.}\ }\textbf {\bibinfo {volume} {126}},\ \bibinfo
  {pages} {13424} (\bibinfo {year} {2004})}\BibitemShut {NoStop}%
\bibitem [{\citenamefont {Howse}\ \emph {et~al.}(2007)\citenamefont {Howse},
  \citenamefont {Jones}, \citenamefont {Ryan}, \citenamefont {Gough},
  \citenamefont {Vafabakhsh},\ and\ \citenamefont
  {Golestanian}}]{howse_self_2007}%
  \BibitemOpen
  \bibfield  {author} {\bibinfo {author} {\bibfnamefont {J.~R.}\ \bibnamefont
  {Howse}}, \bibinfo {author} {\bibfnamefont {R.~A.~L.}\ \bibnamefont {Jones}},
  \bibinfo {author} {\bibfnamefont {A.~J.}\ \bibnamefont {Ryan}}, \bibinfo
  {author} {\bibfnamefont {T.}~\bibnamefont {Gough}}, \bibinfo {author}
  {\bibfnamefont {R.}~\bibnamefont {Vafabakhsh}},\ and\ \bibinfo {author}
  {\bibfnamefont {R.}~\bibnamefont {Golestanian}},\ }\bibfield  {title}
  {\bibinfo {title} {Self-motile colloidal particles: {F}rom directed
  propulsion to random walk},\ }\href
  {https://doi.org/10.1103/PhysRevLett.99.048102} {\bibfield  {journal}
  {\bibinfo  {journal} {Phys. Rev. Lett.}\ }\textbf {\bibinfo {volume} {99}},\
  \bibinfo {pages} {048102} (\bibinfo {year} {2007})}\BibitemShut {NoStop}%
\bibitem [{\citenamefont {Buttinoni}\ \emph {et~al.}(2012)\citenamefont
  {Buttinoni}, \citenamefont {Volpe}, \citenamefont {Kümmel}, \citenamefont
  {Volpe},\ and\ \citenamefont {Bechinger}}]{buttinoni_active_2012}%
  \BibitemOpen
  \bibfield  {author} {\bibinfo {author} {\bibfnamefont {I.}~\bibnamefont
  {Buttinoni}}, \bibinfo {author} {\bibfnamefont {G.}~\bibnamefont {Volpe}},
  \bibinfo {author} {\bibfnamefont {F.}~\bibnamefont {Kümmel}}, \bibinfo
  {author} {\bibfnamefont {G.}~\bibnamefont {Volpe}},\ and\ \bibinfo {author}
  {\bibfnamefont {C.}~\bibnamefont {Bechinger}},\ }\bibfield  {title} {\bibinfo
  {title} {Active {B}rownian motion tunable by light},\ }\href
  {https://doi.org/10.1088/0953-8984/24/28/284129} {\bibfield  {journal}
  {\bibinfo  {journal} {J. Phys.: Condens. Matter}\ }\textbf {\bibinfo {volume}
  {24}},\ \bibinfo {pages} {284129} (\bibinfo {year} {2012})}\BibitemShut
  {NoStop}%
\bibitem [{\citenamefont {Aranson}(2013)}]{aranson_active_2013}%
  \BibitemOpen
  \bibfield  {author} {\bibinfo {author} {\bibfnamefont {I.~S.}\ \bibnamefont
  {Aranson}},\ }\bibfield  {title} {\bibinfo {title} {Active colloids},\ }\href
  {https://doi.org/10.3367/UFNe.0183.201301e.0087} {\bibfield  {journal}
  {\bibinfo  {journal} {Phys.-Uspekhi}\ }\textbf {\bibinfo {volume} {56}},\
  \bibinfo {pages} {79} (\bibinfo {year} {2013})}\BibitemShut {NoStop}%
\bibitem [{\citenamefont {Muraveva}\ \emph {et~al.}(2022)\citenamefont
  {Muraveva}, \citenamefont {Bekir}, \citenamefont {Lomadze}, \citenamefont
  {Gro{\ss}mann}, \citenamefont {Beta},\ and\ \citenamefont
  {Santer}}]{muraveva2022interplay}%
  \BibitemOpen
  \bibfield  {author} {\bibinfo {author} {\bibfnamefont {V.}~\bibnamefont
  {Muraveva}}, \bibinfo {author} {\bibfnamefont {M.}~\bibnamefont {Bekir}},
  \bibinfo {author} {\bibfnamefont {N.}~\bibnamefont {Lomadze}}, \bibinfo
  {author} {\bibfnamefont {R.}~\bibnamefont {Gro{\ss}mann}}, \bibinfo {author}
  {\bibfnamefont {C.}~\bibnamefont {Beta}},\ and\ \bibinfo {author}
  {\bibfnamefont {S.}~\bibnamefont {Santer}},\ }\bibfield  {title} {\bibinfo
  {title} {Interplay of diffusio-and thermo-osmotic flows generated by single
  light stimulus},\ }\href {https://doi.org/10.1063/5.0090229} {\bibfield
  {journal} {\bibinfo  {journal} {Appl. Phys. Lett.}\ }\textbf {\bibinfo
  {volume} {120}} (\bibinfo {year} {2022})}\BibitemShut {NoStop}%
\bibitem [{\citenamefont {Bricard}\ \emph {et~al.}(2013)\citenamefont
  {Bricard}, \citenamefont {Caussin}, \citenamefont {Desreumaux}, \citenamefont
  {Dauchot},\ and\ \citenamefont {Bartolo}}]{bricard2013emergence}%
  \BibitemOpen
  \bibfield  {author} {\bibinfo {author} {\bibfnamefont {A.}~\bibnamefont
  {Bricard}}, \bibinfo {author} {\bibfnamefont {J.-B.}\ \bibnamefont
  {Caussin}}, \bibinfo {author} {\bibfnamefont {N.}~\bibnamefont {Desreumaux}},
  \bibinfo {author} {\bibfnamefont {O.}~\bibnamefont {Dauchot}},\ and\ \bibinfo
  {author} {\bibfnamefont {D.}~\bibnamefont {Bartolo}},\ }\bibfield  {title}
  {\bibinfo {title} {Emergence of macroscopic directed motion in populations of
  motile colloids},\ }\href {https://doi.org/10.1038/nature12673} {\bibfield
  {journal} {\bibinfo  {journal} {Nature}\ }\textbf {\bibinfo {volume} {503}},\
  \bibinfo {pages} {95} (\bibinfo {year} {2013})}\BibitemShut {NoStop}%
\bibitem [{\citenamefont {Pradillo}\ \emph {et~al.}(2019)\citenamefont
  {Pradillo}, \citenamefont {Karani},\ and\ \citenamefont
  {Vlahovska}}]{pradillo2019quincke}%
  \BibitemOpen
  \bibfield  {author} {\bibinfo {author} {\bibfnamefont {G.~E.}\ \bibnamefont
  {Pradillo}}, \bibinfo {author} {\bibfnamefont {H.}~\bibnamefont {Karani}},\
  and\ \bibinfo {author} {\bibfnamefont {P.~M.}\ \bibnamefont {Vlahovska}},\
  }\bibfield  {title} {\bibinfo {title} {Quincke rotor dynamics in confinement:
  Rolling and hovering},\ }\href {https://doi.org/10.1039/c9sm01163c}
  {\bibfield  {journal} {\bibinfo  {journal} {Soft Matter}\ }\textbf {\bibinfo
  {volume} {15}},\ \bibinfo {pages} {6564} (\bibinfo {year}
  {2019})}\BibitemShut {NoStop}%
\bibitem [{\citenamefont {Kato}\ \emph {et~al.}(2022)\citenamefont {Kato},
  \citenamefont {Takeuchi},\ and\ \citenamefont {Sano}}]{kato2022active}%
  \BibitemOpen
  \bibfield  {author} {\bibinfo {author} {\bibfnamefont {A.~N.}\ \bibnamefont
  {Kato}}, \bibinfo {author} {\bibfnamefont {K.~A.}\ \bibnamefont {Takeuchi}},\
  and\ \bibinfo {author} {\bibfnamefont {M.}~\bibnamefont {Sano}},\ }\bibfield
  {title} {\bibinfo {title} {Active colloid with externally induced periodic
  bipolar motility and its cooperative motion},\ }\href
  {https://doi.org/10.1039/D2SM00363E} {\bibfield  {journal} {\bibinfo
  {journal} {Soft Matter}\ }\textbf {\bibinfo {volume} {18}},\ \bibinfo {pages}
  {5435} (\bibinfo {year} {2022})}\BibitemShut {NoStop}%
\bibitem [{\citenamefont {Santra}\ \emph {et~al.}(2020)\citenamefont {Santra},
  \citenamefont {Basu},\ and\ \citenamefont {Sabhapandit}}]{santra_run_2020}%
  \BibitemOpen
  \bibfield  {author} {\bibinfo {author} {\bibfnamefont {I.}~\bibnamefont
  {Santra}}, \bibinfo {author} {\bibfnamefont {U.}~\bibnamefont {Basu}},\ and\
  \bibinfo {author} {\bibfnamefont {S.}~\bibnamefont {Sabhapandit}},\
  }\bibfield  {title} {\bibinfo {title} {Run-and-tumble particles in two
  dimensions: Marginal position distributions},\ }\href
  {https://doi.org/10.1103/PhysRevE.101.062120} {\bibfield  {journal} {\bibinfo
   {journal} {Phys. Rev. E}\ }\textbf {\bibinfo {volume} {101}},\ \bibinfo
  {pages} {062120} (\bibinfo {year} {2020})}\BibitemShut {NoStop}%
\bibitem [{\citenamefont {Kurzthaler}\ \emph {et~al.}(2024)\citenamefont
  {Kurzthaler}, \citenamefont {Zhao}, \citenamefont {Zhou}, \citenamefont
  {Schwarz-Linek}, \citenamefont {Devailly}, \citenamefont {Arlt},
  \citenamefont {Huang}, \citenamefont {Poon}, \citenamefont {Franosch},
  \citenamefont {Tailleur},\ and\ \citenamefont
  {Martinez}}]{kurzthaler_characterization_2024}%
  \BibitemOpen
  \bibfield  {author} {\bibinfo {author} {\bibfnamefont {C.}~\bibnamefont
  {Kurzthaler}}, \bibinfo {author} {\bibfnamefont {Y.}~\bibnamefont {Zhao}},
  \bibinfo {author} {\bibfnamefont {N.}~\bibnamefont {Zhou}}, \bibinfo {author}
  {\bibfnamefont {J.}~\bibnamefont {Schwarz-Linek}}, \bibinfo {author}
  {\bibfnamefont {C.}~\bibnamefont {Devailly}}, \bibinfo {author}
  {\bibfnamefont {J.}~\bibnamefont {Arlt}}, \bibinfo {author} {\bibfnamefont
  {J.-D.}\ \bibnamefont {Huang}}, \bibinfo {author} {\bibfnamefont {W.~C.~K.}\
  \bibnamefont {Poon}}, \bibinfo {author} {\bibfnamefont {T.}~\bibnamefont
  {Franosch}}, \bibinfo {author} {\bibfnamefont {J.}~\bibnamefont {Tailleur}},\
  and\ \bibinfo {author} {\bibfnamefont {V.~A.}\ \bibnamefont {Martinez}},\
  }\bibfield  {title} {\bibinfo {title} {Characterization and control of the
  run-and-tumble dynamics of \textit{Escherichia coli}},\ }\href
  {https://doi.org/10.1103/PhysRevLett.132.038302} {\bibfield  {journal}
  {\bibinfo  {journal} {Phys. Rev. Lett.}\ }\textbf {\bibinfo {volume} {132}},\
  \bibinfo {pages} {038302} (\bibinfo {year} {2024})}\BibitemShut {NoStop}%
\bibitem [{lor()}]{lorentzgas}%
  \BibitemOpen
  \bibinfo {note} {In this regard, we point out that similarities and crucial
  differences of run-and-tumble motion of bacteria, potentially in disordered
  media, and the Lorentz model were discussed in
  Refs~\cite{martens2012probability,chepizhko2013diffusion,zeitz2017active,salgado2022active}.
  Moreover, trajectories of run-and-tumble particles are similar in their
  structural properties to the statistical properties of
  polymers~\cite{doi1988theory}.}\BibitemShut {Stop}%
\bibitem [{\citenamefont {K\"{u}hn}\ \emph {et~al.}(2017)\citenamefont
  {K\"{u}hn}, \citenamefont {Schmidt}, \citenamefont {Eckhardt},\ and\
  \citenamefont {Thormann}}]{kuehn_bacteria_2017}%
  \BibitemOpen
  \bibfield  {author} {\bibinfo {author} {\bibfnamefont {M.~J.}\ \bibnamefont
  {K\"{u}hn}}, \bibinfo {author} {\bibfnamefont {F.~K.}\ \bibnamefont
  {Schmidt}}, \bibinfo {author} {\bibfnamefont {B.}~\bibnamefont {Eckhardt}},\
  and\ \bibinfo {author} {\bibfnamefont {K.~M.}\ \bibnamefont {Thormann}},\
  }\bibfield  {title} {\bibinfo {title} {Bacteria exploit a polymorphic
  instability of the flagellar filament to escape from traps},\ }\href
  {https://doi.org/10.1073/pnas.1701644114} {\bibfield  {journal} {\bibinfo
  {journal} {Proc. Natl. Acad. Sci. USA}\ }\textbf {\bibinfo {volume} {114}},\
  \bibinfo {pages} {6340} (\bibinfo {year} {2017})}\BibitemShut {NoStop}%
\bibitem [{\citenamefont {Hintsche}\ \emph {et~al.}(2017)\citenamefont
  {Hintsche}, \citenamefont {Waljor}, \citenamefont {Gro{\ss}mann},
  \citenamefont {K{\"u}hn}, \citenamefont {Thormann}, \citenamefont {Peruani},\
  and\ \citenamefont {Beta}}]{hintsche2017polar}%
  \BibitemOpen
  \bibfield  {author} {\bibinfo {author} {\bibfnamefont {M.}~\bibnamefont
  {Hintsche}}, \bibinfo {author} {\bibfnamefont {V.}~\bibnamefont {Waljor}},
  \bibinfo {author} {\bibfnamefont {R.}~\bibnamefont {Gro{\ss}mann}}, \bibinfo
  {author} {\bibfnamefont {M.~J.}\ \bibnamefont {K{\"u}hn}}, \bibinfo {author}
  {\bibfnamefont {K.~M.}\ \bibnamefont {Thormann}}, \bibinfo {author}
  {\bibfnamefont {F.}~\bibnamefont {Peruani}},\ and\ \bibinfo {author}
  {\bibfnamefont {C.}~\bibnamefont {Beta}},\ }\bibfield  {title} {\bibinfo
  {title} {A polar bundle of flagella can drive bacterial swimming by pushing,
  pulling, or coiling around the cell body},\ }\href
  {https://doi.org/10.1038/s41598-017-16428-9} {\bibfield  {journal} {\bibinfo
  {journal} {Sci. Rep.}\ }\textbf {\bibinfo {volume} {7}},\ \bibinfo {pages}
  {16771} (\bibinfo {year} {2017})}\BibitemShut {NoStop}%
\bibitem [{\citenamefont {Grognot}\ and\ \citenamefont
  {Taute}(2021)}]{grognot_more_2021}%
  \BibitemOpen
  \bibfield  {author} {\bibinfo {author} {\bibfnamefont {M.}~\bibnamefont
  {Grognot}}\ and\ \bibinfo {author} {\bibfnamefont {K.~M.}\ \bibnamefont
  {Taute}},\ }\bibfield  {title} {\bibinfo {title} {More than propellers: {H}ow
  flagella shape bacterial motility behaviors},\ }\href
  {https://doi.org/10.1016/j.mib.2021.02.005} {\bibfield  {journal} {\bibinfo
  {journal} {Curr. Opin. Microbiol.}\ }\textbf {\bibinfo {volume} {61}},\
  \bibinfo {pages} {73} (\bibinfo {year} {2021})}\BibitemShut {NoStop}%
\bibitem [{\citenamefont {Thormann}\ \emph {et~al.}(2022)\citenamefont
  {Thormann}, \citenamefont {Beta},\ and\ \citenamefont
  {K\"uhn}}]{thormann_wrapped_2022}%
  \BibitemOpen
  \bibfield  {author} {\bibinfo {author} {\bibfnamefont {K.~M.}\ \bibnamefont
  {Thormann}}, \bibinfo {author} {\bibfnamefont {C.}~\bibnamefont {Beta}},\
  and\ \bibinfo {author} {\bibfnamefont {M.~J.}\ \bibnamefont {K\"uhn}},\
  }\bibfield  {title} {\bibinfo {title} {Wrapped up: The motility of polarly
  flagellated bacteria},\ }\href
  {https://doi.org/10.1146/annurev-micro-041122-101032} {\bibfield  {journal}
  {\bibinfo  {journal} {Annu. Rev. Microbiol.}\ }\textbf {\bibinfo {volume}
  {76}},\ \bibinfo {pages} {349} (\bibinfo {year} {2022})}\BibitemShut
  {NoStop}%
\bibitem [{\citenamefont {Moreno}\ \emph {et~al.}(2020)\citenamefont {Moreno},
  \citenamefont {Flemming}, \citenamefont {Font}, \citenamefont {Holschneider},
  \citenamefont {Beta},\ and\ \citenamefont {Alonso}}]{moreno2020modeling}%
  \BibitemOpen
  \bibfield  {author} {\bibinfo {author} {\bibfnamefont {E.}~\bibnamefont
  {Moreno}}, \bibinfo {author} {\bibfnamefont {S.}~\bibnamefont {Flemming}},
  \bibinfo {author} {\bibfnamefont {F.}~\bibnamefont {Font}}, \bibinfo {author}
  {\bibfnamefont {M.}~\bibnamefont {Holschneider}}, \bibinfo {author}
  {\bibfnamefont {C.}~\bibnamefont {Beta}},\ and\ \bibinfo {author}
  {\bibfnamefont {S.}~\bibnamefont {Alonso}},\ }\bibfield  {title} {\bibinfo
  {title} {Modeling cell crawling strategies with a bistable model: From
  amoeboid to fan-shaped cell motion},\ }\href
  {https://doi.org/10.1016/j.physd.2020.132591} {\bibfield  {journal} {\bibinfo
   {journal} {Physica D}\ }\textbf {\bibinfo {volume} {412}},\ \bibinfo {pages}
  {132591} (\bibinfo {year} {2020})}\BibitemShut {NoStop}%
\bibitem [{\citenamefont {Moldenhawer}\ \emph {et~al.}(2022)\citenamefont
  {Moldenhawer}, \citenamefont {Moreno}, \citenamefont {Schindler},
  \citenamefont {Flemming}, \citenamefont {Holschneider}, \citenamefont
  {Huisinga}, \citenamefont {Alonso},\ and\ \citenamefont
  {Beta}}]{moldenhawer_spontaneous_2022}%
  \BibitemOpen
  \bibfield  {author} {\bibinfo {author} {\bibfnamefont {T.}~\bibnamefont
  {Moldenhawer}}, \bibinfo {author} {\bibfnamefont {E.}~\bibnamefont {Moreno}},
  \bibinfo {author} {\bibfnamefont {D.}~\bibnamefont {Schindler}}, \bibinfo
  {author} {\bibfnamefont {S.}~\bibnamefont {Flemming}}, \bibinfo {author}
  {\bibfnamefont {M.}~\bibnamefont {Holschneider}}, \bibinfo {author}
  {\bibfnamefont {W.}~\bibnamefont {Huisinga}}, \bibinfo {author}
  {\bibfnamefont {S.}~\bibnamefont {Alonso}},\ and\ \bibinfo {author}
  {\bibfnamefont {C.}~\bibnamefont {Beta}},\ }\bibfield  {title} {\bibinfo
  {title} {Spontaneous transitions between amoeboid and keratocyte-like modes
  of migration},\ }\href
  {https://www.frontiersin.org/articles/10.3389/fcell.2022.898351} {\bibfield
  {journal} {\bibinfo  {journal} {Front. Cell Dev. Biol.}\ }\textbf {\bibinfo
  {volume} {10}} (\bibinfo {year} {2022})}\BibitemShut {NoStop}%
\bibitem [{\citenamefont {Lepro}\ \emph {et~al.}(2022)\citenamefont {Lepro},
  \citenamefont {Gro\ss{}mann}, \citenamefont {Sharifi~Panah}, \citenamefont
  {Nagel}, \citenamefont {Klumpp}, \citenamefont {Lipowsky},\ and\
  \citenamefont {Beta}}]{lepro_optimal_2022}%
  \BibitemOpen
  \bibfield  {author} {\bibinfo {author} {\bibfnamefont {V.}~\bibnamefont
  {Lepro}}, \bibinfo {author} {\bibfnamefont {R.}~\bibnamefont {Gro\ss{}mann}},
  \bibinfo {author} {\bibfnamefont {S.}~\bibnamefont {Sharifi~Panah}}, \bibinfo
  {author} {\bibfnamefont {O.}~\bibnamefont {Nagel}}, \bibinfo {author}
  {\bibfnamefont {S.}~\bibnamefont {Klumpp}}, \bibinfo {author} {\bibfnamefont
  {R.}~\bibnamefont {Lipowsky}},\ and\ \bibinfo {author} {\bibfnamefont
  {C.}~\bibnamefont {Beta}},\ }\bibfield  {title} {\bibinfo {title} {Optimal
  cargo size for active diffusion of biohybrid microcarriers},\ }\href
  {https://doi.org/10.1103/PhysRevApplied.18.034014} {\bibfield  {journal}
  {\bibinfo  {journal} {Phys. Rev. Appl.}\ }\textbf {\bibinfo {volume} {18}},\
  \bibinfo {pages} {034014} (\bibinfo {year} {2022})}\BibitemShut {NoStop}%
\bibitem [{\citenamefont {Zaferani}\ and\ \citenamefont
  {Abbaspourrad}(2023)}]{zaferani2023biphasic}%
  \BibitemOpen
  \bibfield  {author} {\bibinfo {author} {\bibfnamefont {M.}~\bibnamefont
  {Zaferani}}\ and\ \bibinfo {author} {\bibfnamefont {A.}~\bibnamefont
  {Abbaspourrad}},\ }\bibfield  {title} {\bibinfo {title} {Biphasic
  chemokinesis of mammalian sperm},\ }\href
  {https://link.aps.org/doi/10.1103/PhysRevLett.130.248401} {\bibfield
  {journal} {\bibinfo  {journal} {Phys. Rev. Lett.}\ }\textbf {\bibinfo
  {volume} {130}},\ \bibinfo {pages} {248401} (\bibinfo {year}
  {2023})}\BibitemShut {NoStop}%
\bibitem [{\citenamefont {Strefler}\ \emph {et~al.}(2008)\citenamefont
  {Strefler}, \citenamefont {Erdmann},\ and\ \citenamefont
  {Schimansky-Geier}}]{strefler2008swarming}%
  \BibitemOpen
  \bibfield  {author} {\bibinfo {author} {\bibfnamefont {J.}~\bibnamefont
  {Strefler}}, \bibinfo {author} {\bibfnamefont {U.}~\bibnamefont {Erdmann}},\
  and\ \bibinfo {author} {\bibfnamefont {L.}~\bibnamefont {Schimansky-Geier}},\
  }\bibfield  {title} {\bibinfo {title} {Swarming in three dimensions},\ }\href
  {https://link.aps.org/doi/10.1103/PhysRevE.78.031927} {\bibfield  {journal}
  {\bibinfo  {journal} {Phys. Rev. E}\ }\textbf {\bibinfo {volume} {78}},\
  \bibinfo {pages} {031927} (\bibinfo {year} {2008})}\BibitemShut {NoStop}%
\bibitem [{\citenamefont {Gro{\ss}mann}\ \emph {et~al.}(2012)\citenamefont
  {Gro{\ss}mann}, \citenamefont {Schimansky-Geier},\ and\ \citenamefont
  {Romanczuk}}]{grossmann2012active}%
  \BibitemOpen
  \bibfield  {author} {\bibinfo {author} {\bibfnamefont {R.}~\bibnamefont
  {Gro{\ss}mann}}, \bibinfo {author} {\bibfnamefont {L.}~\bibnamefont
  {Schimansky-Geier}},\ and\ \bibinfo {author} {\bibfnamefont {P.}~\bibnamefont
  {Romanczuk}},\ }\bibfield  {title} {\bibinfo {title} {Active {B}rownian
  particles with velocity-alignment and active fluctuations},\ }\href
  {https://doi.org/10.1088/1367-2630/14/7/073033} {\bibfield  {journal}
  {\bibinfo  {journal} {New J. Phys.}\ }\textbf {\bibinfo {volume} {14}},\
  \bibinfo {pages} {073033} (\bibinfo {year} {2012})}\BibitemShut {NoStop}%
\bibitem [{\citenamefont {Kramer}\ and\ \citenamefont
  {McLaughlin}(2015)}]{kramer2015behavioral}%
  \BibitemOpen
  \bibfield  {author} {\bibinfo {author} {\bibfnamefont {D.~L.}\ \bibnamefont
  {Kramer}}\ and\ \bibinfo {author} {\bibfnamefont {R.~L.}\ \bibnamefont
  {McLaughlin}},\ }\bibfield  {title} {\bibinfo {title} {The behavioral ecology
  of intermittent locomotion},\ }\href {https://doi.org/10.1093/icb/41.2.137}
  {\bibfield  {journal} {\bibinfo  {journal} {Am. Zool.}\ }\textbf {\bibinfo
  {volume} {41}},\ \bibinfo {pages} {137} (\bibinfo {year} {2015})}\BibitemShut
  {NoStop}%
\bibitem [{\citenamefont {G{\'o}mez-Nava}\ \emph {et~al.}(2022)\citenamefont
  {G{\'o}mez-Nava}, \citenamefont {Bon},\ and\ \citenamefont
  {Peruani}}]{gomez2022intermittent}%
  \BibitemOpen
  \bibfield  {author} {\bibinfo {author} {\bibfnamefont {L.}~\bibnamefont
  {G{\'o}mez-Nava}}, \bibinfo {author} {\bibfnamefont {R.}~\bibnamefont
  {Bon}},\ and\ \bibinfo {author} {\bibfnamefont {F.}~\bibnamefont {Peruani}},\
  }\bibfield  {title} {\bibinfo {title} {Intermittent collective motion in
  sheep results from alternating the role of leader and follower},\ }\href
  {https://doi.org/10.1038/s41567-022-01769-8} {\bibfield  {journal} {\bibinfo
  {journal} {Nat. Phys.}\ }\textbf {\bibinfo {volume} {18}},\ \bibinfo {pages}
  {1494} (\bibinfo {year} {2022})}\BibitemShut {NoStop}%
\bibitem [{\citenamefont {Tunstr{\o}m}\ \emph {et~al.}(2013)\citenamefont
  {Tunstr{\o}m}, \citenamefont {Katz}, \citenamefont {Ioannou}, \citenamefont
  {Huepe}, \citenamefont {Lutz},\ and\ \citenamefont
  {Couzin}}]{tunstrom_collective_2013}%
  \BibitemOpen
  \bibfield  {author} {\bibinfo {author} {\bibfnamefont {K.}~\bibnamefont
  {Tunstr{\o}m}}, \bibinfo {author} {\bibfnamefont {Y.}~\bibnamefont {Katz}},
  \bibinfo {author} {\bibfnamefont {C.~C.}\ \bibnamefont {Ioannou}}, \bibinfo
  {author} {\bibfnamefont {C.}~\bibnamefont {Huepe}}, \bibinfo {author}
  {\bibfnamefont {M.~J.}\ \bibnamefont {Lutz}},\ and\ \bibinfo {author}
  {\bibfnamefont {I.~D.}\ \bibnamefont {Couzin}},\ }\bibfield  {title}
  {\bibinfo {title} {Collective states, multistability and transitional
  behavior in schooling fish},\ }\href
  {https://doi.org/10.1371/journal.pcbi.1002915} {\bibfield  {journal}
  {\bibinfo  {journal} {PLoS Comput. Biol.}\ }\textbf {\bibinfo {volume} {9}},\
  \bibinfo {pages} {e1002915} (\bibinfo {year} {2013})}\BibitemShut {NoStop}%
\bibitem [{\citenamefont {Lecheval}\ \emph {et~al.}(2018)\citenamefont
  {Lecheval}, \citenamefont {Jiang}, \citenamefont {Tichit}, \citenamefont
  {Sire}, \citenamefont {Hemelrijk},\ and\ \citenamefont
  {Theraulaz}}]{lecheval_social_2018}%
  \BibitemOpen
  \bibfield  {author} {\bibinfo {author} {\bibfnamefont {V.}~\bibnamefont
  {Lecheval}}, \bibinfo {author} {\bibfnamefont {L.}~\bibnamefont {Jiang}},
  \bibinfo {author} {\bibfnamefont {P.}~\bibnamefont {Tichit}}, \bibinfo
  {author} {\bibfnamefont {C.}~\bibnamefont {Sire}}, \bibinfo {author}
  {\bibfnamefont {C.~K.}\ \bibnamefont {Hemelrijk}},\ and\ \bibinfo {author}
  {\bibfnamefont {G.}~\bibnamefont {Theraulaz}},\ }\bibfield  {title} {\bibinfo
  {title} {Social conformity and propagation of information in collective
  {U}-turns of fish schools},\ }\href
  {https://doi.org/https://doi.org/10.1098/rspb.2018.0251} {\bibfield
  {journal} {\bibinfo  {journal} {Proc. R. Soc. B}\ }\textbf {\bibinfo {volume}
  {285}},\ \bibinfo {pages} {20180251} (\bibinfo {year} {2018})}\BibitemShut
  {NoStop}%
\bibitem [{\citenamefont {Chepizhko}\ and\ \citenamefont
  {Peruani}(2013)}]{chepizhko2013diffusion}%
  \BibitemOpen
  \bibfield  {author} {\bibinfo {author} {\bibfnamefont {O.}~\bibnamefont
  {Chepizhko}}\ and\ \bibinfo {author} {\bibfnamefont {F.}~\bibnamefont
  {Peruani}},\ }\bibfield  {title} {\bibinfo {title} {Diffusion, subdiffusion,
  and trapping of active particles in heterogeneous media},\ }\href
  {https://doi.org/10.1103/PhysRevLett.111.160604} {\bibfield  {journal}
  {\bibinfo  {journal} {Phys. Rev. Lett.}\ }\textbf {\bibinfo {volume} {111}},\
  \bibinfo {pages} {160604} (\bibinfo {year} {2013})}\BibitemShut {NoStop}%
\bibitem [{\citenamefont {Bhattacharjee}\ and\ \citenamefont
  {Datta}(2019)}]{bhattacharjee2019bacterial}%
  \BibitemOpen
  \bibfield  {author} {\bibinfo {author} {\bibfnamefont {T.}~\bibnamefont
  {Bhattacharjee}}\ and\ \bibinfo {author} {\bibfnamefont {S.~S.}\ \bibnamefont
  {Datta}},\ }\bibfield  {title} {\bibinfo {title} {Bacterial hopping and
  trapping in porous media},\ }\href
  {https://doi.org/10.1038/s41467-019-10115-1} {\bibfield  {journal} {\bibinfo
  {journal} {Nat. Commun.}\ }\textbf {\bibinfo {volume} {10}},\ \bibinfo
  {pages} {2075} (\bibinfo {year} {2019})}\BibitemShut {NoStop}%
\bibitem [{\citenamefont {Perez~Ipi{\~n}a}\ \emph {et~al.}(2019)\citenamefont
  {Perez~Ipi{\~n}a}, \citenamefont {Otte}, \citenamefont {Pontier-Bres},
  \citenamefont {Czerucka},\ and\ \citenamefont {Peruani}}]{perez2019bacteria}%
  \BibitemOpen
  \bibfield  {author} {\bibinfo {author} {\bibfnamefont {E.}~\bibnamefont
  {Perez~Ipi{\~n}a}}, \bibinfo {author} {\bibfnamefont {S.}~\bibnamefont
  {Otte}}, \bibinfo {author} {\bibfnamefont {R.}~\bibnamefont {Pontier-Bres}},
  \bibinfo {author} {\bibfnamefont {D.}~\bibnamefont {Czerucka}},\ and\
  \bibinfo {author} {\bibfnamefont {F.}~\bibnamefont {Peruani}},\ }\bibfield
  {title} {\bibinfo {title} {Bacteria display optimal transport near
  surfaces},\ }\href {https://doi.org/10.1038/s41567-019-0460-5} {\bibfield
  {journal} {\bibinfo  {journal} {Nat. Phys.}\ }\textbf {\bibinfo {volume}
  {15}},\ \bibinfo {pages} {610} (\bibinfo {year} {2019})}\BibitemShut
  {NoStop}%
\bibitem [{\citenamefont {Figueroa-Morales}\ \emph {et~al.}(2020)\citenamefont
  {Figueroa-Morales}, \citenamefont {Soto}, \citenamefont {Junot},
  \citenamefont {Darnige}, \citenamefont {Douarche}, \citenamefont {Martinez},
  \citenamefont {Lindner},\ and\ \citenamefont {Cl\'ement}}]{figueroa20203d}%
  \BibitemOpen
  \bibfield  {author} {\bibinfo {author} {\bibfnamefont {N.}~\bibnamefont
  {Figueroa-Morales}}, \bibinfo {author} {\bibfnamefont {R.}~\bibnamefont
  {Soto}}, \bibinfo {author} {\bibfnamefont {G.}~\bibnamefont {Junot}},
  \bibinfo {author} {\bibfnamefont {T.}~\bibnamefont {Darnige}}, \bibinfo
  {author} {\bibfnamefont {C.}~\bibnamefont {Douarche}}, \bibinfo {author}
  {\bibfnamefont {V.~A.}\ \bibnamefont {Martinez}}, \bibinfo {author}
  {\bibfnamefont {A.}~\bibnamefont {Lindner}},\ and\ \bibinfo {author}
  {\bibfnamefont {E.}~\bibnamefont {Cl\'ement}},\ }\bibfield  {title} {\bibinfo
  {title} {3d spatial exploration by \textit{E. coli} echoes motor temporal
  variability},\ }\href {https://link.aps.org/doi/10.1103/PhysRevX.10.021004}
  {\bibfield  {journal} {\bibinfo  {journal} {Phys. Rev. X}\ }\textbf {\bibinfo
  {volume} {10}},\ \bibinfo {pages} {021004} (\bibinfo {year}
  {2020})}\BibitemShut {NoStop}%
\bibitem [{\citenamefont {Raza}\ \emph {et~al.}(2023)\citenamefont {Raza},
  \citenamefont {George}, \citenamefont {Kumari}, \citenamefont {Mitra},\ and\
  \citenamefont {Paul}}]{raza2022anomalous}%
  \BibitemOpen
  \bibfield  {author} {\bibinfo {author} {\bibfnamefont {M.~R.}\ \bibnamefont
  {Raza}}, \bibinfo {author} {\bibfnamefont {J.~E.}\ \bibnamefont {George}},
  \bibinfo {author} {\bibfnamefont {S.}~\bibnamefont {Kumari}}, \bibinfo
  {author} {\bibfnamefont {M.~K.}\ \bibnamefont {Mitra}},\ and\ \bibinfo
  {author} {\bibfnamefont {D.}~\bibnamefont {Paul}},\ }\bibfield  {title}
  {\bibinfo {title} {Anomalous diffusion of \textit{E. coli} under microfluidic
  confinement and chemical gradient},\ }\href
  {https://pubs.rsc.org/en/content/articlelanding/2023/sm/d3sm00286a}
  {\bibfield  {journal} {\bibinfo  {journal} {Soft Matter}\ }\textbf {\bibinfo
  {volume} {19}},\ \bibinfo {pages} {6446} (\bibinfo {year}
  {2023})}\BibitemShut {NoStop}%
\bibitem [{\citenamefont {Raatz}\ \emph {et~al.}(2015)\citenamefont {Raatz},
  \citenamefont {Hintsche}, \citenamefont {Bahrs}, \citenamefont {Theves},\
  and\ \citenamefont {Beta}}]{raatz2015swimming}%
  \BibitemOpen
  \bibfield  {author} {\bibinfo {author} {\bibfnamefont {M.}~\bibnamefont
  {Raatz}}, \bibinfo {author} {\bibfnamefont {M.}~\bibnamefont {Hintsche}},
  \bibinfo {author} {\bibfnamefont {M.}~\bibnamefont {Bahrs}}, \bibinfo
  {author} {\bibfnamefont {M.}~\bibnamefont {Theves}},\ and\ \bibinfo {author}
  {\bibfnamefont {C.}~\bibnamefont {Beta}},\ }\bibfield  {title} {\bibinfo
  {title} {Swimming patterns of a polarly flagellated bacterium in environments
  of increasing complexity},\ }\href
  {https://link.springer.com/article/10.1140/epjst/e2015-02454-3} {\bibfield
  {journal} {\bibinfo  {journal} {Eur. Phys. J.: Spec. Top.}\ }\textbf
  {\bibinfo {volume} {224}},\ \bibinfo {pages} {1185} (\bibinfo {year}
  {2015})}\BibitemShut {NoStop}%
\bibitem [{\citenamefont {Jakuszeit}\ and\ \citenamefont
  {Croze}(2024)}]{jakuszeit2024role}%
  \BibitemOpen
  \bibfield  {author} {\bibinfo {author} {\bibfnamefont {T.}~\bibnamefont
  {Jakuszeit}}\ and\ \bibinfo {author} {\bibfnamefont {O.~A.}\ \bibnamefont
  {Croze}},\ }\bibfield  {title} {\bibinfo {title} {Role of tumbling in
  bacterial scattering at convex obstacles},\ }\href
  {https://doi.org/10.1103/PhysRevE.109.044405} {\bibfield  {journal} {\bibinfo
   {journal} {Phys. Rev. E}\ }\textbf {\bibinfo {volume} {109}},\ \bibinfo
  {pages} {044405} (\bibinfo {year} {2024})}\BibitemShut {NoStop}%
\bibitem [{\citenamefont {Karani}\ \emph {et~al.}(2019)\citenamefont {Karani},
  \citenamefont {Pradillo},\ and\ \citenamefont
  {Vlahovska}}]{karani2019tuning}%
  \BibitemOpen
  \bibfield  {author} {\bibinfo {author} {\bibfnamefont {H.}~\bibnamefont
  {Karani}}, \bibinfo {author} {\bibfnamefont {G.~E.}\ \bibnamefont
  {Pradillo}},\ and\ \bibinfo {author} {\bibfnamefont {P.~M.}\ \bibnamefont
  {Vlahovska}},\ }\bibfield  {title} {\bibinfo {title} {Tuning the random walk
  of active colloids: From individual run-and-tumble to dynamic clustering},\
  }\href {https://doi.org/10.1103/PhysRevLett.123.208002} {\bibfield  {journal}
  {\bibinfo  {journal} {Phys. Rev. Lett.}\ }\textbf {\bibinfo {volume} {123}},\
  \bibinfo {pages} {208002} (\bibinfo {year} {2019})}\BibitemShut {NoStop}%
\bibitem [{\citenamefont {Mora}\ and\ \citenamefont
  {Pomeau}(2018)}]{mora2018brownian}%
  \BibitemOpen
  \bibfield  {author} {\bibinfo {author} {\bibfnamefont {S.}~\bibnamefont
  {Mora}}\ and\ \bibinfo {author} {\bibfnamefont {Y.}~\bibnamefont {Pomeau}},\
  }\bibfield  {title} {\bibinfo {title} {Brownian diffusion in a dilute field
  of traps is {F}ickean but non-{G}aussian},\ }\href
  {https://link.aps.org/doi/10.1103/PhysRevE.98.040101} {\bibfield  {journal}
  {\bibinfo  {journal} {Phys. Rev. E}\ }\textbf {\bibinfo {volume} {98}},\
  \bibinfo {pages} {040101} (\bibinfo {year} {2018})}\BibitemShut {NoStop}%
\bibitem [{\citenamefont {Doerries}\ \emph
  {et~al.}(2022{\natexlab{a}})\citenamefont {Doerries}, \citenamefont
  {Chechkin}, \citenamefont {Schumer},\ and\ \citenamefont
  {Metzler}}]{doerries2022rate}%
  \BibitemOpen
  \bibfield  {author} {\bibinfo {author} {\bibfnamefont {T.~J.}\ \bibnamefont
  {Doerries}}, \bibinfo {author} {\bibfnamefont {A.~V.}\ \bibnamefont
  {Chechkin}}, \bibinfo {author} {\bibfnamefont {R.}~\bibnamefont {Schumer}},\
  and\ \bibinfo {author} {\bibfnamefont {R.}~\bibnamefont {Metzler}},\
  }\bibfield  {title} {\bibinfo {title} {Rate equations, spatial moments, and
  concentration profiles for mobile-immobile models with power-law and mixed
  waiting time distributions},\ }\href
  {https://link.aps.org/doi/10.1103/PhysRevE.105.014105} {\bibfield  {journal}
  {\bibinfo  {journal} {Phys. Rev. E}\ }\textbf {\bibinfo {volume} {105}},\
  \bibinfo {pages} {014105} (\bibinfo {year} {2022}{\natexlab{a}})}\BibitemShut
  {NoStop}%
\bibitem [{\citenamefont {Doerries}\ \emph
  {et~al.}(2022{\natexlab{b}})\citenamefont {Doerries}, \citenamefont
  {Chechkin},\ and\ \citenamefont {Metzler}}]{doerries2022apparent}%
  \BibitemOpen
  \bibfield  {author} {\bibinfo {author} {\bibfnamefont {T.~J.}\ \bibnamefont
  {Doerries}}, \bibinfo {author} {\bibfnamefont {A.~V.}\ \bibnamefont
  {Chechkin}},\ and\ \bibinfo {author} {\bibfnamefont {R.}~\bibnamefont
  {Metzler}},\ }\bibfield  {title} {\bibinfo {title} {Apparent anomalous
  diffusion and non-{G}aussian distributions in a simple mobile-immobile
  transport model with {P}oissonian switching},\ }\href
  {https://royalsocietypublishing.org/doi/10.1098/rsif.2022.0233} {\bibfield
  {journal} {\bibinfo  {journal} {J. R. Soc. Interface}\ }\textbf {\bibinfo
  {volume} {19}},\ \bibinfo {pages} {20220233} (\bibinfo {year}
  {2022}{\natexlab{b}})}\BibitemShut {NoStop}%
\bibitem [{\citenamefont {Doerries}\ \emph {et~al.}(2023)\citenamefont
  {Doerries}, \citenamefont {Metzler},\ and\ \citenamefont
  {Chechkin}}]{doerries2023emergent}%
  \BibitemOpen
  \bibfield  {author} {\bibinfo {author} {\bibfnamefont {T.~J.}\ \bibnamefont
  {Doerries}}, \bibinfo {author} {\bibfnamefont {R.}~\bibnamefont {Metzler}},\
  and\ \bibinfo {author} {\bibfnamefont {A.~V.}\ \bibnamefont {Chechkin}},\
  }\bibfield  {title} {\bibinfo {title} {Emergent anomalous transport and
  non-{G}aussianity in a simple mobile-immobile model: The role of advection},\
  }\href {https://doi.org/10.1088/1367-2630/acd950} {\bibfield  {journal}
  {\bibinfo  {journal} {New J. Phys.}\ }\textbf {\bibinfo {volume} {25}},\
  \bibinfo {pages} {063009} (\bibinfo {year} {2023})}\BibitemShut {NoStop}%
\bibitem [{\citenamefont {Kurilovich}\ \emph {et~al.}(2022)\citenamefont
  {Kurilovich}, \citenamefont {Mantsevich}, \citenamefont {Mardoukhi},
  \citenamefont {Stevenson}, \citenamefont {Chechkin},\ and\ \citenamefont
  {Palyulin}}]{kurilovich2022nonmarkovian}%
  \BibitemOpen
  \bibfield  {author} {\bibinfo {author} {\bibfnamefont {A.~A.}\ \bibnamefont
  {Kurilovich}}, \bibinfo {author} {\bibfnamefont {V.~N.}\ \bibnamefont
  {Mantsevich}}, \bibinfo {author} {\bibfnamefont {Y.}~\bibnamefont
  {Mardoukhi}}, \bibinfo {author} {\bibfnamefont {K.~J.}\ \bibnamefont
  {Stevenson}}, \bibinfo {author} {\bibfnamefont {A.~V.}\ \bibnamefont
  {Chechkin}},\ and\ \bibinfo {author} {\bibfnamefont {V.~V.}\ \bibnamefont
  {Palyulin}},\ }\bibfield  {title} {\bibinfo {title} {Non-markovian diffusion
  of excitons in layered perovskites and transition metal dichalcogenides},\
  }\href {https://doi.org/10.1039/D2CP00557C} {\bibfield  {journal} {\bibinfo
  {journal} {Phys. Chem. Chem. Phys.}\ }\textbf {\bibinfo {volume} {24}},\
  \bibinfo {pages} {13941} (\bibinfo {year} {2022})}\BibitemShut {NoStop}%
\bibitem [{\citenamefont {Igaev}\ \emph {et~al.}(2014)\citenamefont {Igaev},
  \citenamefont {Janning}, \citenamefont {Sündermann}, \citenamefont
  {Niewidok}, \citenamefont {Brandt},\ and\ \citenamefont
  {Junge}}]{igaev2014refined}%
  \BibitemOpen
  \bibfield  {author} {\bibinfo {author} {\bibfnamefont {M.}~\bibnamefont
  {Igaev}}, \bibinfo {author} {\bibfnamefont {D.}~\bibnamefont {Janning}},
  \bibinfo {author} {\bibfnamefont {F.}~\bibnamefont {Sündermann}}, \bibinfo
  {author} {\bibfnamefont {B.}~\bibnamefont {Niewidok}}, \bibinfo {author}
  {\bibfnamefont {R.}~\bibnamefont {Brandt}},\ and\ \bibinfo {author}
  {\bibfnamefont {W.}~\bibnamefont {Junge}},\ }\bibfield  {title} {\bibinfo
  {title} {A refined reaction-diffusion model of tau-microtubule dynamics and
  its application in {FDAP} analysis},\ }\href
  {https://doi.org/https://doi.org/10.1016/j.bpj.2014.09.016} {\bibfield
  {journal} {\bibinfo  {journal} {Biophys. J.}\ }\textbf {\bibinfo {volume}
  {107}},\ \bibinfo {pages} {2567} (\bibinfo {year} {2014})}\BibitemShut
  {NoStop}%
\bibitem [{per()}]{peruani2023active}%
  \BibitemOpen
  \href@noop {} {}\bibinfo {note} {F. Peruani and D. Chaudhuri, Active stop and
  go motion: A strategy to improve spatial exploration?,
  \href{https://arxiv.org/abs/2306.05647}{arXiv:2306.05647 (2023)}}\BibitemShut
  {NoStop}%
\bibitem [{\citenamefont {Perez}\ \emph {et~al.}(2021)\citenamefont {Perez},
  \citenamefont {Bhattacharjee}, \citenamefont {Datta}, \citenamefont
  {Parashar},\ and\ \citenamefont {Sund}}]{perez_impact_2021}%
  \BibitemOpen
  \bibfield  {author} {\bibinfo {author} {\bibfnamefont {L.~J.}\ \bibnamefont
  {Perez}}, \bibinfo {author} {\bibfnamefont {T.}~\bibnamefont
  {Bhattacharjee}}, \bibinfo {author} {\bibfnamefont {S.~S.}\ \bibnamefont
  {Datta}}, \bibinfo {author} {\bibfnamefont {R.}~\bibnamefont {Parashar}},\
  and\ \bibinfo {author} {\bibfnamefont {N.~L.}\ \bibnamefont {Sund}},\
  }\bibfield  {title} {\bibinfo {title} {Impact of confined geometries on
  hopping and trapping of motile bacteria in porous media},\ }\href
  {https://doi.org/10.1103/PhysRevE.103.012611} {\bibfield  {journal} {\bibinfo
   {journal} {Phys. Rev. E}\ }\textbf {\bibinfo {volume} {103}},\ \bibinfo
  {pages} {012611} (\bibinfo {year} {2021})}\BibitemShut {NoStop}%
\bibitem [{mat()}]{mattingly2023bacterial}%
  \BibitemOpen
  \href@noop {} {}\bibinfo {note} {H. H. Mattingly, Bacterial diffusion in
  disordered media, by forgetting the media,
  \href{https://arxiv.org/abs/2311.10612}{arXiv:2311.10612 (2023)}}\BibitemShut
  {NoStop}%
\bibitem [{\citenamefont {Saintillan}(2023)}]{saintillan2023dispersion}%
  \BibitemOpen
  \bibfield  {author} {\bibinfo {author} {\bibfnamefont {D.}~\bibnamefont
  {Saintillan}},\ }\bibfield  {title} {\bibinfo {title} {Dispersion of
  run-and-tumble microswimmers through disordered media},\ }\href
  {https://doi.org/10.1103/PhysRevE.108.064608} {\bibfield  {journal} {\bibinfo
   {journal} {Phys. Rev. E}\ }\textbf {\bibinfo {volume} {108}},\ \bibinfo
  {pages} {064608} (\bibinfo {year} {2023})}\BibitemShut {NoStop}%
\bibitem [{\citenamefont {Kurzthaler}\ \emph {et~al.}(2021)\citenamefont
  {Kurzthaler}, \citenamefont {Mandal}, \citenamefont {Bhattacharjee},
  \citenamefont {L{\"o}wen}, \citenamefont {Datta},\ and\ \citenamefont
  {Stone}}]{kurzthaler2021geometric}%
  \BibitemOpen
  \bibfield  {author} {\bibinfo {author} {\bibfnamefont {C.}~\bibnamefont
  {Kurzthaler}}, \bibinfo {author} {\bibfnamefont {S.}~\bibnamefont {Mandal}},
  \bibinfo {author} {\bibfnamefont {T.}~\bibnamefont {Bhattacharjee}}, \bibinfo
  {author} {\bibfnamefont {H.}~\bibnamefont {L{\"o}wen}}, \bibinfo {author}
  {\bibfnamefont {S.~S.}\ \bibnamefont {Datta}},\ and\ \bibinfo {author}
  {\bibfnamefont {H.~A.}\ \bibnamefont {Stone}},\ }\bibfield  {title} {\bibinfo
  {title} {A geometric criterion for the optimal spreading of active polymers
  in porous media},\ }\href {https://doi.org/10.1038/s41467-021-26942-0}
  {\bibfield  {journal} {\bibinfo  {journal} {Nat. Commun.}\ }\textbf {\bibinfo
  {volume} {12}},\ \bibinfo {pages} {7088} (\bibinfo {year}
  {2021})}\BibitemShut {NoStop}%
\bibitem [{\citenamefont {Lohrmann}\ and\ \citenamefont
  {Holm}(2023)}]{lohrmann_optimal_2023}%
  \BibitemOpen
  \bibfield  {author} {\bibinfo {author} {\bibfnamefont {C.}~\bibnamefont
  {Lohrmann}}\ and\ \bibinfo {author} {\bibfnamefont {C.}~\bibnamefont
  {Holm}},\ }\bibfield  {title} {\bibinfo {title} {Optimal motility strategies
  for self-propelled agents to explore porous media},\ }\href
  {https://doi.org/10.1103/PhysRevE.108.054401} {\bibfield  {journal} {\bibinfo
   {journal} {Phys. Rev. E}\ }\textbf {\bibinfo {volume} {108}},\ \bibinfo
  {pages} {054401} (\bibinfo {year} {2023})}\BibitemShut {NoStop}%
\bibitem [{\citenamefont {Portillo}\ \emph {et~al.}(2011)\citenamefont
  {Portillo}, \citenamefont {Campos},\ and\ \citenamefont
  {M{\'e}ndez}}]{portillo2011intermittent}%
  \BibitemOpen
  \bibfield  {author} {\bibinfo {author} {\bibfnamefont {I.~G.}\ \bibnamefont
  {Portillo}}, \bibinfo {author} {\bibfnamefont {D.}~\bibnamefont {Campos}},\
  and\ \bibinfo {author} {\bibfnamefont {V.}~\bibnamefont {M{\'e}ndez}},\
  }\bibfield  {title} {\bibinfo {title} {Intermittent random walks: Transport
  regimes and implications on search strategies},\ }\href
  {https://iopscience.iop.org/article/10.1088/1742-5468/2011/02/P02033/meta}
  {\bibfield  {journal} {\bibinfo  {journal} {J. Stat. Mech.}\ }\textbf
  {\bibinfo {volume} {2011}},\ \bibinfo {pages} {P02033} (\bibinfo {year}
  {2011})}\BibitemShut {NoStop}%
\bibitem [{\citenamefont {Thiel}\ \emph {et~al.}(2012)\citenamefont {Thiel},
  \citenamefont {Schimansky-Geier},\ and\ \citenamefont
  {Sokolov}}]{thiel2012anomalous}%
  \BibitemOpen
  \bibfield  {author} {\bibinfo {author} {\bibfnamefont {F.}~\bibnamefont
  {Thiel}}, \bibinfo {author} {\bibfnamefont {L.}~\bibnamefont
  {Schimansky-Geier}},\ and\ \bibinfo {author} {\bibfnamefont {I.~M.}\
  \bibnamefont {Sokolov}},\ }\bibfield  {title} {\bibinfo {title} {Anomalous
  diffusion in run-and-tumble motion},\ }\href
  {https://link.aps.org/doi/10.1103/PhysRevE.86.021117} {\bibfield  {journal}
  {\bibinfo  {journal} {Phys. Rev. E}\ }\textbf {\bibinfo {volume} {86}},\
  \bibinfo {pages} {021117} (\bibinfo {year} {2012})}\BibitemShut {NoStop}%
\bibitem [{\citenamefont {Zaburdaev}\ \emph {et~al.}(2015)\citenamefont
  {Zaburdaev}, \citenamefont {Denisov},\ and\ \citenamefont
  {Klafter}}]{zaburdaev_levy_2015}%
  \BibitemOpen
  \bibfield  {author} {\bibinfo {author} {\bibfnamefont {V.}~\bibnamefont
  {Zaburdaev}}, \bibinfo {author} {\bibfnamefont {S.}~\bibnamefont {Denisov}},\
  and\ \bibinfo {author} {\bibfnamefont {J.}~\bibnamefont {Klafter}},\
  }\bibfield  {title} {\bibinfo {title} {L\'evy walks},\ }\href
  {https://doi.org/10.1103/RevModPhys.87.483} {\bibfield  {journal} {\bibinfo
  {journal} {Rev. Mod. Phys.}\ }\textbf {\bibinfo {volume} {87}},\ \bibinfo
  {pages} {483} (\bibinfo {year} {2015})}\BibitemShut {NoStop}%
\bibitem [{\citenamefont {Detcheverry}(2017)}]{detcheverry2017generalized}%
  \BibitemOpen
  \bibfield  {author} {\bibinfo {author} {\bibfnamefont {F.}~\bibnamefont
  {Detcheverry}},\ }\bibfield  {title} {\bibinfo {title} {Generalized
  run-and-turn motions: From bacteria to {L}\'evy walks},\ }\href
  {https://link.aps.org/doi/10.1103/PhysRevE.96.012415} {\bibfield  {journal}
  {\bibinfo  {journal} {Phys. Rev. E}\ }\textbf {\bibinfo {volume} {96}},\
  \bibinfo {pages} {012415} (\bibinfo {year} {2017})}\BibitemShut {NoStop}%
\bibitem [{\citenamefont {Salgado-Garc\'{i}a}(2022)}]{salgado2022active}%
  \BibitemOpen
  \bibfield  {author} {\bibinfo {author} {\bibfnamefont {R.}~\bibnamefont
  {Salgado-Garc\'{i}a}},\ }\bibfield  {title} {\bibinfo {title} {Active
  particles in reactive disordered media: How does adsorption affect
  diffusion?},\ }\href
  {https://doi.org/https://doi.org/10.1016/j.physa.2022.127702} {\bibfield
  {journal} {\bibinfo  {journal} {Phys. A: Stat. Mech. Appl.}\ }\textbf
  {\bibinfo {volume} {603}},\ \bibinfo {pages} {127702} (\bibinfo {year}
  {2022})}\BibitemShut {NoStop}%
\bibitem [{jun()}]{jung2023hyperdiffusion}%
  \BibitemOpen
  \href@noop {} {}\bibinfo {note} {Y. Jung, Hyperdiffusion of Poissonian
  run-and-tumble particles in two dimensions,
  \href{https://arxiv.org/abs/2308.00554}{arXiv:2308.00554 (2023)}}\BibitemShut
  {NoStop}%
\bibitem [{\citenamefont {Zhao}\ \emph {et~al.}(2024)\citenamefont {Zhao},
  \citenamefont {Kurzthaler}, \citenamefont {Zhou}, \citenamefont
  {Schwarz-Linek}, \citenamefont {Devailly}, \citenamefont {Arlt},
  \citenamefont {Huang}, \citenamefont {Poon}, \citenamefont {Franosch},
  \citenamefont {Martinez},\ and\ \citenamefont
  {Tailleur}}]{zhao_quantitative_2024}%
  \BibitemOpen
  \bibfield  {author} {\bibinfo {author} {\bibfnamefont {Y.}~\bibnamefont
  {Zhao}}, \bibinfo {author} {\bibfnamefont {C.}~\bibnamefont {Kurzthaler}},
  \bibinfo {author} {\bibfnamefont {N.}~\bibnamefont {Zhou}}, \bibinfo {author}
  {\bibfnamefont {J.}~\bibnamefont {Schwarz-Linek}}, \bibinfo {author}
  {\bibfnamefont {C.}~\bibnamefont {Devailly}}, \bibinfo {author}
  {\bibfnamefont {J.}~\bibnamefont {Arlt}}, \bibinfo {author} {\bibfnamefont
  {J.-D.}\ \bibnamefont {Huang}}, \bibinfo {author} {\bibfnamefont {W.~C.~K.}\
  \bibnamefont {Poon}}, \bibinfo {author} {\bibfnamefont {T.}~\bibnamefont
  {Franosch}}, \bibinfo {author} {\bibfnamefont {V.~A.}\ \bibnamefont
  {Martinez}},\ and\ \bibinfo {author} {\bibfnamefont {J.}~\bibnamefont
  {Tailleur}},\ }\bibfield  {title} {\bibinfo {title} {Quantitative
  characterization of run-and-tumble statistics in bulk bacterial
  suspensions},\ }\href {https://doi.org/10.1103/PhysRevE.109.014612}
  {\bibfield  {journal} {\bibinfo  {journal} {Phys. Rev. E}\ }\textbf {\bibinfo
  {volume} {109}},\ \bibinfo {pages} {014612} (\bibinfo {year}
  {2024})}\BibitemShut {NoStop}%
\bibitem [{ang()}]{angelani2024anomalous}%
  \BibitemOpen
  \href@noop {} {}\bibinfo {note} {L. Angelani, A. De~Gregorio, R. Garra, and
  F. Iafrate, Anomalous random flights and time-fractional run-and-tumble
  equations, \href{https://arxiv.org/abs/2404.15941}{arXiv:2404.15941
  (2024)}}\BibitemShut {NoStop}%
\bibitem [{\citenamefont {Celani}\ and\ \citenamefont
  {Vergassola}(2010)}]{celani_bacterial_2010}%
  \BibitemOpen
  \bibfield  {author} {\bibinfo {author} {\bibfnamefont {A.}~\bibnamefont
  {Celani}}\ and\ \bibinfo {author} {\bibfnamefont {M.}~\bibnamefont
  {Vergassola}},\ }\bibfield  {title} {\bibinfo {title} {Bacterial strategies
  for chemotaxis response},\ }\href {https://doi.org/10.1073/pnas.0909673107}
  {\bibfield  {journal} {\bibinfo  {journal} {Proc. Natl. Acad. Sci. USA}\
  }\textbf {\bibinfo {volume} {107}},\ \bibinfo {pages} {1391} (\bibinfo {year}
  {2010})}\BibitemShut {NoStop}%
\bibitem [{\citenamefont {Nava}\ \emph {et~al.}(2018)\citenamefont {Nava},
  \citenamefont {Gro\ss{}mann},\ and\ \citenamefont
  {Peruani}}]{nava_markovian_2018}%
  \BibitemOpen
  \bibfield  {author} {\bibinfo {author} {\bibfnamefont {L.~G.}\ \bibnamefont
  {Nava}}, \bibinfo {author} {\bibfnamefont {R.}~\bibnamefont {Gro\ss{}mann}},\
  and\ \bibinfo {author} {\bibfnamefont {F.}~\bibnamefont {Peruani}},\
  }\bibfield  {title} {\bibinfo {title} {Markovian robots: Minimal navigation
  strategies for active particles},\ }\href
  {https://doi.org/10.1103/PhysRevE.97.042604} {\bibfield  {journal} {\bibinfo
  {journal} {Phys. Rev. E}\ }\textbf {\bibinfo {volume} {97}},\ \bibinfo
  {pages} {042604} (\bibinfo {year} {2018})}\BibitemShut {NoStop}%
\bibitem [{\citenamefont {Nava}\ \emph {et~al.}(2020)\citenamefont {Nava},
  \citenamefont {Gro\ss{}mann}, \citenamefont {Hintsche}, \citenamefont
  {Beta},\ and\ \citenamefont {Peruani}}]{nava_novel_2020}%
  \BibitemOpen
  \bibfield  {author} {\bibinfo {author} {\bibfnamefont {L.~G.}\ \bibnamefont
  {Nava}}, \bibinfo {author} {\bibfnamefont {R.}~\bibnamefont {Gro\ss{}mann}},
  \bibinfo {author} {\bibfnamefont {M.}~\bibnamefont {Hintsche}}, \bibinfo
  {author} {\bibfnamefont {C.}~\bibnamefont {Beta}},\ and\ \bibinfo {author}
  {\bibfnamefont {F.}~\bibnamefont {Peruani}},\ }\bibfield  {title} {\bibinfo
  {title} {A novel approach to chemotaxis: Active particles guided by internal
  clocks},\ }\href {https://doi.org/10.1209/0295-5075/130/68002} {\bibfield
  {journal} {\bibinfo  {journal} {EPL}\ }\textbf {\bibinfo {volume} {130}},\
  \bibinfo {pages} {68002} (\bibinfo {year} {2020})}\BibitemShut {NoStop}%
\bibitem [{\citenamefont {Pohl}\ \emph {et~al.}(2017)\citenamefont {Pohl},
  \citenamefont {Hintsche}, \citenamefont {Alirezaeizanjani}, \citenamefont
  {Seyrich}, \citenamefont {Beta},\ and\ \citenamefont
  {Stark}}]{pohl2017inferring}%
  \BibitemOpen
  \bibfield  {author} {\bibinfo {author} {\bibfnamefont {O.}~\bibnamefont
  {Pohl}}, \bibinfo {author} {\bibfnamefont {M.}~\bibnamefont {Hintsche}},
  \bibinfo {author} {\bibfnamefont {Z.}~\bibnamefont {Alirezaeizanjani}},
  \bibinfo {author} {\bibfnamefont {M.}~\bibnamefont {Seyrich}}, \bibinfo
  {author} {\bibfnamefont {C.}~\bibnamefont {Beta}},\ and\ \bibinfo {author}
  {\bibfnamefont {H.}~\bibnamefont {Stark}},\ }\bibfield  {title} {\bibinfo
  {title} {Inferring the chemotactic strategy of \textit{{P.~putida}} and
  \textit{{E.~coli}} using modified {K}ramers-{M}oyal coefficients},\ }\href
  {https://doi.org/10.1371/journal.pcbi.1005329} {\bibfield  {journal}
  {\bibinfo  {journal} {PLoS Comput. Biol.}\ }\textbf {\bibinfo {volume}
  {13}},\ \bibinfo {pages} {e1005329} (\bibinfo {year} {2017})}\BibitemShut
  {NoStop}%
\bibitem [{\citenamefont {Seyrich}\ \emph {et~al.}(2018)\citenamefont
  {Seyrich}, \citenamefont {Alirezaeizanjani}, \citenamefont {Beta},\ and\
  \citenamefont {Stark}}]{seyrich2018statistical}%
  \BibitemOpen
  \bibfield  {author} {\bibinfo {author} {\bibfnamefont {M.}~\bibnamefont
  {Seyrich}}, \bibinfo {author} {\bibfnamefont {Z.}~\bibnamefont
  {Alirezaeizanjani}}, \bibinfo {author} {\bibfnamefont {C.}~\bibnamefont
  {Beta}},\ and\ \bibinfo {author} {\bibfnamefont {H.}~\bibnamefont {Stark}},\
  }\bibfield  {title} {\bibinfo {title} {Statistical parameter inference of
  bacterial swimming strategies},\ }\href
  {https://doi.org/10.1088/1367-2630/aae72c} {\bibfield  {journal} {\bibinfo
  {journal} {New J. Phys.}\ }\textbf {\bibinfo {volume} {20}},\ \bibinfo
  {pages} {103033} (\bibinfo {year} {2018})}\BibitemShut {NoStop}%
\bibitem [{\citenamefont {Schnitzer}(1993)}]{schnitzer_theory_1993}%
  \BibitemOpen
  \bibfield  {author} {\bibinfo {author} {\bibfnamefont {M.~J.}\ \bibnamefont
  {Schnitzer}},\ }\bibfield  {title} {\bibinfo {title} {Theory of continuum
  random walks and application to chemotaxis},\ }\href
  {https://doi.org/10.1103/PhysRevE.48.2553} {\bibfield  {journal} {\bibinfo
  {journal} {Phys. Rev. E}\ }\textbf {\bibinfo {volume} {48}},\ \bibinfo
  {pages} {2553} (\bibinfo {year} {1993})}\BibitemShut {NoStop}%
\bibitem [{\citenamefont {Saragosti}\ \emph {et~al.}(2012)\citenamefont
  {Saragosti}, \citenamefont {Silberzan},\ and\ \citenamefont
  {Buguin}}]{saragosti_modeling_2012}%
  \BibitemOpen
  \bibfield  {author} {\bibinfo {author} {\bibfnamefont {J.}~\bibnamefont
  {Saragosti}}, \bibinfo {author} {\bibfnamefont {P.}~\bibnamefont
  {Silberzan}},\ and\ \bibinfo {author} {\bibfnamefont {A.}~\bibnamefont
  {Buguin}},\ }\bibfield  {title} {\bibinfo {title} {Modeling \textit{E. coli}
  tumbles by rotational diffusion. {I}mplications for chemotaxis},\ }\href
  {https://doi.org/10.1371/journal.pone.0035412} {\bibfield  {journal}
  {\bibinfo  {journal} {PLOS ONE}\ }\textbf {\bibinfo {volume} {7}},\ \bibinfo
  {pages} {1} (\bibinfo {year} {2012})}\BibitemShut {NoStop}%
\bibitem [{\citenamefont {Gro{\ss}mann}\ \emph {et~al.}(2016)\citenamefont
  {Gro{\ss}mann}, \citenamefont {Peruani},\ and\ \citenamefont
  {B{\"a}r}}]{grossmann2016diffusion}%
  \BibitemOpen
  \bibfield  {author} {\bibinfo {author} {\bibfnamefont {R.}~\bibnamefont
  {Gro{\ss}mann}}, \bibinfo {author} {\bibfnamefont {F.}~\bibnamefont
  {Peruani}},\ and\ \bibinfo {author} {\bibfnamefont {M.}~\bibnamefont
  {B{\"a}r}},\ }\bibfield  {title} {\bibinfo {title} {Diffusion properties of
  active particles with directional reversal},\ }\href
  {https://iopscience.iop.org/article/10.1088/1367-2630/18/4/043009} {\bibfield
   {journal} {\bibinfo  {journal} {New J. Phys.}\ }\textbf {\bibinfo {volume}
  {18}},\ \bibinfo {pages} {043009} (\bibinfo {year} {2016})}\BibitemShut
  {NoStop}%
\bibitem [{\citenamefont {Fedotov}(2016)}]{fedotov2016single}%
  \BibitemOpen
  \bibfield  {author} {\bibinfo {author} {\bibfnamefont {S.}~\bibnamefont
  {Fedotov}},\ }\bibfield  {title} {\bibinfo {title} {Single
  integrodifferential wave equation for a {L}\'evy walk},\ }\href
  {https://link.aps.org/doi/10.1103/PhysRevE.93.020101} {\bibfield  {journal}
  {\bibinfo  {journal} {Phys. Rev. E}\ }\textbf {\bibinfo {volume} {93}},\
  \bibinfo {pages} {020101} (\bibinfo {year} {2016})}\BibitemShut {NoStop}%
\bibitem [{\citenamefont {Giona}\ \emph {et~al.}(2022)\citenamefont {Giona},
  \citenamefont {Cairoli},\ and\ \citenamefont {Klages}}]{giona2022extended}%
  \BibitemOpen
  \bibfield  {author} {\bibinfo {author} {\bibfnamefont {M.}~\bibnamefont
  {Giona}}, \bibinfo {author} {\bibfnamefont {A.}~\bibnamefont {Cairoli}},\
  and\ \bibinfo {author} {\bibfnamefont {R.}~\bibnamefont {Klages}},\
  }\bibfield  {title} {\bibinfo {title} {Extended poisson-kac theory: A
  unifying framework for stochastic processes with finite propagation
  velocity},\ }\href {https://doi.org/10.1103/PhysRevX.12.021004} {\bibfield
  {journal} {\bibinfo  {journal} {Phys. Rev. X}\ }\textbf {\bibinfo {volume}
  {12}},\ \bibinfo {pages} {021004} (\bibinfo {year} {2022})}\BibitemShut
  {NoStop}%
\bibitem [{not({\natexlab{a}})}]{noteSI}%
  \BibitemOpen
  \bibinfo {note} {See Supplemental Material, which includes
  Refs.~\cite{doetsch_tabellen_1947,godreche_statistics_2001,klafter_first_2011,dubner1968numerical,glasserman2006computing,cohen2000convergence},
  at http://link.aps.org/supplemental/XXX for complementary information as well
  as technical details.}\BibitemShut {Stop}%
\bibitem [{\citenamefont {Feller}(2008)}]{feller_an2_2008}%
  \BibitemOpen
  \bibfield  {author} {\bibinfo {author} {\bibfnamefont {W.}~\bibnamefont
  {Feller}},\ }\href@noop {} {\emph {\bibinfo {title} {An Introduction to
  Probability Theory and Its Applications}}},\ Vol.~\bibinfo {volume} {2}\
  (\bibinfo  {publisher} {John Wiley \& Sons},\ \bibinfo {year}
  {2008})\BibitemShut {NoStop}%
\bibitem [{\citenamefont {Klafter}\ and\ \citenamefont
  {Sokolov}(2011)}]{klafter_first_2011}%
  \BibitemOpen
  \bibfield  {author} {\bibinfo {author} {\bibfnamefont {J.}~\bibnamefont
  {Klafter}}\ and\ \bibinfo {author} {\bibfnamefont {I.~M.}\ \bibnamefont
  {Sokolov}},\ }\href
  {https://doi.org/10.1093/acprof:oso/9780199234868.001.0001} {\emph {\bibinfo
  {title} {First steps in random walks: {F}rom tools to applications}}}\
  (\bibinfo  {publisher} {Oxford University Press},\ \bibinfo {year}
  {2011})\BibitemShut {NoStop}%
\bibitem [{\citenamefont {Peruani}\ and\ \citenamefont
  {Morelli}(2007)}]{peruani2007self}%
  \BibitemOpen
  \bibfield  {author} {\bibinfo {author} {\bibfnamefont {F.}~\bibnamefont
  {Peruani}}\ and\ \bibinfo {author} {\bibfnamefont {L.~G.}\ \bibnamefont
  {Morelli}},\ }\bibfield  {title} {\bibinfo {title} {Self-propelled particles
  with fluctuating speed and direction of motion in two dimensions},\ }\href
  {https://doi.org/10.1103/PhysRevLett.99.010602} {\bibfield  {journal}
  {\bibinfo  {journal} {Phys. Rev. Lett.}\ }\textbf {\bibinfo {volume} {99}},\
  \bibinfo {pages} {010602} (\bibinfo {year} {2007})}\BibitemShut {NoStop}%
\bibitem [{\citenamefont {Romanczuk}\ and\ \citenamefont
  {Schimansky-Geier}(2011)}]{romanczuk2011brownian}%
  \BibitemOpen
  \bibfield  {author} {\bibinfo {author} {\bibfnamefont {P.}~\bibnamefont
  {Romanczuk}}\ and\ \bibinfo {author} {\bibfnamefont {L.}~\bibnamefont
  {Schimansky-Geier}},\ }\bibfield  {title} {\bibinfo {title} {Brownian motion
  with active fluctuations},\ }\href
  {https://doi.org/10.1103/PhysRevLett.106.230601} {\bibfield  {journal}
  {\bibinfo  {journal} {Phys. Rev. Lett.}\ }\textbf {\bibinfo {volume} {106}},\
  \bibinfo {pages} {230601} (\bibinfo {year} {2011})}\BibitemShut {NoStop}%
\bibitem [{\citenamefont {Sevilla}\ and\ \citenamefont
  {G\'omez~Nava}(2014)}]{sevilla_theory_2014}%
  \BibitemOpen
  \bibfield  {author} {\bibinfo {author} {\bibfnamefont {F.~J.}\ \bibnamefont
  {Sevilla}}\ and\ \bibinfo {author} {\bibfnamefont {L.~A.}\ \bibnamefont
  {G\'omez~Nava}},\ }\bibfield  {title} {\bibinfo {title} {Theory of diffusion
  of active particles that move at constant speed in two dimensions},\ }\href
  {https://doi.org/10.1103/PhysRevE.90.022130} {\bibfield  {journal} {\bibinfo
  {journal} {Phys. Rev. E}\ }\textbf {\bibinfo {volume} {90}},\ \bibinfo
  {pages} {022130} (\bibinfo {year} {2014})}\BibitemShut {NoStop}%
\bibitem [{\citenamefont {Taylor}\ and\ \citenamefont
  {Koshland}(1974)}]{taylor_reversal_1974}%
  \BibitemOpen
  \bibfield  {author} {\bibinfo {author} {\bibfnamefont {B.~L.}\ \bibnamefont
  {Taylor}}\ and\ \bibinfo {author} {\bibfnamefont {D.}~\bibnamefont
  {Koshland}},\ }\bibfield  {title} {\bibinfo {title} {Reversal of flagellar
  rotation in monotrichous and peritrichous bacteria: Generation of changes in
  direction},\ }\href {https://doi.org/10.1128/jb.119.2.640-642.1974}
  {\bibfield  {journal} {\bibinfo  {journal} {J. Bacteriol.}\ }\textbf
  {\bibinfo {volume} {119}},\ \bibinfo {pages} {640} (\bibinfo {year}
  {1974})}\BibitemShut {NoStop}%
\bibitem [{\citenamefont {Johansen}\ \emph {et~al.}(2002)\citenamefont
  {Johansen}, \citenamefont {Pinhassi}, \citenamefont {Blackburn},
  \citenamefont {Zweifel},\ and\ \citenamefont
  {Hagstr\"om}}]{johansen_variability_2002}%
  \BibitemOpen
  \bibfield  {author} {\bibinfo {author} {\bibfnamefont {J.~E.}\ \bibnamefont
  {Johansen}}, \bibinfo {author} {\bibfnamefont {J.}~\bibnamefont {Pinhassi}},
  \bibinfo {author} {\bibfnamefont {N.}~\bibnamefont {Blackburn}}, \bibinfo
  {author} {\bibfnamefont {U.~L.}\ \bibnamefont {Zweifel}},\ and\ \bibinfo
  {author} {\bibfnamefont {A.}~\bibnamefont {Hagstr\"om}},\ }\bibfield  {title}
  {\bibinfo {title} {Variability in motility characteristics among marine
  bacteria},\ }\href {https://doi.org/10.3354/ame028229} {\bibfield  {journal}
  {\bibinfo  {journal} {Aquat. Microb. Ecol.}\ }\textbf {\bibinfo {volume}
  {28}},\ \bibinfo {pages} {229} (\bibinfo {year} {2002})}\BibitemShut
  {NoStop}%
\bibitem [{\citenamefont {Leonardy}\ \emph {et~al.}(2008)\citenamefont
  {Leonardy}, \citenamefont {Bulyha},\ and\ \citenamefont
  {S{\o}gaard-Andersen}}]{leonardy_reversing_2008}%
  \BibitemOpen
  \bibfield  {author} {\bibinfo {author} {\bibfnamefont {S.}~\bibnamefont
  {Leonardy}}, \bibinfo {author} {\bibfnamefont {I.}~\bibnamefont {Bulyha}},\
  and\ \bibinfo {author} {\bibfnamefont {L.}~\bibnamefont
  {S{\o}gaard-Andersen}},\ }\bibfield  {title} {\bibinfo {title} {Reversing
  cells and oscillating motility proteins},\ }\href
  {https://doi.org/10.1039/B806640J} {\bibfield  {journal} {\bibinfo  {journal}
  {Mol. BioSyst.}\ }\textbf {\bibinfo {volume} {4}},\ \bibinfo {pages} {1009}
  (\bibinfo {year} {2008})}\BibitemShut {NoStop}%
\bibitem [{\citenamefont {Rashkov}\ \emph {et~al.}(2012)\citenamefont
  {Rashkov}, \citenamefont {Schmitt}, \citenamefont {S{\o}gaard-Andersen},
  \citenamefont {Lenz},\ and\ \citenamefont {Dahlke}}]{rashkov_model_2012}%
  \BibitemOpen
  \bibfield  {author} {\bibinfo {author} {\bibfnamefont {P.}~\bibnamefont
  {Rashkov}}, \bibinfo {author} {\bibfnamefont {B.~A.}\ \bibnamefont
  {Schmitt}}, \bibinfo {author} {\bibfnamefont {L.}~\bibnamefont
  {S{\o}gaard-Andersen}}, \bibinfo {author} {\bibfnamefont {P.}~\bibnamefont
  {Lenz}},\ and\ \bibinfo {author} {\bibfnamefont {S.}~\bibnamefont {Dahlke}},\
  }\bibfield  {title} {\bibinfo {title} {A model of oscillatory protein
  dynamics in bacteria},\ }\href {https://doi.org/10.1007/s11538-012-9752-y}
  {\bibfield  {journal} {\bibinfo  {journal} {B. Math. Biol.}\ }\textbf
  {\bibinfo {volume} {74}},\ \bibinfo {pages} {2183} (\bibinfo {year}
  {2012})}\BibitemShut {NoStop}%
\bibitem [{\citenamefont {Wu}\ \emph {et~al.}(2009)\citenamefont {Wu},
  \citenamefont {Kaiser}, \citenamefont {Jiang},\ and\ \citenamefont
  {Alber}}]{wu_periodic_2009}%
  \BibitemOpen
  \bibfield  {author} {\bibinfo {author} {\bibfnamefont {Y.}~\bibnamefont
  {Wu}}, \bibinfo {author} {\bibfnamefont {A.~D.}\ \bibnamefont {Kaiser}},
  \bibinfo {author} {\bibfnamefont {Y.}~\bibnamefont {Jiang}},\ and\ \bibinfo
  {author} {\bibfnamefont {M.~S.}\ \bibnamefont {Alber}},\ }\bibfield  {title}
  {\bibinfo {title} {Periodic reversal of direction allows myxobacteria to
  swarm},\ }\href {https://doi.org/10.1073/pnas.0811662106} {\bibfield
  {journal} {\bibinfo  {journal} {Proc. Natl. Acad. Sci. USA}\ }\textbf
  {\bibinfo {volume} {106}},\ \bibinfo {pages} {1222} (\bibinfo {year}
  {2009})}\BibitemShut {NoStop}%
\bibitem [{\citenamefont {Thutupalli}\ \emph {et~al.}(2015)\citenamefont
  {Thutupalli}, \citenamefont {Sun}, \citenamefont {Bunyak}, \citenamefont
  {Palaniappan},\ and\ \citenamefont {Shaevitz}}]{thutupalli_directional_2015}%
  \BibitemOpen
  \bibfield  {author} {\bibinfo {author} {\bibfnamefont {S.}~\bibnamefont
  {Thutupalli}}, \bibinfo {author} {\bibfnamefont {M.}~\bibnamefont {Sun}},
  \bibinfo {author} {\bibfnamefont {F.}~\bibnamefont {Bunyak}}, \bibinfo
  {author} {\bibfnamefont {K.}~\bibnamefont {Palaniappan}},\ and\ \bibinfo
  {author} {\bibfnamefont {J.~W.}\ \bibnamefont {Shaevitz}},\ }\bibfield
  {title} {\bibinfo {title} {Directional reversals enable \textit{Myxococcus
  xanthus} cells to produce collective one-dimensional streams during
  fruiting-body formation},\ }\href {https://doi.org/10.1098/rsif.2015.0049}
  {\bibfield  {journal} {\bibinfo  {journal} {J. R. Soc. Interface}\ }\textbf
  {\bibinfo {volume} {12}} (\bibinfo {year} {2015})}\BibitemShut {NoStop}%
\bibitem [{\citenamefont {Mu{\~n}oz-Dorado}\ \emph {et~al.}(2016)\citenamefont
  {Mu{\~n}oz-Dorado}, \citenamefont {Marcos-Torres}, \citenamefont
  {Garc{\'\i}a-Bravo}, \citenamefont {Moraleda-Mu{\~n}oz},\ and\ \citenamefont
  {P{\'e}rez}}]{munoz2016myxobacteria}%
  \BibitemOpen
  \bibfield  {author} {\bibinfo {author} {\bibfnamefont {J.}~\bibnamefont
  {Mu{\~n}oz-Dorado}}, \bibinfo {author} {\bibfnamefont {F.~J.}\ \bibnamefont
  {Marcos-Torres}}, \bibinfo {author} {\bibfnamefont {E.}~\bibnamefont
  {Garc{\'\i}a-Bravo}}, \bibinfo {author} {\bibfnamefont {A.}~\bibnamefont
  {Moraleda-Mu{\~n}oz}},\ and\ \bibinfo {author} {\bibfnamefont
  {J.}~\bibnamefont {P{\'e}rez}},\ }\bibfield  {title} {\bibinfo {title}
  {Myxobacteria: Moving, killing, feeding, and surviving together},\ }\href
  {https://doi.org/10.3389/fmicb.2016.00781} {\bibfield  {journal} {\bibinfo
  {journal} {Front. Microbiol.}\ }\textbf {\bibinfo {volume} {7}},\ \bibinfo
  {pages} {203017} (\bibinfo {year} {2016})}\BibitemShut {NoStop}%
\bibitem [{\citenamefont {Barbara}\ and\ \citenamefont
  {Mitchell}(2003)}]{barbara_bacterial_2003}%
  \BibitemOpen
  \bibfield  {author} {\bibinfo {author} {\bibfnamefont {G.~M.}\ \bibnamefont
  {Barbara}}\ and\ \bibinfo {author} {\bibfnamefont {J.~G.}\ \bibnamefont
  {Mitchell}},\ }\bibfield  {title} {\bibinfo {title} {Bacterial tracking of
  motile algae},\ }\href
  {https://doi.org/http://dx.doi.org/10.1016/S0168-6496(02)00452-X} {\bibfield
  {journal} {\bibinfo  {journal} {FEMS Microbiol. Ecol.}\ }\textbf {\bibinfo
  {volume} {44}},\ \bibinfo {pages} {79} (\bibinfo {year} {2003})}\BibitemShut
  {NoStop}%
\bibitem [{\citenamefont {Be'er}\ \emph {et~al.}(2013)\citenamefont {Be'er},
  \citenamefont {Strain}, \citenamefont {Hern{\'a}ndez}, \citenamefont
  {Ben-Jacob},\ and\ \citenamefont {Florin}}]{beer_periodic_2013}%
  \BibitemOpen
  \bibfield  {author} {\bibinfo {author} {\bibfnamefont {A.}~\bibnamefont
  {Be'er}}, \bibinfo {author} {\bibfnamefont {S.~K.}\ \bibnamefont {Strain}},
  \bibinfo {author} {\bibfnamefont {R.~A.}\ \bibnamefont {Hern{\'a}ndez}},
  \bibinfo {author} {\bibfnamefont {E.}~\bibnamefont {Ben-Jacob}},\ and\
  \bibinfo {author} {\bibfnamefont {E.-L.}\ \bibnamefont {Florin}},\ }\bibfield
   {title} {\bibinfo {title} {Periodic reversals in \textit{Paenibacillus
  dendritiformis} swarming},\ }\href {https://doi.org/10.1128/jb.00080-13}
  {\bibfield  {journal} {\bibinfo  {journal} {J. Bacteriol.}\ }\textbf
  {\bibinfo {volume} {195}},\ \bibinfo {pages} {2709} (\bibinfo {year}
  {2013})}\BibitemShut {NoStop}%
\bibitem [{\citenamefont {Duffy}\ and\ \citenamefont
  {Ford}(1997)}]{duffy_turn_1997}%
  \BibitemOpen
  \bibfield  {author} {\bibinfo {author} {\bibfnamefont {K.~J.}\ \bibnamefont
  {Duffy}}\ and\ \bibinfo {author} {\bibfnamefont {R.~M.}\ \bibnamefont
  {Ford}},\ }\bibfield  {title} {\bibinfo {title} {Turn angle and run time
  distributions characterize swimming behavior for \textit{Pseudomonas
  putida}},\ }\href {https://doi.org/10.1128/jb.179.4.1428-1430.1997}
  {\bibfield  {journal} {\bibinfo  {journal} {J. Bacteriol.}\ }\textbf
  {\bibinfo {volume} {179}},\ \bibinfo {pages} {1428} (\bibinfo {year}
  {1997})}\BibitemShut {NoStop}%
\bibitem [{\citenamefont {Davis}\ \emph {et~al.}(2011)\citenamefont {Davis},
  \citenamefont {Mounteer}, \citenamefont {Stevens}, \citenamefont {Miller},\
  and\ \citenamefont {Zhou}}]{davis_2d_2011}%
  \BibitemOpen
  \bibfield  {author} {\bibinfo {author} {\bibfnamefont {M.~L.}\ \bibnamefont
  {Davis}}, \bibinfo {author} {\bibfnamefont {L.~C.}\ \bibnamefont {Mounteer}},
  \bibinfo {author} {\bibfnamefont {L.~K.}\ \bibnamefont {Stevens}}, \bibinfo
  {author} {\bibfnamefont {C.~D.}\ \bibnamefont {Miller}},\ and\ \bibinfo
  {author} {\bibfnamefont {A.}~\bibnamefont {Zhou}},\ }\bibfield  {title}
  {\bibinfo {title} {2d motility tracking of \textit{Pseudomonas putida}
  {KT2440} in growth phases using video microscopy},\ }\href
  {https://doi.org/10.1016/j.jbiosc.2011.01.007} {\bibfield  {journal}
  {\bibinfo  {journal} {J. Biosci. Bioeng.}\ }\textbf {\bibinfo {volume}
  {111}},\ \bibinfo {pages} {605} (\bibinfo {year} {2011})}\BibitemShut
  {NoStop}%
\bibitem [{\citenamefont {Theves}\ \emph {et~al.}(2013)\citenamefont {Theves},
  \citenamefont {Taktikos}, \citenamefont {Zaburdaev}, \citenamefont {Stark},\
  and\ \citenamefont {Beta}}]{theves_bacterial_2013}%
  \BibitemOpen
  \bibfield  {author} {\bibinfo {author} {\bibfnamefont {M.}~\bibnamefont
  {Theves}}, \bibinfo {author} {\bibfnamefont {J.}~\bibnamefont {Taktikos}},
  \bibinfo {author} {\bibfnamefont {V.}~\bibnamefont {Zaburdaev}}, \bibinfo
  {author} {\bibfnamefont {H.}~\bibnamefont {Stark}},\ and\ \bibinfo {author}
  {\bibfnamefont {C.}~\bibnamefont {Beta}},\ }\bibfield  {title} {\bibinfo
  {title} {A bacterial swimmer with two alternating speeds of propagation},\
  }\href {https://doi.org/http://dx.doi.org/10.1016/j.bpj.2013.08.047}
  {\bibfield  {journal} {\bibinfo  {journal} {Biophys. J.}\ }\textbf {\bibinfo
  {volume} {105}},\ \bibinfo {pages} {1915} (\bibinfo {year}
  {2013})}\BibitemShut {NoStop}%
\bibitem [{\citenamefont {Alirezaeizanjani}\ \emph {et~al.}(2020)\citenamefont
  {Alirezaeizanjani}, \citenamefont {Gro\ss{}mann}, \citenamefont {Pfeifer},
  \citenamefont {Hintsche},\ and\ \citenamefont
  {Beta}}]{alirezaei_chemotaxis_2020}%
  \BibitemOpen
  \bibfield  {author} {\bibinfo {author} {\bibfnamefont {Z.}~\bibnamefont
  {Alirezaeizanjani}}, \bibinfo {author} {\bibfnamefont {R.}~\bibnamefont
  {Gro\ss{}mann}}, \bibinfo {author} {\bibfnamefont {V.}~\bibnamefont
  {Pfeifer}}, \bibinfo {author} {\bibfnamefont {M.}~\bibnamefont {Hintsche}},\
  and\ \bibinfo {author} {\bibfnamefont {C.}~\bibnamefont {Beta}},\ }\bibfield
  {title} {\bibinfo {title} {Chemotaxis strategies of bacteria with multiple
  run modes},\ }\href {https://doi.org/10.1126/sciadv.aaz6153} {\bibfield
  {journal} {\bibinfo  {journal} {Sci. Adv.}\ }\textbf {\bibinfo {volume}
  {6}},\ \bibinfo {pages} {eaaz6153} (\bibinfo {year} {2020})}\BibitemShut
  {NoStop}%
\bibitem [{\citenamefont {Bhattacharjee}\ \emph {et~al.}(2021)\citenamefont
  {Bhattacharjee}, \citenamefont {Amchin}, \citenamefont {Ott}, \citenamefont
  {Kratz},\ and\ \citenamefont {Datta}}]{bhattacharjee2021chemotactic}%
  \BibitemOpen
  \bibfield  {author} {\bibinfo {author} {\bibfnamefont {T.}~\bibnamefont
  {Bhattacharjee}}, \bibinfo {author} {\bibfnamefont {D.~B.}\ \bibnamefont
  {Amchin}}, \bibinfo {author} {\bibfnamefont {J.~A.}\ \bibnamefont {Ott}},
  \bibinfo {author} {\bibfnamefont {F.}~\bibnamefont {Kratz}},\ and\ \bibinfo
  {author} {\bibfnamefont {S.~S.}\ \bibnamefont {Datta}},\ }\bibfield  {title}
  {\bibinfo {title} {Chemotactic migration of bacteria in porous media},\
  }\href {https://doi.org/https://doi.org/10.1016/j.bpj.2021.05.012} {\bibfield
   {journal} {\bibinfo  {journal} {Biophys. J.}\ }\textbf {\bibinfo {volume}
  {120}},\ \bibinfo {pages} {3483} (\bibinfo {year} {2021})}\BibitemShut
  {NoStop}%
\bibitem [{\citenamefont {Miyaguchi}\ \emph {et~al.}(2016)\citenamefont
  {Miyaguchi}, \citenamefont {Akimoto},\ and\ \citenamefont
  {Yamamoto}}]{miyaguchi2016langevin}%
  \BibitemOpen
  \bibfield  {author} {\bibinfo {author} {\bibfnamefont {T.}~\bibnamefont
  {Miyaguchi}}, \bibinfo {author} {\bibfnamefont {T.}~\bibnamefont {Akimoto}},\
  and\ \bibinfo {author} {\bibfnamefont {E.}~\bibnamefont {Yamamoto}},\
  }\bibfield  {title} {\bibinfo {title} {Langevin equation with fluctuating
  diffusivity: A two-state model},\ }\href
  {https://link.aps.org/doi/10.1103/PhysRevE.94.012109} {\bibfield  {journal}
  {\bibinfo  {journal} {Phys. Rev. E}\ }\textbf {\bibinfo {volume} {94}},\
  \bibinfo {pages} {012109} (\bibinfo {year} {2016})}\BibitemShut {NoStop}%
\bibitem [{\citenamefont {Miyaguchi}\ \emph {et~al.}(2019)\citenamefont
  {Miyaguchi}, \citenamefont {Uneyama},\ and\ \citenamefont
  {Akimoto}}]{miyaguchi2019brownian}%
  \BibitemOpen
  \bibfield  {author} {\bibinfo {author} {\bibfnamefont {T.}~\bibnamefont
  {Miyaguchi}}, \bibinfo {author} {\bibfnamefont {T.}~\bibnamefont {Uneyama}},\
  and\ \bibinfo {author} {\bibfnamefont {T.}~\bibnamefont {Akimoto}},\
  }\bibfield  {title} {\bibinfo {title} {Brownian motion with alternately
  fluctuating diffusivity: Stretched-exponential and power-law relaxation},\
  }\href {https://doi.org/10.1103/PhysRevE.100.012116} {\bibfield  {journal}
  {\bibinfo  {journal} {Phys. Rev. E}\ }\textbf {\bibinfo {volume} {100}},\
  \bibinfo {pages} {012116} (\bibinfo {year} {2019})}\BibitemShut {NoStop}%
\bibitem [{\citenamefont {Metzler}\ \emph {et~al.}(2014)\citenamefont
  {Metzler}, \citenamefont {Jeon}, \citenamefont {Cherstvy},\ and\
  \citenamefont {Barkai}}]{metzler2014anomalous}%
  \BibitemOpen
  \bibfield  {author} {\bibinfo {author} {\bibfnamefont {R.}~\bibnamefont
  {Metzler}}, \bibinfo {author} {\bibfnamefont {J.-H.}\ \bibnamefont {Jeon}},
  \bibinfo {author} {\bibfnamefont {A.~G.}\ \bibnamefont {Cherstvy}},\ and\
  \bibinfo {author} {\bibfnamefont {E.}~\bibnamefont {Barkai}},\ }\bibfield
  {title} {\bibinfo {title} {Anomalous diffusion models and their properties:
  non-stationarity, non-ergodicity, and ageing at the centenary of single
  particle tracking},\ }\href {https://doi.org/10.1039/C4CP03465A} {\bibfield
  {journal} {\bibinfo  {journal} {Phys. Chem. Chem. Phys.}\ }\textbf {\bibinfo
  {volume} {16}},\ \bibinfo {pages} {24128} (\bibinfo {year}
  {2014})}\BibitemShut {NoStop}%
\bibitem [{\citenamefont {Doetsch}(1947)}]{doetsch_tabellen_1947}%
  \BibitemOpen
  \bibfield  {author} {\bibinfo {author} {\bibfnamefont {G.}~\bibnamefont
  {Doetsch}},\ }\href@noop {} {\emph {\bibinfo {title} {Tabellen zur
  Laplace-Transformation und Anleitung zum Gebrauch}}},\ Die Grundlehren der
  mathematischen Wissenschaften in Einzeldarstellungen mit besonderer
  Ber{\"u}cksichtigung der Anwendungsgebiete\ (\bibinfo  {publisher}
  {Springer},\ \bibinfo {year} {1947})\BibitemShut {NoStop}%
\bibitem [{\citenamefont {Burov}\ \emph {et~al.}(2010)\citenamefont {Burov},
  \citenamefont {Metzler},\ and\ \citenamefont {Barkai}}]{burov_aging_2010}%
  \BibitemOpen
  \bibfield  {author} {\bibinfo {author} {\bibfnamefont {S.}~\bibnamefont
  {Burov}}, \bibinfo {author} {\bibfnamefont {R.}~\bibnamefont {Metzler}},\
  and\ \bibinfo {author} {\bibfnamefont {E.}~\bibnamefont {Barkai}},\
  }\bibfield  {title} {\bibinfo {title} {Aging and nonergodicity beyond the
  {K}hinchin theorem},\ }\href {https://doi.org/10.1073/pnas.1003693107}
  {\bibfield  {journal} {\bibinfo  {journal} {Proc. Natl. Acad. Sci. USA}\
  }\textbf {\bibinfo {volume} {107}},\ \bibinfo {pages} {13228} (\bibinfo
  {year} {2010})}\BibitemShut {NoStop}%
\bibitem [{\citenamefont {Barkai}\ \emph {et~al.}(2012)\citenamefont {Barkai},
  \citenamefont {Garini},\ and\ \citenamefont {Metzler}}]{barkai_2012_strange}%
  \BibitemOpen
  \bibfield  {author} {\bibinfo {author} {\bibfnamefont {E.}~\bibnamefont
  {Barkai}}, \bibinfo {author} {\bibfnamefont {Y.}~\bibnamefont {Garini}},\
  and\ \bibinfo {author} {\bibfnamefont {R.}~\bibnamefont {Metzler}},\
  }\bibfield  {title} {\bibinfo {title} {{Strange kinetics of single molecules
  in living cells}},\ }\href {https://doi.org/10.1063/PT.3.1677} {\bibfield
  {journal} {\bibinfo  {journal} {Phys. Today}\ }\textbf {\bibinfo {volume}
  {65}},\ \bibinfo {pages} {29} (\bibinfo {year} {2012})}\BibitemShut {NoStop}%
\bibitem [{\citenamefont {Mikhailov}\ and\ \citenamefont
  {Meink\"ohn}(1997)}]{mikhailov_self_1997}%
  \BibitemOpen
  \bibfield  {author} {\bibinfo {author} {\bibfnamefont {A.}~\bibnamefont
  {Mikhailov}}\ and\ \bibinfo {author} {\bibfnamefont {D.}~\bibnamefont
  {Meink\"ohn}},\ }\bibfield  {title} {\bibinfo {title} {Self-motion in
  physico-chemical systems far from thermal equilibrium},\ }in\ \href
  {https://doi.org/10.1007/BFb0105621} {\emph {\bibinfo {booktitle} {Stochastic
  Dynamics}}},\ \bibinfo {series} {Lecture Notes in Physics}, Vol.\ \bibinfo
  {volume} {484},\ \bibinfo {editor} {edited by\ \bibinfo {editor}
  {\bibfnamefont {L.}~\bibnamefont {Schimansky-Geier}}\ and\ \bibinfo {editor}
  {\bibfnamefont {T.}~\bibnamefont {P\"oschel}}}\ (\bibinfo  {publisher}
  {Springer Berlin Heidelberg},\ \bibinfo {year} {1997})\ pp.\ \bibinfo {pages}
  {334--345}\BibitemShut {NoStop}%
\bibitem [{\citenamefont {Gro\ss{}mann}\ \emph {et~al.}(2015)\citenamefont
  {Gro\ss{}mann}, \citenamefont {Peruani},\ and\ \citenamefont
  {B{\"a}r}}]{grossmann_anistropic_2015}%
  \BibitemOpen
  \bibfield  {author} {\bibinfo {author} {\bibfnamefont {R.}~\bibnamefont
  {Gro\ss{}mann}}, \bibinfo {author} {\bibfnamefont {F.}~\bibnamefont
  {Peruani}},\ and\ \bibinfo {author} {\bibfnamefont {M.}~\bibnamefont
  {B{\"a}r}},\ }\bibfield  {title} {\bibinfo {title} {A geometric approach to
  self-propelled motion in isotropic \& anisotropic environments},\ }\href
  {https://doi.org/10.1140/epjst/e2015-02465-0} {\bibfield  {journal} {\bibinfo
   {journal} {Eur. Phys. J.: Spec. Top.}\ }\textbf {\bibinfo {volume} {224}},\
  \bibinfo {pages} {1377} (\bibinfo {year} {2015})}\BibitemShut {NoStop}%
\bibitem [{Note1()}]{Note1}%
  \BibitemOpen
  \bibinfo {note} {Higher order moments, like the kurtosis of the displacement
  distribution, will, in contrast, depend on details of the distribution of
  reorientation angles; similar arguments apply to the structure of
  trajectories.}\BibitemShut {Stop}%
\bibitem [{\citenamefont {Gardiner}(2009)}]{gardiner_stochastic_2009}%
  \BibitemOpen
  \bibfield  {author} {\bibinfo {author} {\bibfnamefont {C.}~\bibnamefont
  {Gardiner}},\ }\href@noop {} {\emph {\bibinfo {title} {Stochastic Methods: A
  Handbook for the Natural and Social Sciences}}},\ Springer Series in
  Synergetics\ (\bibinfo  {publisher} {Springer},\ \bibinfo {year}
  {2009})\BibitemShut {NoStop}%
\bibitem [{\citenamefont {Lovely}\ and\ \citenamefont
  {Dahlquist}(1975)}]{lovely1975statistical}%
  \BibitemOpen
  \bibfield  {author} {\bibinfo {author} {\bibfnamefont {P.~S.}\ \bibnamefont
  {Lovely}}\ and\ \bibinfo {author} {\bibfnamefont {F.}~\bibnamefont
  {Dahlquist}},\ }\bibfield  {title} {\bibinfo {title} {Statistical measures of
  bacterial motility and chemotaxis},\ }\href
  {https://doi.org/10.1016/0022-5193(75)90094-6} {\bibfield  {journal}
  {\bibinfo  {journal} {J. Theor. Biol.}\ }\textbf {\bibinfo {volume} {50}},\
  \bibinfo {pages} {477} (\bibinfo {year} {1975})}\BibitemShut {NoStop}%
\bibitem [{\citenamefont {F{\"u}rth}(1920)}]{fuerth_brownsche_1920}%
  \BibitemOpen
  \bibfield  {author} {\bibinfo {author} {\bibfnamefont {R.}~\bibnamefont
  {F{\"u}rth}},\ }\bibfield  {title} {\bibinfo {title} {Die {B}rownsche
  {B}ewegung bei {B}er{\"u}cksichtigung einer {P}ersistenz der
  {B}ewegungsrichtung. {M}it {A}nwendungen auf die {B}ewegung lebender
  {I}nfusorien},\ }\href {https://doi.org/10.1007/BF01328731} {\bibfield
  {journal} {\bibinfo  {journal} {Z. Phys.}\ }\textbf {\bibinfo {volume} {2}},\
  \bibinfo {pages} {244} (\bibinfo {year} {1920})}\BibitemShut {NoStop}%
\bibitem [{Note2()}]{Note2}%
  \BibitemOpen
  \bibinfo {note} {This argument presumes small rotational
  diffusion~$D_{\varphi }^{(r)}$.}\BibitemShut {Stop}%
\bibitem [{not({\natexlab{b}})}]{noteStrato}%
  \BibitemOpen
  \bibinfo {note} {The Langevin equation~\eqref{eqn:e_nd} with multiplicative
  noise has to be interpreted in the sense of
  Stratonovich~\cite{grossmann_anistropic_2015}.}\BibitemShut {Stop}%
\bibitem [{bei()}]{beier2024mot}%
  \BibitemOpen
  \href@noop {} {}\bibinfo {note} {A. Datta, S. Beier, V. Pfeifer, R.
  Gro\ss{}mann, and C. Beta, Intermittent run motility of bacteria in gels
  exhibits power-law distributed dwell times,
  \href{https://arxiv.org/abs/2408.02317}{arXiv:2408.02317 (2024)}}\BibitemShut
  {NoStop}%
\bibitem [{\citenamefont {He}\ \emph {et~al.}(2016)\citenamefont {He},
  \citenamefont {Song}, \citenamefont {Su}, \citenamefont {Geng}, \citenamefont
  {Ackerson}, \citenamefont {Peng},\ and\ \citenamefont
  {Tong}}]{he_dynamic_2016}%
  \BibitemOpen
  \bibfield  {author} {\bibinfo {author} {\bibfnamefont {W.}~\bibnamefont
  {He}}, \bibinfo {author} {\bibfnamefont {H.}~\bibnamefont {Song}}, \bibinfo
  {author} {\bibfnamefont {Y.}~\bibnamefont {Su}}, \bibinfo {author}
  {\bibfnamefont {L.}~\bibnamefont {Geng}}, \bibinfo {author} {\bibfnamefont
  {B.~J.}\ \bibnamefont {Ackerson}}, \bibinfo {author} {\bibfnamefont
  {H.}~\bibnamefont {Peng}},\ and\ \bibinfo {author} {\bibfnamefont
  {P.}~\bibnamefont {Tong}},\ }\bibfield  {title} {\bibinfo {title} {Dynamic
  heterogeneity and non-{G}aussian statistics for acetylcholine receptors on
  live cell membrane},\ }\href {https://doi.org/10.1038/ncomms11701} {\bibfield
   {journal} {\bibinfo  {journal} {Nat. Commun.}\ }\textbf {\bibinfo {volume}
  {7}},\ \bibinfo {pages} {11701} (\bibinfo {year} {2016})}\BibitemShut
  {NoStop}%
\bibitem [{\citenamefont {Lampo}\ \emph {et~al.}(2017)\citenamefont {Lampo},
  \citenamefont {Stylianidou}, \citenamefont {Backlund}, \citenamefont
  {Wiggins},\ and\ \citenamefont {Spakowitz}}]{lamp_cytoplasmic_2017}%
  \BibitemOpen
  \bibfield  {author} {\bibinfo {author} {\bibfnamefont {T.~J.}\ \bibnamefont
  {Lampo}}, \bibinfo {author} {\bibfnamefont {S.}~\bibnamefont {Stylianidou}},
  \bibinfo {author} {\bibfnamefont {M.~P.}\ \bibnamefont {Backlund}}, \bibinfo
  {author} {\bibfnamefont {P.~A.}\ \bibnamefont {Wiggins}},\ and\ \bibinfo
  {author} {\bibfnamefont {A.~J.}\ \bibnamefont {Spakowitz}},\ }\bibfield
  {title} {\bibinfo {title} {Cytoplasmic {RNA}-protein particles exhibit
  non-{G}aussian subdiffusive behavior},\ }\href
  {https://doi.org/https://doi.org/10.1016/j.bpj.2016.11.3208} {\bibfield
  {journal} {\bibinfo  {journal} {Biophys. J.}\ }\textbf {\bibinfo {volume}
  {112}},\ \bibinfo {pages} {532} (\bibinfo {year} {2017})}\BibitemShut
  {NoStop}%
\bibitem [{\citenamefont {Gro\ss{}mann}\ \emph {et~al.}(2024)\citenamefont
  {Gro\ss{}mann}, \citenamefont {Bort}, \citenamefont {Moldenhawer},
  \citenamefont {Stange}, \citenamefont {Panah}, \citenamefont {Metzler},\ and\
  \citenamefont {Beta}}]{grossmann2024nongaussian}%
  \BibitemOpen
  \bibfield  {author} {\bibinfo {author} {\bibfnamefont {R.}~\bibnamefont
  {Gro\ss{}mann}}, \bibinfo {author} {\bibfnamefont {L.~S.}\ \bibnamefont
  {Bort}}, \bibinfo {author} {\bibfnamefont {T.}~\bibnamefont {Moldenhawer}},
  \bibinfo {author} {\bibfnamefont {M.}~\bibnamefont {Stange}}, \bibinfo
  {author} {\bibfnamefont {S.~S.}\ \bibnamefont {Panah}}, \bibinfo {author}
  {\bibfnamefont {R.}~\bibnamefont {Metzler}},\ and\ \bibinfo {author}
  {\bibfnamefont {C.}~\bibnamefont {Beta}},\ }\bibfield  {title} {\bibinfo
  {title} {Non-{G}aussian displacements in active transport on a carpet of
  motile cells},\ }\href {https://doi.org/10.1103/PhysRevLett.132.088301}
  {\bibfield  {journal} {\bibinfo  {journal} {Phys. Rev. Lett.}\ }\textbf
  {\bibinfo {volume} {132}},\ \bibinfo {pages} {088301} (\bibinfo {year}
  {2024})}\BibitemShut {NoStop}%
\bibitem [{\citenamefont {Martens}\ \emph {et~al.}(2012)\citenamefont
  {Martens}, \citenamefont {Angelani}, \citenamefont {Di~Leonardo},\ and\
  \citenamefont {Bocquet}}]{martens2012probability}%
  \BibitemOpen
  \bibfield  {author} {\bibinfo {author} {\bibfnamefont {K.}~\bibnamefont
  {Martens}}, \bibinfo {author} {\bibfnamefont {L.}~\bibnamefont {Angelani}},
  \bibinfo {author} {\bibfnamefont {R.}~\bibnamefont {Di~Leonardo}},\ and\
  \bibinfo {author} {\bibfnamefont {L.}~\bibnamefont {Bocquet}},\ }\bibfield
  {title} {\bibinfo {title} {Probability distributions for the run-and-tumble
  bacterial dynamics: {A}n analogy to the {L}orentz model},\ }\href
  {https://doi.org/10.1140/epje/i2012-12084-y} {\bibfield  {journal} {\bibinfo
  {journal} {Eur. Phys. J. E}\ }\textbf {\bibinfo {volume} {35}},\ \bibinfo
  {pages} {84} (\bibinfo {year} {2012})}\BibitemShut {NoStop}%
\bibitem [{\citenamefont {Zeitz}\ \emph {et~al.}(2017)\citenamefont {Zeitz},
  \citenamefont {Wolff},\ and\ \citenamefont {Stark}}]{zeitz2017active}%
  \BibitemOpen
  \bibfield  {author} {\bibinfo {author} {\bibfnamefont {M.}~\bibnamefont
  {Zeitz}}, \bibinfo {author} {\bibfnamefont {K.}~\bibnamefont {Wolff}},\ and\
  \bibinfo {author} {\bibfnamefont {H.}~\bibnamefont {Stark}},\ }\bibfield
  {title} {\bibinfo {title} {Active {B}rownian particles moving in a random
  {L}orentz gas},\ }\href {https://doi.org/10.1140/epje/i2017-11510-0}
  {\bibfield  {journal} {\bibinfo  {journal} {Eur. Phys. J. E}\ }\textbf
  {\bibinfo {volume} {40}},\ \bibinfo {pages} {23} (\bibinfo {year}
  {2017})}\BibitemShut {NoStop}%
\bibitem [{\citenamefont {Doi}\ and\ \citenamefont
  {Edwards}(1988)}]{doi1988theory}%
  \BibitemOpen
  \bibfield  {author} {\bibinfo {author} {\bibfnamefont {M.}~\bibnamefont
  {Doi}}\ and\ \bibinfo {author} {\bibfnamefont {S.~F.}\ \bibnamefont
  {Edwards}},\ }\href@noop {} {\emph {\bibinfo {title} {The theory of polymer
  dynamics}}},\ \bibinfo {series} {International Series of Monographs on
  Physics}, Vol.~\bibinfo {volume} {73}\ (\bibinfo  {publisher} {Oxford
  University Press},\ \bibinfo {year} {1988})\BibitemShut {NoStop}%
\bibitem [{\citenamefont {Godr\`{e}che}\ and\ \citenamefont
  {Luck}(2001)}]{godreche_statistics_2001}%
  \BibitemOpen
  \bibfield  {author} {\bibinfo {author} {\bibfnamefont {C.}~\bibnamefont
  {Godr\`{e}che}}\ and\ \bibinfo {author} {\bibfnamefont {J.}~\bibnamefont
  {Luck}},\ }\bibfield  {title} {\bibinfo {title} {Statistics of the occupation
  time of renewal processes},\ }\href {https://doi.org/10.1023/A:1010364003250}
  {\bibfield  {journal} {\bibinfo  {journal} {J. Stat. Phys.}\ }\textbf
  {\bibinfo {volume} {104}},\ \bibinfo {pages} {489} (\bibinfo {year}
  {2001})}\BibitemShut {NoStop}%
\bibitem [{\citenamefont {Dubner}\ and\ \citenamefont
  {Abate}(1968)}]{dubner1968numerical}%
  \BibitemOpen
  \bibfield  {author} {\bibinfo {author} {\bibfnamefont {H.}~\bibnamefont
  {Dubner}}\ and\ \bibinfo {author} {\bibfnamefont {J.}~\bibnamefont {Abate}},\
  }\bibfield  {title} {\bibinfo {title} {Numerical inversion of {L}aplace
  transforms by relating them to the finite {F}ourier cosine transform},\
  }\href {https://doi.org/10.1145/321439.321446} {\bibfield  {journal}
  {\bibinfo  {journal} {J. ACM}\ }\textbf {\bibinfo {volume} {15}},\ \bibinfo
  {pages} {115} (\bibinfo {year} {1968})}\BibitemShut {NoStop}%
\bibitem [{\citenamefont {Glasserman}\ and\ \citenamefont
  {Ruiz-Mata}(2006)}]{glasserman2006computing}%
  \BibitemOpen
  \bibfield  {author} {\bibinfo {author} {\bibfnamefont {P.}~\bibnamefont
  {Glasserman}}\ and\ \bibinfo {author} {\bibfnamefont {J.}~\bibnamefont
  {Ruiz-Mata}},\ }\bibfield  {title} {\bibinfo {title} {Computing the credit
  loss distribution in the {G}aussian copula model: A comparison of methods},\
  }\href {https://doi.org/10.21314/JCR.2006.057} {\bibfield  {journal}
  {\bibinfo  {journal} {J. Credit Risk}\ }\textbf {\bibinfo {volume} {2}},\
  \bibinfo {pages} {33} (\bibinfo {year} {2006})}\BibitemShut {NoStop}%
\bibitem [{\citenamefont {Cohen}\ \emph {et~al.}(2000)\citenamefont {Cohen},
  \citenamefont {Villegas},\ and\ \citenamefont
  {Zagier}}]{cohen2000convergence}%
  \BibitemOpen
  \bibfield  {author} {\bibinfo {author} {\bibfnamefont {H.}~\bibnamefont
  {Cohen}}, \bibinfo {author} {\bibfnamefont {F.~R.}\ \bibnamefont
  {Villegas}},\ and\ \bibinfo {author} {\bibfnamefont {D.}~\bibnamefont
  {Zagier}},\ }\bibfield  {title} {\bibinfo {title} {Convergence acceleration
  of alternating series},\ }\href
  {https://doi.org/10.1080/10586458.2000.10504632} {\bibfield  {journal}
  {\bibinfo  {journal} {Exp. Math.}\ }\textbf {\bibinfo {volume} {9}},\
  \bibinfo {pages} {3} (\bibinfo {year} {2000})}\BibitemShut {NoStop}%
\end{thebibliography}
%

\end{document}